\documentclass[]{svjour3}                     

\smartqed  
\usepackage{graphicx}
\usepackage{enumerate}

\newcommand{\chandra}{{\it Chandra}}
\newcommand{\xmm}{{\it XMM-Newton}}
\newcommand{\rosat}{{\it ROSAT}}
\newcommand{\planck}{{\it Planck}}
\newcommand{\asca}{{\it ASCA}}
\newcommand{\sax}{{\it BeppoSAX}}
\newcommand{\suzaku}{{\it Suzaku}}
\newcommand{\wmap}{{\it WMAP}}
\newcommand{\aap}{{A\&A}}
\newcommand{\aaps}{{A\&A Supp}}
\newcommand{\apj}{{ApJ}}
\newcommand{\apjl}{{ApJ Lett}}
\newcommand{\mnras}{{MNRAS}}
\newcommand{\pasj}{{PASJ}}

\begin{document}

\title{Mass profiles of Galaxy Clusters from X-ray analysis}

\author{Stefano Ettori \and Annamaria Donnarumma \and Etienne Pointecouteau \and Thomas H. Reiprich 
\and Stefania Giodini \and Lorenzo Lovisari \and Robert W. Schmidt 
}

\authorrunning{S. Ettori et al.} 

\institute{S. Ettori \at
INAF-Osservatorio Astronomico, via Ranzani 1, 40127 Bologna, Italy \\
INFN, Sezione di Bologna, viale Berti Pichat 6/2, 40127 Bologna, Italy
\and
A. Donnarumma \at
INAF-Osservatorio Astronomico, via Ranzani 1, 40127 Bologna, Italy \\
Dipartimento di Astronomia, Universit\`a di Bologna, via Ranzani 1, I-40127 Bologna, Italy
\and 
E. Pointecouteau  \at
Universit\'e de Toulouse, UPS-OMP, IRAP, F-31028 Toulouse cedex 4, France \\
CNRS, IRAP, 9 Av. colonel Roche, BP 44346, F-31028 Toulouse 9	cedex 4, France
\and
T. H. Reiprich,  L. Lovisari \at
Argelander Institute for Astronomy, Bonn University, Auf dem H\"ugel 71, 53121 Bonn, Germany
\and
S. Giodini \at 
Leiden Observatory, Leiden University, PO Box 9513, 2300 RA Leiden, the Netherlands
\and
 R. W. Schmidt \at
Astronomisches Rechen-Institut, Zentrum f\"ur Astronomie der Universit\"at Heidelberg, M\"onchhofstrasse 12-14, 69120 Heidelberg, Germany
}

\date{Accepted: March 2013}

\maketitle

\begin{abstract}
We review the methods adopted to reconstruct the mass profiles in X-ray luminous galaxy clusters. 
We discuss the limitations and the biases affecting these measurements and how these mass profiles can be used as cosmological proxies.
\end{abstract}

\section{Introduction to the Clusters of Galaxies}

It was in the early thirties that the role of a {\it missing mass}
to explain the gravitational effect observed in rich clusters
of galaxies was highlighted from Zwicky (1933, 1937), opening
the still-debated issue on how to relate their bounding
mass to their observables.

Since 1950s (Abell 1958, Zwicky et al. 1961-68, Abell et al. 1989),
galaxy clusters have been characterized from
galaxies overdensities in the optical bands.
Therefore, the {\it Richness} in number of galaxies,
the total luminosity of cluster galaxies $L_{\rm opt}$,
the velocity dispersion of member galaxies,
and the shear and strong lensing features
induced from the mass distribution of intervening galaxy clusters
on background galaxies have been the tools
to measure the mass and the distribution of galaxy clusters
in systems at relatively low redshift ($z<0.3$;
see, e.g., the MaxBCG catalog from the Sloan Digital Sky Survey
and the constraints it provided on the cosmological
parameters in Rozo et al. 2010).

At higher redshifts, the field galaxy population overwhelms
galaxy overdensities associated with clusters, in particular
when a single filter is adopted for the detection.
An efficient way to counteract this effect is
to observe in the near-infrared bands ($>1 \mu m$).
The cores of the clusters of galaxies are dominated by red, early-type galaxies
(at least out to $z\sim1.4$). Moreover, being
the number counts of the field galaxy population flatter
in the near-IR bands than in the optical, by moving to z, J, H, and K bands,
one can progressively compensate the strong $K$-correction
and enhance the contrast of (red) cluster galaxies against
the background (blue) galaxy distribution
(as clearly demonstrated firstly by Stanford et al. 1997).

Another efficient technique to discover and characterize galaxy clusters is by
mapping the distortion of the Cosmic Microwave Background spectrum
due to the inverse Compton scattering induced from the high-energy
electrons present in the hot intra-cluster medium (ICM).
Clear detections of these features (named Sunyaev-Zeldovich SZ effect; Sunyaev \& Zeldovich 1972)
occurred in the late 90s (see the review by Carlstrom et al. 2002)
and many surveys over wide areas of the sky have started to produce results
(e.g., the South Pole Telescope reported the first SZ-discovered clusters
in Staniszewski et al. 2009; the Atacama Cosmology Telescope has also
recently reported their initial catalog of SZ-discovered clusters
in Marriage et al. 2011; the \planck\ collaboration has presented
the first sample of 189 high signal-to-noise clusters in January 2011 --Planck Collaboration 2011).
The integrated SZ signal, being proportional to the ICM pressure
along the line-of-sight, can be used as proxy of the total
cluster mass.

In the X-ray band, the gas luminosity, temperature and mass
are the direct observables used to infer the cluster total mass.
X-ray observations occurred to be particularly successful because
galaxy clusters appear as well resolved extended emission with a total
luminosity that is proportional to the square of the gas density (see next section).
X-ray detections have started in the 70s with {\it Uhuru} and
{\it HEAO-1} and provided well defined samples thanks
to the improved spectral and spatial resolution capabilities
available to the following generation of satellites,
like {\it Einstein} (Gioia et al. 1990),
{\it EXOSAT} (Edge et al. 1990), \rosat\
(based both on the All-Sky Survey --e.g. Bright Cluster Sample / BCS
in Ebeling et al. 1998, the Northern \rosat\ All-Sky Survey / NORAS
in B\"ohringer et al. 2000, the \rosat-ESO flux limited X-ray / REFLEX
\footnote{with 452 objects selected in the southern hemisphere
with the nominal flux limit of $3 \times 10^{-12}$ erg/s/cm$^{2}$
in the \rosat\ energy band ($0.1-2.4$) keV, it is
the largest compilation of X-ray galaxy clusters to date.}
in B\"ohringer et al. 2001, the Massive Cluster Survey / MACS in Ebeling et al. 2001,
the North Ecliptic Pole / NEP survey in Henry et al. 2001,
the Highest X-ray flux Galaxy Cluster Sample / HIFLUGCS in
Reiprich \& B\"ohringer 2002--
and on archival pointed PSPC observations
--e.g. the RIXOS survey in Castander et al. 1995, the \rosat\
Deep Cluster Survey / RDCS in Rosati et al. 1998, the Serendipitous
High-Redshift Archival \rosat\ Cluster survey / SHARC in Collins et al. 1997,
the Wide Angle \rosat\ Pointed X-Ray Survey of clusters / WARPS in Scharf
et al. 1997, the \rosat\ Optical X-Ray Survey ROXS in Donahue et al. 2001,
the 400 deg$^2$ survey in Burenin et al. 2007).
More recently, the new generation of X-ray observatories with improved sensitivity and
angular resolution, like \xmm\ and \chandra, has allowed deeper studies
of the cluster emission both over large areas of the sky (e.g. XMM-Large Scale Structure
survey, Pacaud et al. 2007), and of serendipitous sources identified in public
exposures (e.g. XMM Cluster survey / XCS, Romer et al. 2001),
and of previously known objects (e.g. Vikhlinin et al. 2009, Mantz et al. 2010).

\subsection{Galaxy clusters in the X-ray band}

The primordial cosmic gas is composed by hydrogen (about 75 per cent by mass),
helium ($\sim 24$ per cent), and traces of other light elements, like
deuterium, helium-3, lithium and berillium.
When this gas collapses into the dark matter halos typical of galaxy clusters
($> 10^{14} M_{\odot}$), it undergoes shocks and
adiabatic compression, reaching densities of about $10^{-3}$ particles
cm$^{-3}$ and temperatures of the order of $10^8$ K.
The density drops at larger radii $r$ approximately as $r^{-2}$. Hence,
under these physical conditions, the plasma is optically thin and 
in ionization equilibrium, where the ionization and emission
processes result mainly from collisions of ions with electrons
(see details in the reviews from Sarazin 1988, Peterson \& Fabian 2006,
B\"ohringer \& Werner 2010).

The continuum intensity is the combination of 3 main processes:
\begin{enumerate}
\item Thermal bremsstrahlung (free-free emission; $\epsilon \sim T_{\rm gas}^{0.5}$)
\item Recombination (free-bound)
\item Two-photon decay of metastable levels
\end{enumerate}
Once the emission from collisionally excited lines ($\epsilon \sim T_{\rm gas}^{-0.5}$)
is considered, the total X-ray emission is
\begin{equation}
\epsilon_{\nu} = \sum_i \Lambda_{\nu}(X_i, T_{\rm gas}) \; n(X_i) n_{\rm e}
\end{equation}
where $ \Lambda_{\nu}(X_i, T_{\rm gas})$ is the cooling function that depends
on the abundance of the ion of the element $X_i$ with density $n(X_i)$ and $n_{\rm e}$ is the electron density.
At $T>3 \times 10^7$ K, where bremsstrahlung dominates, 
$\epsilon \sim 4.4\times 10^{-27}  n_{\rm e}^2 T_{\rm gas}^{0.5}$ erg s$^{-1}$ cm$^{-3}$, whereas
at $10^5 < T < 3 \times 10^7$ K, where line cooling is more relevant,
$\epsilon \sim 9 \times 10^{-19} n_{\rm e}^2 T_{\rm gas}^{-0.5}$ erg s$^{-1}$ cm$^{-3}$.

\begin{figure}
\centerline{
 \hbox{
  \includegraphics[height=.25\textheight]{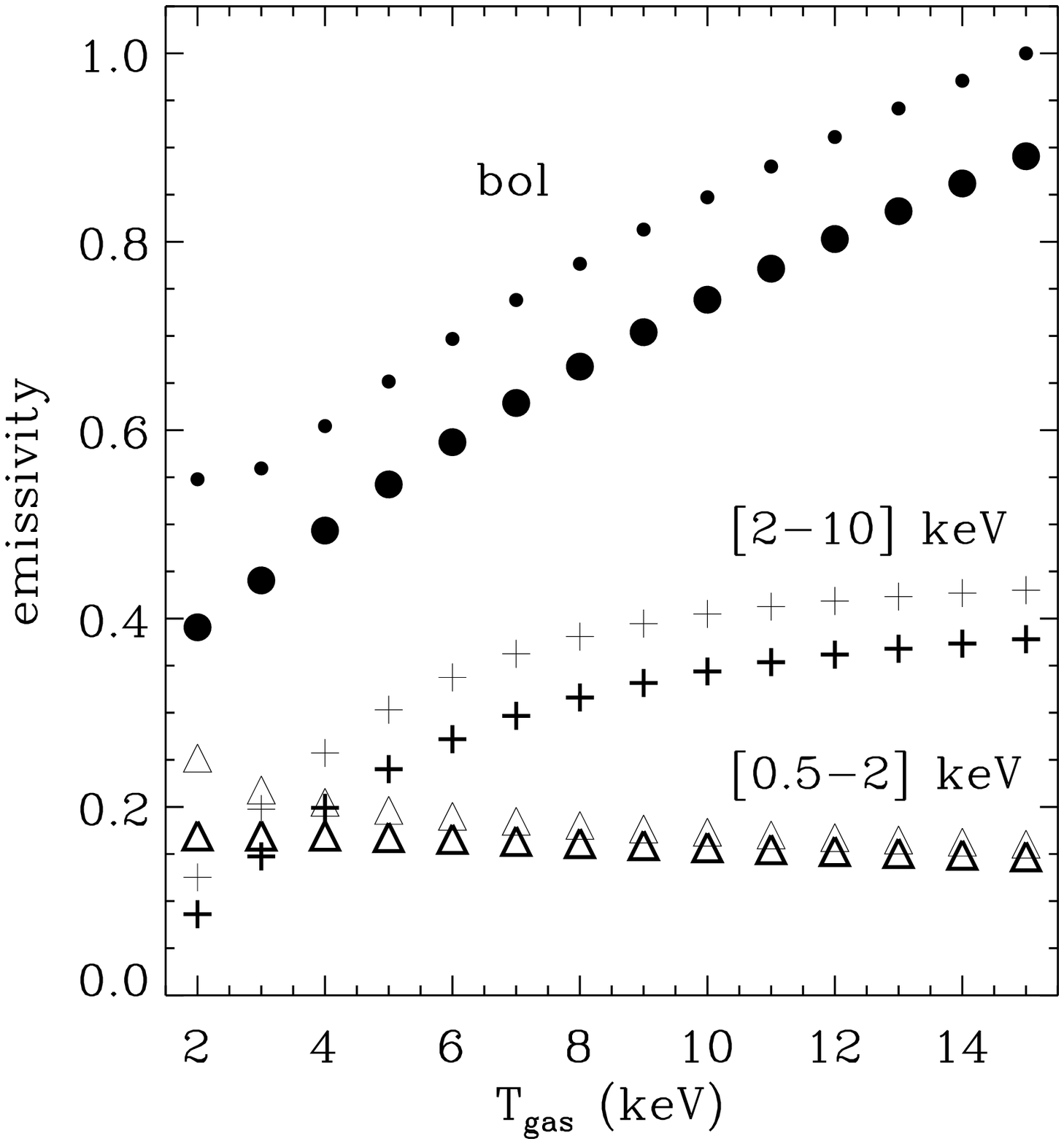}
\hspace*{1.0cm}
  \includegraphics[height=.25\textheight]{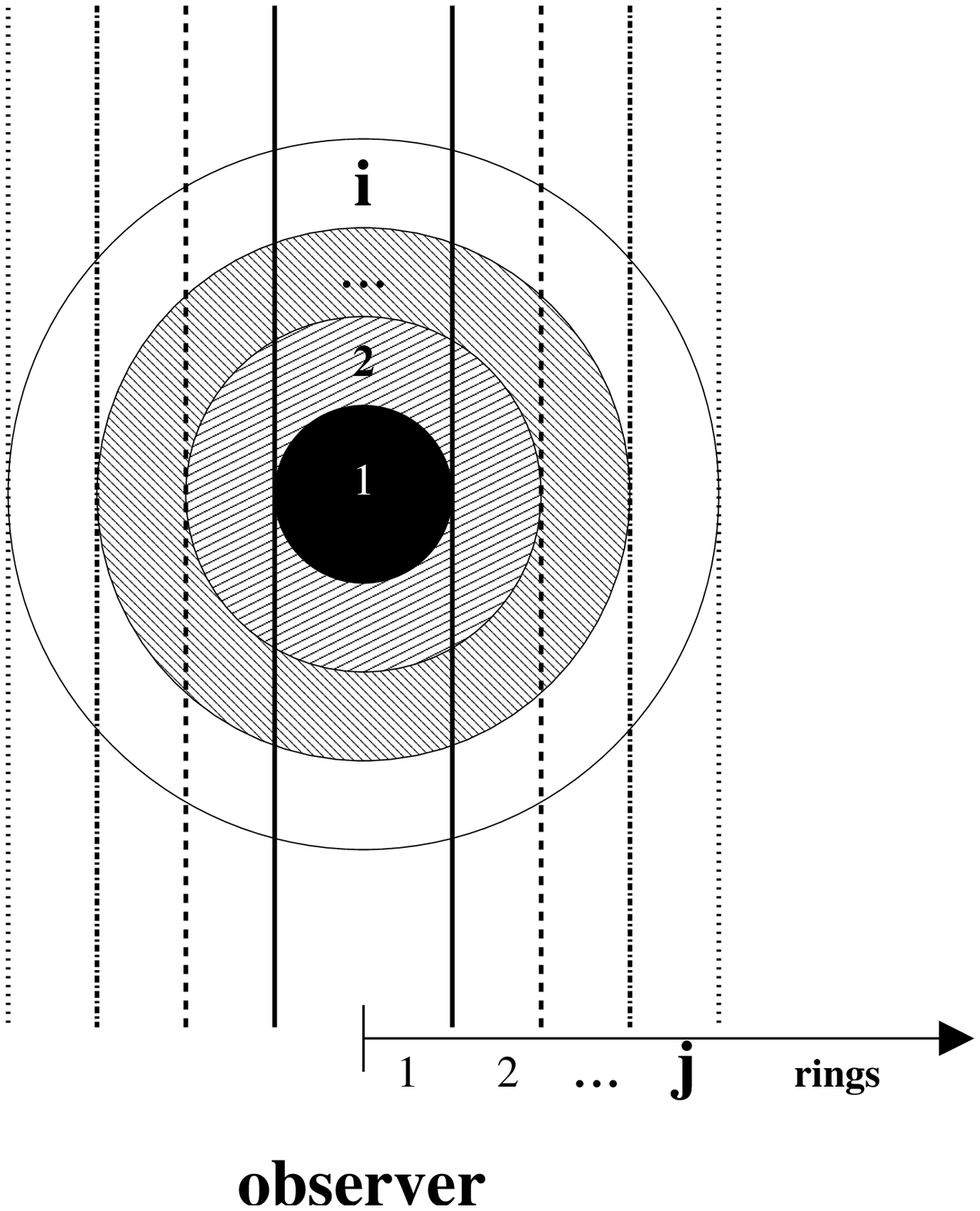}
 }
}
\caption{
{\bf (Left)} Total dependence of the emissivity upon the
temperature in different energy window [{\it bolometric}, 2-10 keV,
0.5-2 keV] and for a cluster whose temperature ranges between 2 and
15 keV (where the bremsstrahlung emission dominates) at redshift 0.1.
The curves are calculated using a MEKAL model (Kaastra 1992,  Liedahl
et al. 1995) in XSPEC (Arnaud 1996)
for two different values of metallicity: 0.3 (thickest symbols) and 1.0
times the solar abundance as in Anders \& Grevesse (1989). 
They are normalized to the bolometric
emissivity at $T_{\rm gas} = 15$ keV with $Z = 1 Z_{\odot}$.
When the free-free radiation dominates, these curves can be calculated
analytically  integrating over the window energy [$E1-E2$] the Gaunt
factor, assumed equal to $0.9 (E/kT)^{-0.3}$, multiplied by
$(kT)^{-0.5} e^{-(E/kT)}$.
{\bf (Right)} 
Relation between the shells and the rings in the geometrical deprojection
(for details see, e.g., Fabian et al. 1981, Kriss et al. 1983, McLaughlin 1999, Buote 2000, Ettori et al. 2002).
Defined $V_{ij}$ as the amount of the volume, $V_i$, of the shell $i$
observed through the ring $j$ adopted in the spectral/spatial analysis, 
a flux $F_j$ is measured and modelled in each radial ring with area $A_j$ as
$F_j = \sum_{i,{\rm shell}} \epsilon_i V_{ij} / A_j =
 \sum_{i,{\rm shell}} n_i^2 \Lambda(T_i) V_{ij} / A_j$, 
where $\epsilon_i$ is the emissivity in shell $i$, $n_i$ 
is the gas density and $\Lambda(T_i)$ is the cooling function.
} \label{fig:ew} \end{figure}
 
The dependence of the emissivity upon the temperature in different 
energy window of the observer's frame is shown in Fig.~\ref{fig:ew}.
Note that the emissivity $\epsilon = \int_{\nu} \epsilon_{\nu} d\nu$ is almost 
independent from the gas temperature in the soft X-ray band (e.g. 0.5--2 keV), 
where also the signal-to-noise ratio is maximized because of the efficiency of the present X-ray mirrors
and detectors and of the properties of the cosmic and particle background (see discussion in, e.g. Ettori \& Molendi 2011).
At a given electron temperature $T$, the continuum intensity is proportional to the emission measure $EM = \int n_{\rm e} n_{\rm H} dV$.
The line intensity depends on $n_{\rm e} n_{\rm X}$, where $X$ is a given heavy element.
The line-to-continuum ratio is thus proportional to the metallicity of the ICM, 
$n_{\rm X}/n_{\rm H} = Z$.

\begin{figure}
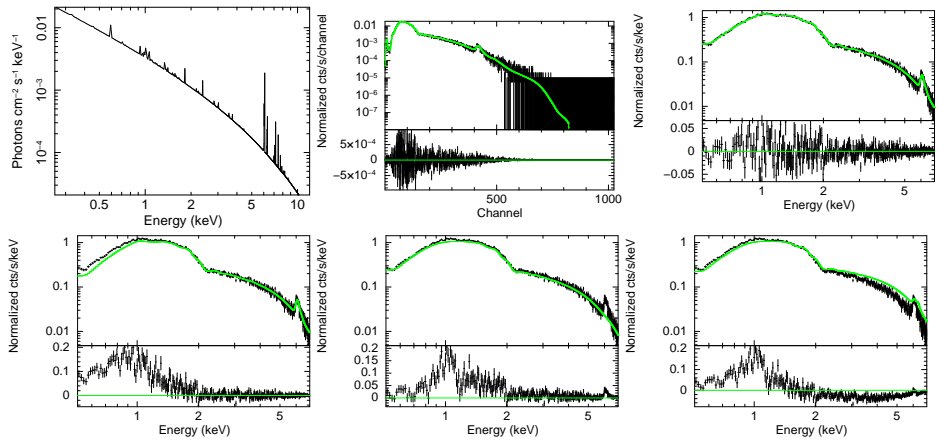

 \vbox{
  \hbox{
  \includegraphics[height=.2\textheight,angle=-90]{mod_theor}
  \includegraphics[height=.205\textheight,angle=-90]{mod_chan}
  \includegraphics[height=.2\textheight,angle=-90]{mod_eneok}
  }
  \hbox{
  \includegraphics[height=.2\textheight,angle=-90]{mod_eneok_nh05}
  \includegraphics[height=.2\textheight,angle=-90]{mod_eneok_a0}
  \includegraphics[height=.2\textheight,angle=-90]{mod_eneok_t10}
  }
 }
\caption{
Examples of X-ray spectra. {\bf From left to right, top}:
(i) model of the X-ray emission associated to a 5 keV cluster with
bolometric $L_X = 6\times 10^{44}$ erg/s at $z=0.1$; (ii) the model
is convolved with the response of a CCD instrument;
(iii) spectrum as a function of the rest-frame energy.
{\bf Bottom}: changing some input values, from 0 to
$5 \times 10^{20}$ cm$^{-2}$ in galactic absorption;
from 0.3 times solar to 0 in the ICM metalliticy;
from 5 to 10 keV. 
} \label{fig:xspec} \end{figure}

Such physical processes produce photons that are registered as, e.g., counts 
on the CCDs, in the case of \chandra\ or \xmm\ exposures. These ``events"
are registered in a file that is first processed along with the calibration files to select only
the {\it events} defined {\it good} accordingly to both the time
when they have been registered and their properties.
After this cleaning process, the file contains all, and only, the {\it events} that can be used to do science 
as a function of time, position on the sky, energy. 
By projecting along one of these axes, one can create light curves (photon counts vs time), 
images (counts vs sky position), spectra (counts vs energy; see, e.g., Fig.~\ref{fig:xspec}).

An X-ray emitting source is characterized by its count-rate (cts/s) at the detector that 
is converted to an un-absorbed flux given a source spectrum, an instrument response
and a model for the absorption due to our Galaxy, and then in luminosity, once the distance is known. 
Given the flux (i.e. the luminosity divided by $4 \pi d_{\rm lum}^2$ 
--see eq.~\ref{eq:dang}-- and multiplied by the $K$-correction factor 
that depends mainly on the temperature of the source and converts the flux 
observed in a given energy band to a quantity in the rest-frame of the object),
the surface brightness is then the flux divided by the area covered from the emitting source.

The spatial and spectral analyses in the X-ray waveband provide, thus,  the observational tools needed
to describe the physical state of the ICM.

As we discuss in the following sections, present observations provide routinely reasonable estimates 
of the gas density, $n_{\rm gas}$, and
temperature, $T_{\rm gas}$, up to about $R_{2500}$ ($\approx 0.3 R_{200}$;
$R_{\Delta}$ is defined as the radius of the sphere that encloses a mean mass density
of $\Delta$ times the critical density at the cluster's redshift;
$R_{200}$ defines approximately the virialized region in galaxy clusters).
Only few cases provide meaningful measurements at $R_{500}$ ($\approx 0.7 R_{200}$) and beyond
(e.g. Vikhlinin et al. 2005, Leccardi \& Molendi 2008a,
Neumann 2005, Ettori \& Balestra 2009, Eckert et al. 2012;
see Reiprich et al. in the present volume).
Consequently, more than two-thirds of the typical cluster volume,
just where primordial gas is accreting and dark matter halo is
forming, is still unknown for what concerns both its mass distribution
and its thermodynamical properties.
Indeed the characterization of thermodynamic properties at large radii
would allow us to provide constraints on the virialization process and, thus,
to improve significantly the measures of the gas and total gravitating masses.
This is one of the main request for a more accurate use of galaxy clusters as cosmological probes.

\subsection{Surface brightness and gas density profiles}\label{sec: sb}

Few tens of counts are needed to have a detection and to estimate the gas density through 
the observed surface brightness, whereas few thousands of net counts are required to measure properly $T_{\rm gas}$.
The X-ray surface brightness is thus a quantity much easier to characterize than
the temperature and it is still rich in physical information being proportional
to the emission measure of the emitting source.

In detail, the observed surface brightness profile, $S_{\rm b}$, at the projected radius $r_p$
is the projection on the sky of the plasma emissivity $\epsilon (r)$,
\begin{equation}
S_{\rm b}(r_p) = \int_{r_p^2}^{\infty} \frac{\epsilon(r) \ dr^2}{\sqrt{r^2 - r_p^2}}.
\label{eq:sb} \end{equation}

The gas density is obtained either from the deprojection (see right panel of 
Fig.~\ref{fig:ew}) or from the modelling of $S_{\rm b}$. 
For instance, assuming isothermality and a $\beta$-model for the gas density 
($n_{\rm gas} = n_{\rm 0,gas} (1+x^2)^{-3 \beta /2}$ where $x= r / r_{\rm c}$ and $r_{\rm c}$ is the core radius; 
Cavaliere \& Fusco-Femiano 1976, Sarazin \& Bahcall 1977),
the surface brightness profile has an analytic solution:
\begin{eqnarray}
S_{\rm b} & = & \sqrt{\pi} n_0^2 r_{\rm c} \Lambda(T_{\rm gas}) 
\frac{\Gamma (3\beta -0.5)}{\Gamma (3\beta) } (1 + x^2)^{0.5 -3 \beta} 
\nonumber \\
 & = & S_0 (1+x^2)^{0.5 -3 \beta},
\label{eq:beta}
\end{eqnarray}
that is strictly valid under the condition that $3 \beta > 0.5$ and
that the cooling function $\Lambda(T_{\rm gas})$ does not change radially.
This functional form has only 3 free parameters ($S_0$, $r_{\rm c}$, $\beta$) that can be fitted
to the observed surface brightness profile to provide a direct characterization of the gas density.
Thanks to the higher sensitivity of the present detectors, more complex, 
and therefore flexible, description of the profile of the electron density 
can be adopted.
For example, we can combine few power-laws and $\beta-$models and use up to 10 free parameters
to model the square of the electron density, that defines the integrand of the emission measure, 
as $n_{\rm e}^2 = n_0^2 x^{-\alpha} (1+x^2)^{\alpha/2 -3 \beta}
(1+x_s^{\gamma})^{\epsilon / \gamma} + n_{02}^2 (1+x_2^2)^{-3 \beta_2}$
(Vikhlinin et al. 2006).

$S_{\rm b}$ has been successfully parametrized through the use of one or more
$\beta-$model and is now routinely deprojected to recover the gas density
profile (e.g. Mohr et al. 1999,  Ettori \& Fabian 1999).
The use of a second $\beta-$model allows, for instance, to properly
trace the gas distribution in the cooling core of X-ray bright objects
\footnote{Galaxy clusters where the cooling has been the most effective thermodynamical
process in defining the properties of the X-ray emitting central region, like a decreasing temperature
and a rising gas density and metallicity moving inwards, are defined as
{\it cool core} (CC) objects and are generally considered as the most relaxed systems.
For a recent discussion on the properties of CC, and NCC, clusters see Rossetti \& Molendi 2010, 
Hudson et al. 2010, Ettori \& Brighenti 2008 and references therein.}.
Mohr et al. (1999) presented a systematic analysis of the ICM surface
brightness in a sample of 45 local systems observed with \rosat\ PSPC.
In reconstructing the ICM properties, they addressed several systematic 
uncertainties, including Poisson noise, choice of cluster emission center, 
PSF blurring, X-ray background subtraction, ICM temperature uncertainties, 
luminosity uncertainties, and cool gas associated with central cooling instabilities.
Overall, the most significant errors are induced from very asymmetric X-ray images 
for which a proper centering becomes problematic (like recent merger
systems as Abell754, where they observed a change in the gas mass by 16\%), 
biases in the background subtraction (deviation by 4\%), 
and PSPC absolute flux calibration (implying uncertainties on $M_{\rm gas}$
of about 7.5\%). 
The fit with a $\beta-$model provides results that are strongly correlated in the 
\{core radius, $\beta$\} plane, with typical values (minimum, maximum, median; 
after the exclusion of the very peaked emission of Cygnus-A)
of $r_{\rm c} = (0.03, 0.47, 0.17) h_{70}^{-1}$ Mpc and $\beta = (0.56, 1.0, 0.78)$.
Similar values are obtained also at higher redshifts ($0.4 < z < 1.3$), 
with median $r_{\rm c}$ of $\sim 0.15 h_{70}^{-1}$ Mpc and $\beta$ between 0.63 and 0.78 (Ettori et al. 2004).

In general, a $\beta-$model provides a good representation of the cluster regions 
outside the core once the $\beta$ value is permitted to increase from $\sim 0.65$ to about 0.9 at the virial radius, 
as consequence of the radial steepening of the gas density profile (see e.g. Eckert et al. 2012).
Other analytical descriptions of the gas density radial profile which follow the  
behaviour of a $\beta-$model at large radii, and that is then integrated numerically to recover
a surface brightness model, are described in, e.g., Ettori, Fabian \& White (1998),
Pratt \& Arnaud (2002).

\begin{figure}[ht]
  \centerline{ \hbox{
    \includegraphics[height=.2\textheight]{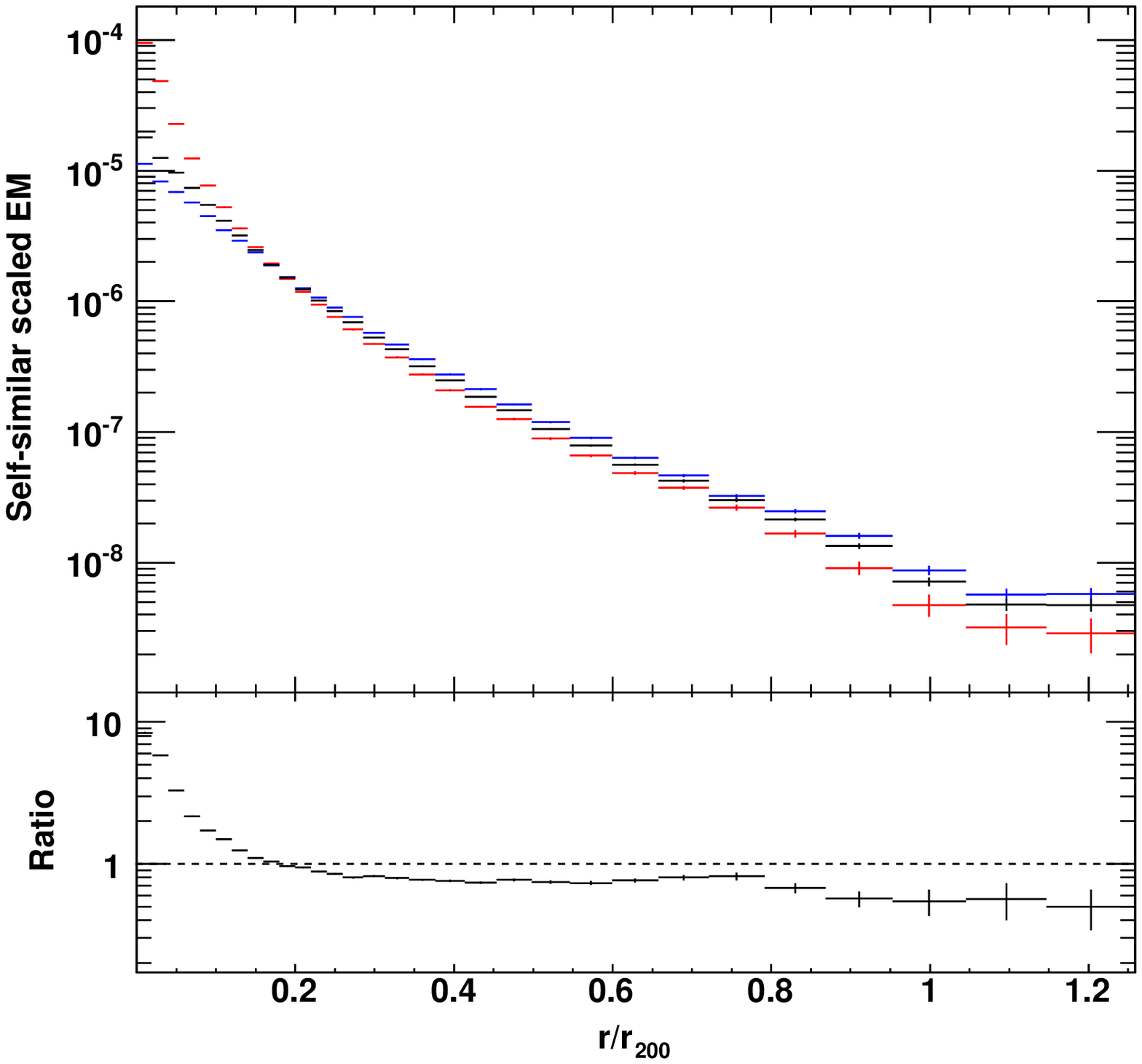} 
    \includegraphics[height=.2\textheight]{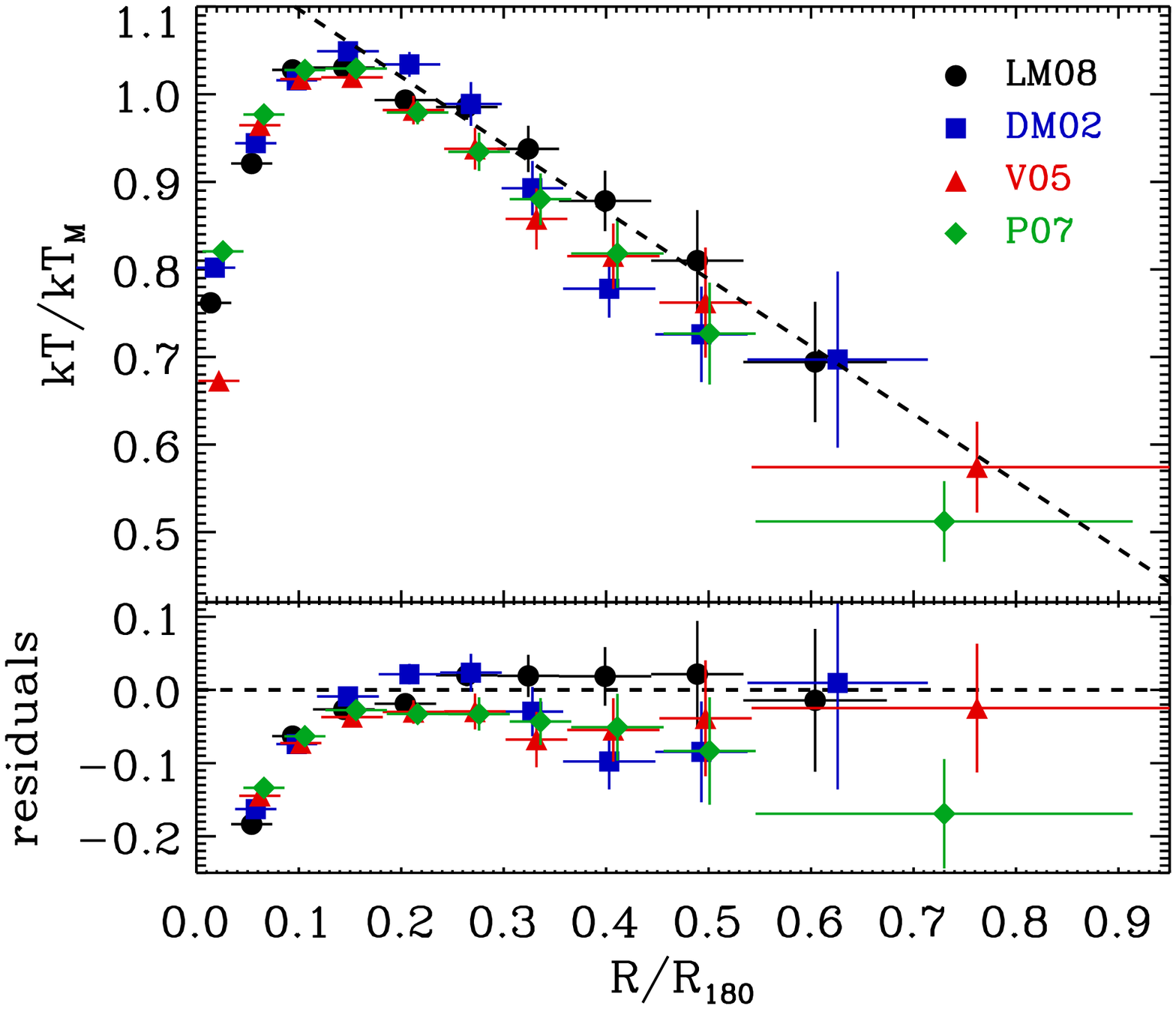}
    \includegraphics[height=.2\textheight]{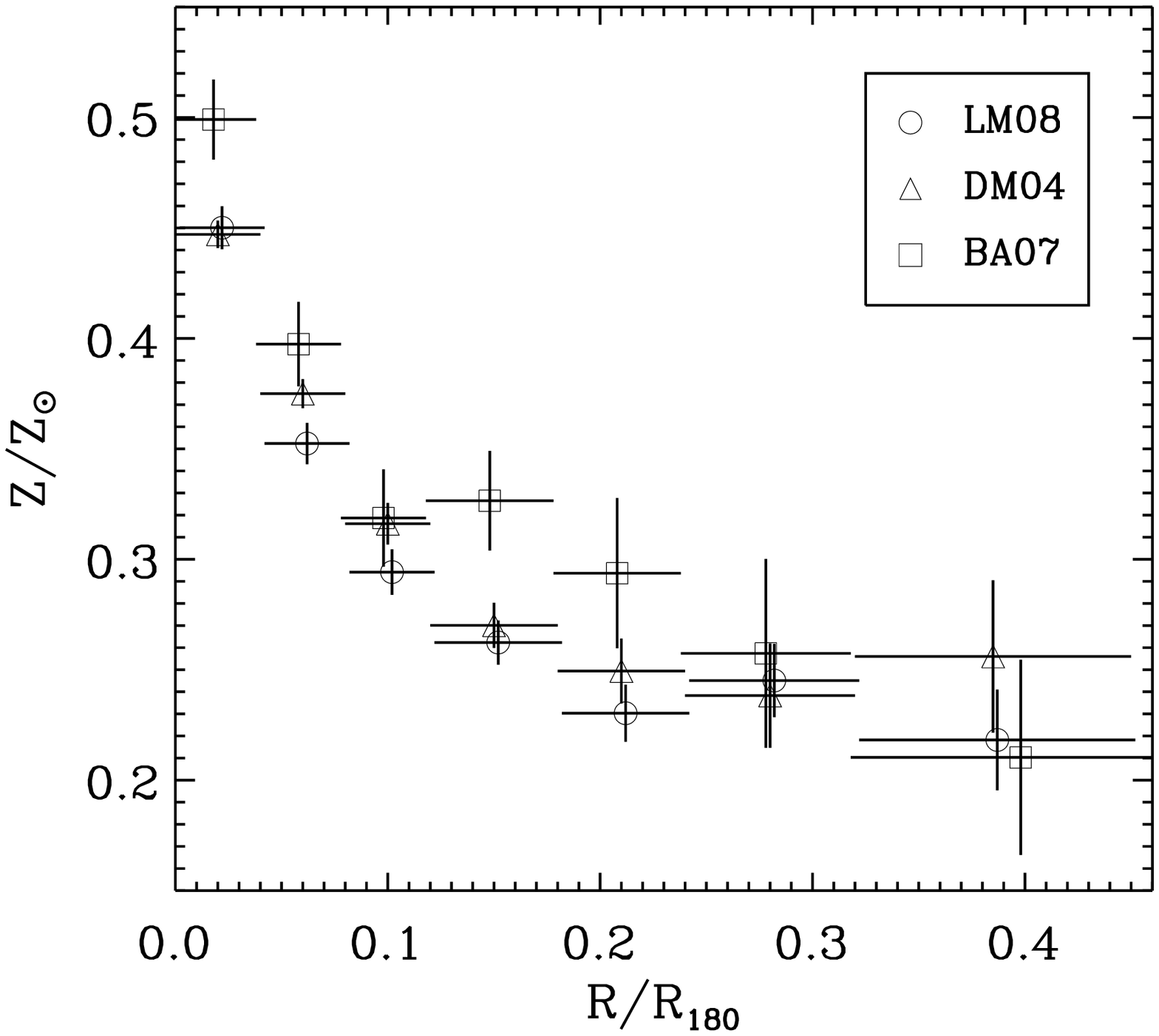}
   } }
\caption{{\bf From left to right}
(i) Stacked emission measure profile (in units of cm$^{-6}$ Mpc) for the entire sample
of 31 nearby galaxy clusters observed with \rosat\ PSPC (black), the Cool Core systems only (red)
and the Non-Cool-Core objects only (blue). From Eckert et al. (2011).
(ii) Mean temperature profiles obtained from Leccardi \& Molendi (2008a;
LM08, black circles), De Grandi \& Molendi (2002; DM02, blue squares),
Vikhlinin et al. (2005; V05, red upward triangles)
and Pratt et al. (2007; P07, green diamonds). All profiles are rescaled by
$\mathrm{k}T_\mathrm{M}$ and $R_{180}$. The dashed line shows the best
fit with a linear model beyond 0.2~$R_{180}$ and is drawn to guide the eye.
The LM08 profile is the flattest one.
(iii) Mean metallicity profiles obtained from Leccardi \& Molendi (2008b; LM08, circles),
De Grandi et al. (2004; DM04, triangle) and Baldi et al. (2007; BA07, squares).
Abundances are expressed in Anders \& Grevesse (1989) solar values.
} \label{fig:sbT} \end{figure}

\subsection{Temperature and metallicity profiles}\label{sec: tprof}

The gas temperature is estimated from the spectrum continuum intensity
(Fig.~\ref{fig:xspec}).

Early attempts to produce temperature profiles  were made with the \rosat\ PSPC,
these were mostly limited to low mass systems (e.g. David et al. 1996) where the
temperatures were within reach of the PSPC soft response.
Resolved spectroscopy of hot systems began with the coming into operation of \asca\ (1994) 
and \sax\ (1996).
Both missions enjoyed a relatively low instrumental background, 
which was a considerable asset when extending measures out to large radii, 
however they both suffered from limited spatially resolution.
The situation was somewhat less severe
with the \sax\ MECS than with the \asca\ GIS since the former had a factor of 2 better angular resolution
and a modest energy dependence in the PSF. 
These difficulties led to substantial differences in temperature measures. 
On the one side, Markevitch et al. (1998) using \asca\ and De Grandi \& Molendi (2002) 
using \sax\ MECS found evidence of declining temperature profiles, while, on the other,
White (2000) using \asca\ and Irwin et al. (1999) using \sax\ data found flat temperature profiles.
The situation was somewhat clearer on abundance profiles were workers using \asca\ 
(e.g. Finoguenov et al. 2000) and \sax\ data (De Grandi \& Molendi 2001) consistently found 
evidence that {\it cool core} (CC) systems featured more centrally peaked profiles than NCC system.
The coming into operation of the second generation of medium energy X-ray telescopes, 
namely \xmm\ and \chandra, both characterized by substantially better spatial resolution, 
allowed more direct measures of the temperature profiles.
The new \chandra\ (Vikhlinin et al. 2005) and \xmm\ measurements (e.g. Pratt et al. 2007, 
Snowden et al. 2008) confirmed the presence of the temperature gradients measured 
with \asca\ and \sax.
In a detailed study of a sample of 44 objects observed with \xmm, Leccardi \& Molendi 
(2008a) found that temperature measurements could be extended out to about 0.7$R_{180}$ 
(see Fig.~\ref{fig:sbT}).

Unfortunately the high orbit of the \xmm\ and \chandra\ satellites, as well as the fact 
that the design of the satellites was driven by scientific objectives other than 
the characterization of low surface brightness regions, led to a substantially higher 
and more variable background than with the previous satellite generation,
thereby limiting the exploration of the temperature and metal abundance profiles to roughly 
the same regions already investigated with \asca\ and \sax\ (see Fig.~\ref{fig:sbT}).
Recently measures of temperature profiles have been made with the \suzaku\ X-ray 
imaging spectrometer (XIS), which benefits from the modest background associated 
to the low earth orbit but suffers from broad and position-dependent PSFs 
with their typical half-power diameter (HPD) of about $110$ arcsec.
These measurements extend beyond the regions explored with \chandra\ and \xmm, even though
only parts of the outermost annuli are explored and both radial bins
and error bars on temperature measurements are still quite large
(see more details and results in Reiprich et al. in the present volume).

Moreover, it has been proved (e.g. Leccardi \& Molendi 2007) that the measure of the ICM temperature from 
an X-ray thermal spectrum extracted from these regions at low surface brightness and in a hard
energy band ($>2$ keV) is biased by the use
of commonly adopted estimators, such as those based on $\chi^2$ and Cash statistics.
An a-posteriori correction can be needed to provide more reliable and robust results.

\section{Total mass, gas mass \& systematics}
\label{sect:mass}

To evaluate the cluster total mass through X-ray observations, one needs to assume that 
the gas is in hydrostatic equilibrium with the gravitational potential. 
This assumptions relies on the facts that (i) the gas can be treated as a 
collisional fluid (in general, the time scales of any heating and/or cooling and/or 
dynamical process is much longer than the elastic collisions time for ions
and electrons, allowing the ICM to be represented by a single kinetic
temperature; the mean free paths of electrons and ions, determined by Coulomb collisions,
are thus shorter than the length scales of interest in galaxy clusters, i.e.
$t_{\rm Coulomb} <<  t_{\rm cooling} \sim t_{\rm heating}$); 
(ii) a sound wave crosses the ICM in a time shorter than the age 
of the cluster itself, i.e. $t_{\rm sound} < t_{\rm age}$, where:
\begin{eqnarray}
t_{\rm Coulomb}  \sim & 0.02 \; T_{\rm gas}^{3/2} \; n_{\rm gas}^{-1} \; {\rm Gyr} \nonumber \\
t_{\rm cooling} \sim & 35 \; T_{\rm gas}^{1/2} \; n_{\rm gas}^{-1/2} \; {\rm Gyr} \nonumber \\
t_{\rm sound} \sim & 4.4 \; T_{\rm gas}^{-1/2} \;  R_{\rm Mpc} \;  {\rm Gyr}  \nonumber \\
t_{\rm age} \sim & H_0^{-1} \sim 13.6 \; {\rm Gyr},
\label{eq:t_scale}
\end{eqnarray}
where $T_{\rm gas}$ is measured in keV and $n_{\rm gas}$ in $10^{-3}$ atoms per cm$^3$.

Satisfied these conditions, the Euler's equation for an ideal fluid (i.e. a fluid in which thermal conductivity and viscosity do not play a relevant role)
in a gravitational potential $\phi$ and with a velocity $\vec{v}$,  pressure $P_{\rm gas}$ and density $\rho_{\rm gas}$ is (Landau \& Lifshitz 1959; Suto et al. 2013)
\begin{equation} 
\frac{\partial \vec{v}}{\partial t} +(\vec{v} \cdot \nabla)\vec{v} = - \frac{1}{\rho_{\rm gas}} \nabla P_{\rm gas} - \nabla \phi.
\label{eq:euler}
\end{equation}
Setting to zero the velocity of the gas and assuming a spherically-symmetric distribution of the gas, 
we ca write the hydrostatic equilibrium equation (HEE) of the ICM as
\begin{equation} 
\frac{1}{\rho_{\rm gas}} \frac{dP_{\rm gas}}{dr} = -\frac{d\phi}{dr} = 
- \frac{G M_{\rm tot}}{r^2},
\label{eq:hee}
\end{equation}
where $G$ is the gravitational constant and the gas mass density and pressure are
related through the perfect gas law, $P_{\rm gas} =\rho_{\rm gas} \, kT_{\rm gas} 
/ (\mu m_{\rm u}) = n_{\rm gas} \, kT_{\rm gas}$, 
$m_{\rm u} = 1.66 \times 10^{-24}$ g is the atomic mass unit, and $\mu$ is  
the mean molecular weight in a.m.u. and is equal to $\approx (2 X +3/4 Y +1/2 Z)^{-1} \sim 0.6$, 
where $X$, $Y$ and $Z$ are the mass fraction in hydrogen, helium and heavier elements, respectively.

The total mass of X-ray luminous galaxy clusters can be estimated
by solving equation~\ref{eq:hee}.
We can rewrite it as a function of the gas density and temperature profiles, that are the quantities observed directly:
\begin{equation}
M_{\rm tot}(<r) = - \frac{kT_{\rm gas}(r) \, r}{\mu m_{\rm u} G} 
 \left( \frac{\partial \log T_{\rm gas}}{\partial \log r} +
   \frac{\partial \log n_{\rm gas}}{\partial \log r} \right).
\label{eq:mhe}
\end{equation}
It is useful to represent a galaxy cluster as a spherical region with a radius $R_{\Delta}$ 
and a mean overdensity $\Delta$
with respect to the critical density at the cluster's redshift $z$, $\rho_{\rm c, z}=3 H_z^2/ (8
\pi G)$ with $H_z = H_0 E(z)$ (see eq.~\ref{eq:ez}):
\begin{equation}
M_{\rm tot}(<R_{\Delta}) = \frac{4}{3} \pi \Delta \rho_{\rm c,z} R_{\Delta}^3.
\label{eq:mdelta}
\end{equation}
The collapsed structure associated to a galaxy cluster is defined at $\Delta=200$, whereas the regions
that can be routinely proved with X-ray observation are typically at lower radii ($\Delta \sim 500$)
(see discussion in the previous section and Fig.~\ref{fig:nfw}).
The mass of the gas within $R_{\Delta}$ is defined as
\begin{equation}
M_{\rm gas}(<R_{\Delta}) = \int_0^{R_{\Delta}} \mu_{\rm e} m_{\rm u} n_{\rm e}(r) \,  4 \pi r^2 dr,
\label{eq:mgas}
\end{equation}
where $n_{\rm e}(r)$ is the electron number density ($n_{\rm gas}$ is the sum of the electron and proton densities $n_{\rm e} +n_{\rm p} \approx 1.826 n_{\rm e}$), $\mu_{\rm e}=\rho_{\rm gas} / (n_{\rm e} \, m_{\rm u}) = 1.155$ is the value associated to a cosmic mix of hydrogen and
helium with 0.3 times solar abundance in the remaining elements with
a relative contribution that follows Grevesse \& Sauval (1998)\footnote{
The solar photospheric abundance in Anders \& Grevesse (1989) have been revised to match 
the meteoritic determinations, as summarized in Grevesse \& Sauval (1998; see further
redeterminations in Asplund et al. 2009), and require
conversion factors of $(0.676, 0.794, 1, 1.321, 1)$ to correct
the original photospheric abundance for Fe, O, Si, S and Ni, respectively.
However, the Anders \& Grevesse values are still largely adopted in the X-ray
clusters community for consistency with previous observational constraints.},
$\mu=0.6$ is the corresponding mean molecular weight, $m_{\rm u}$ is 
the atomic mass unit.
The gas mass fraction is then $f_{\rm gas}(R_{\Delta}) = M_{\rm gas}(<R_{\Delta}) /
M_{\rm tot}(<R_{\Delta})$.

\begin{figure*}
\begin{center}
\hbox{
  \includegraphics[width=0.5\textwidth]{9154fig2.eps}
  \includegraphics[width=0.5\textwidth]{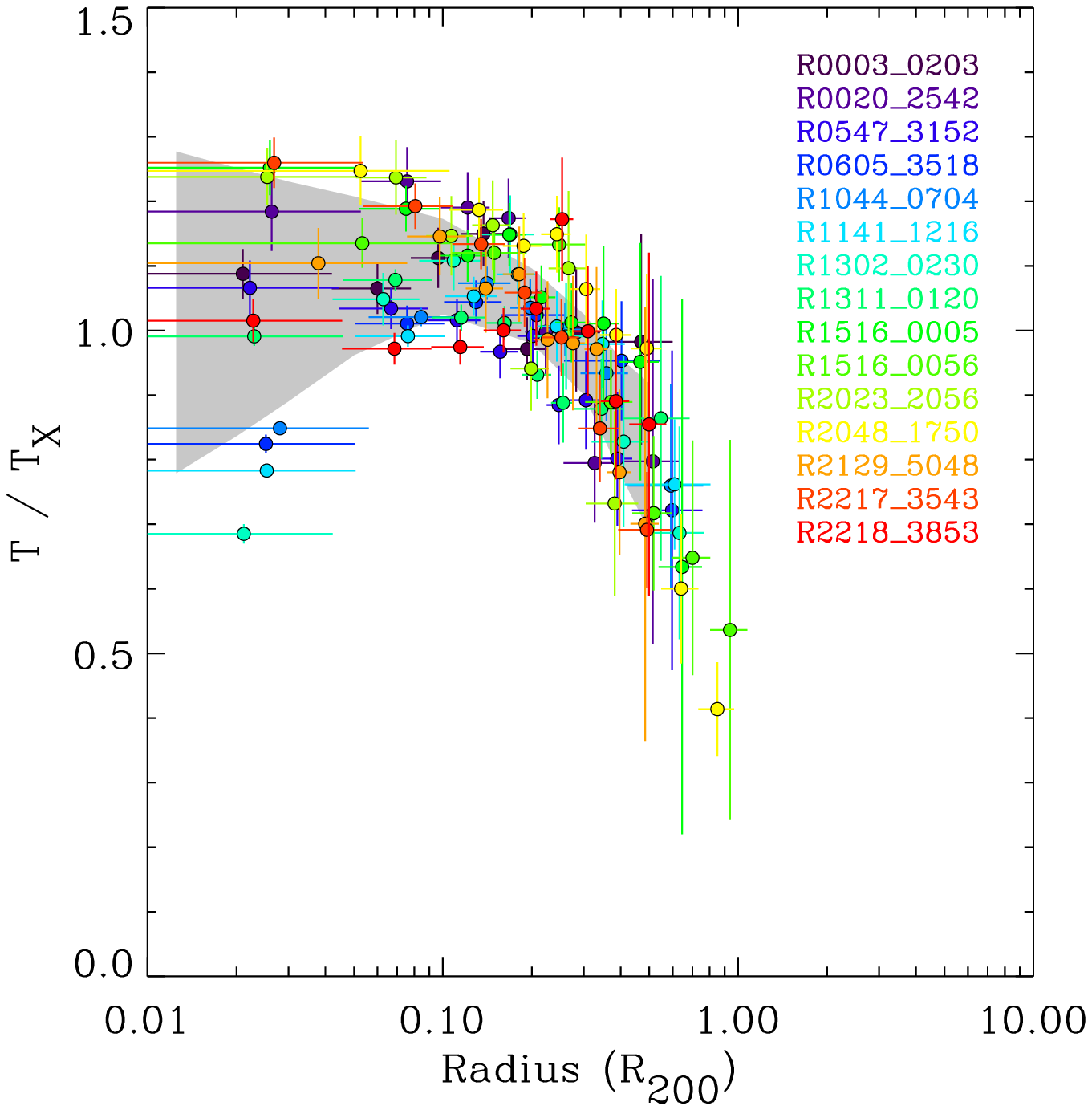}%
}  \includegraphics[width=\textwidth]{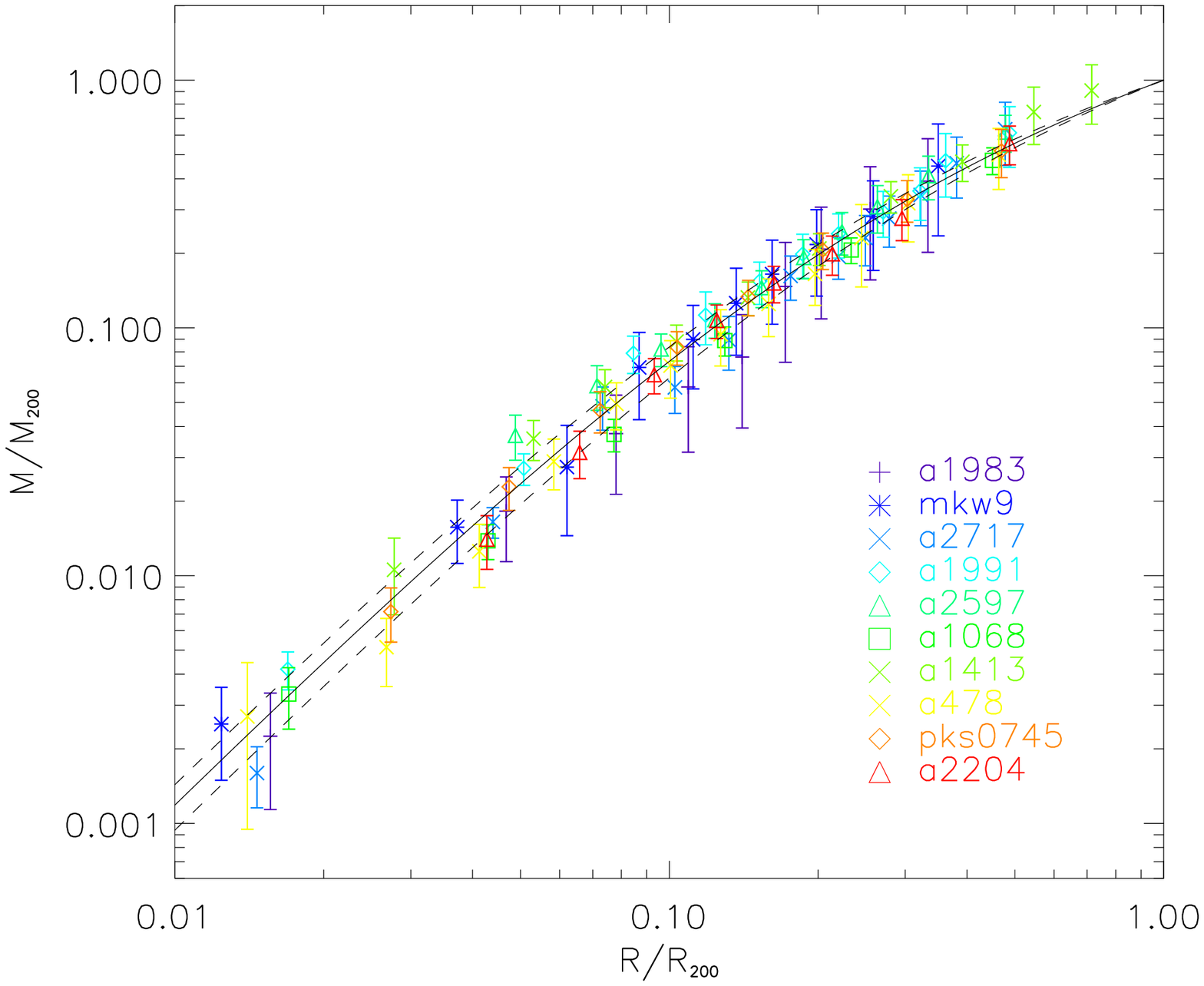}
\end{center}
\caption{{\it (left)} Scaled density profiles of the REXCESS sample (Croston et al. 2008). Using a non-parametric regularised method (Croston et al. 2006), the X-ray surface brightness profile extracted in the low energy band, where the signal-to-noise is higher, is PSF-corrected and deprojected into 3D emission measure profiles. Conversion to a gas density profile is undertaken using the emissivity profile in the energy band under consideration, taking into account the observed temperature profile. {\it (Middle)} Temperature profiles  of  the REXCESS sample from Pratt et al. (2007) normalized to the mean temperature estimated for each cluster. The derived temperature profiles are then deconvolved from the PSF blurring and deprojected  into 3D physical profiles. An analytical functional is then chosen to represent the cluster 3D temperature profile. The deprojected 3D density and temperature profiles are applied to the hydrostatic mass equation (see Democles et al 2010 for details).  {\it (Right)} Scaled mass profiles from 10 relaxed nearby clusters observed with \xmm\ (from Pointecouteau et al. 2005). As for the temperature profile, a Monte Carlo method is used to derive  the errors in each bin were the 3D mass profile is computed, with the prior that cumulative radial 3D mass profile are monotonically increasing. The observed mass profiles can then be fitted by a chosen mass model.
\label{fig:forw}}
\end{figure*}

\begin{figure*}
\hbox{
 \includegraphics[width=0.5\columnwidth]{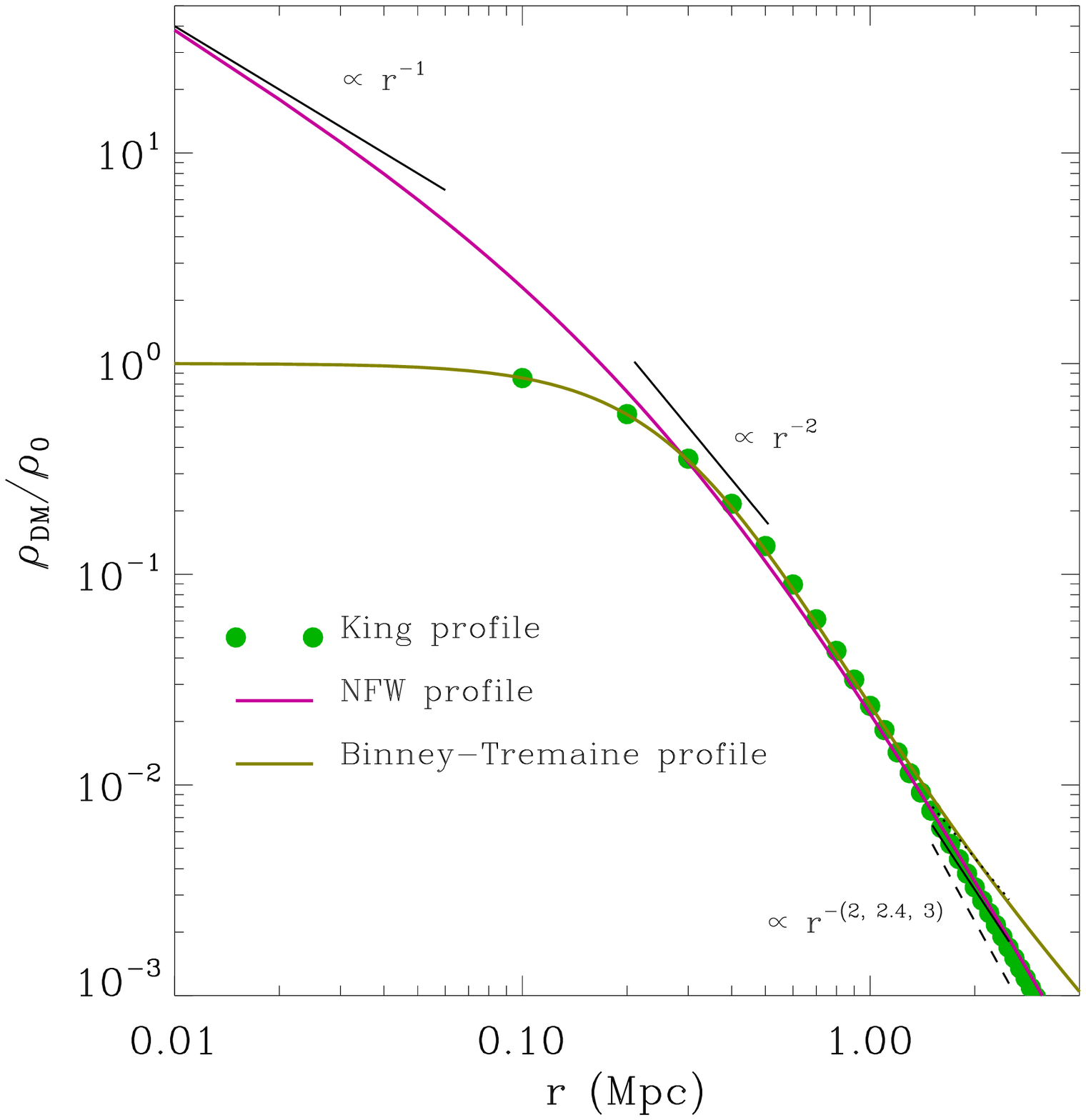}
  \includegraphics[width=0.5\columnwidth]{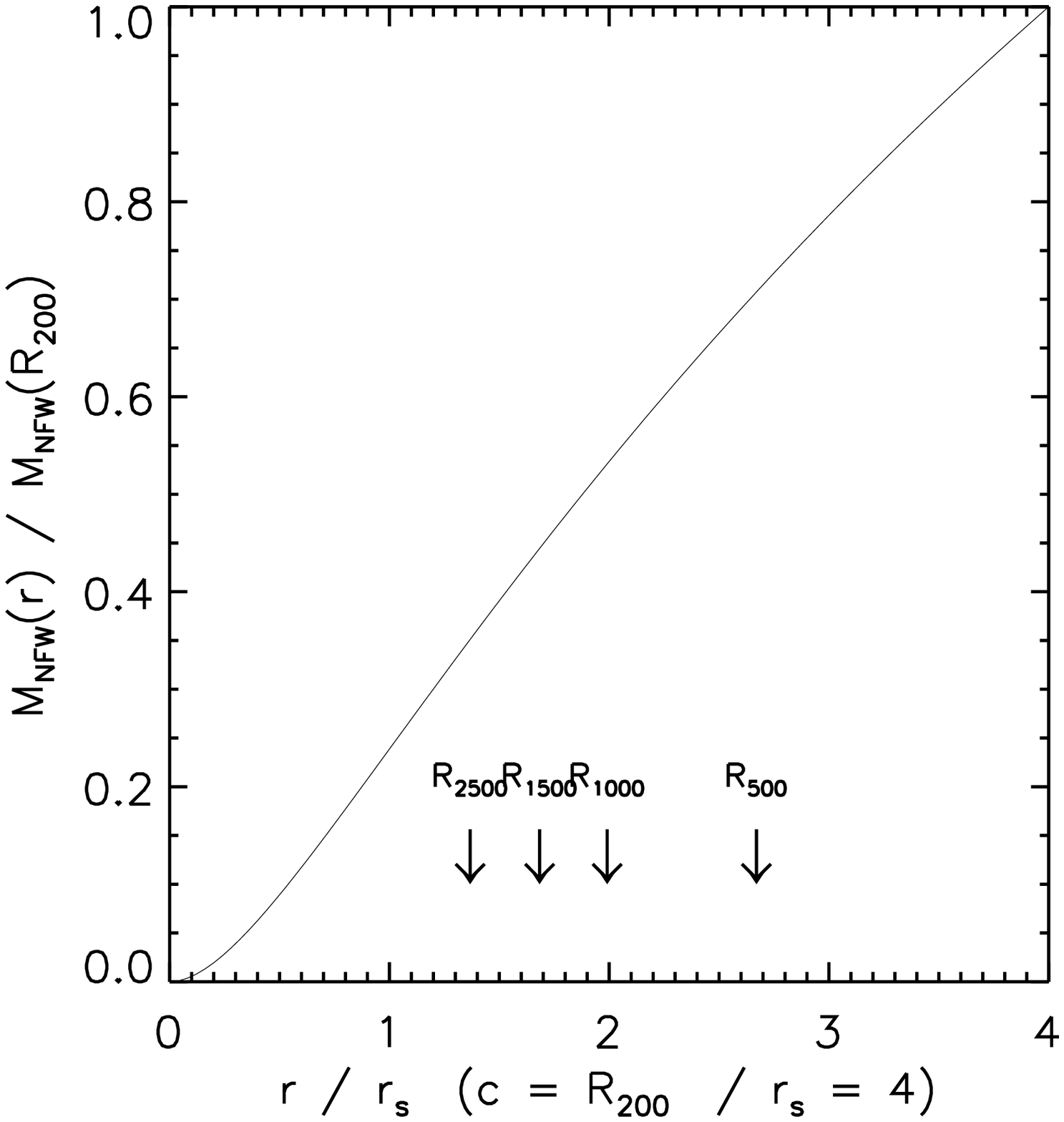}
}
\caption{
{\bf (Left)} The Binney \& Tremaine (from equation 4-125 in Binney \&
Tremaine 1987) dark matter profile for the self-gravitating
isothermal sphere is here compared for different input parameters,
[$\sigma$ (km s$^{-1}$), $r_{\rm c}$ (or $r_{\rm s}$, Mpc)],
to the Navarro-Frenk-White (NFW, 1997) profile that comes from
extended and highly resolved numerical simulations of clusters of galaxies.
Both of these are also compared with the King's approximation (King 1962)
to the inner part of the self-gravitating isothermal sphere.
All of them are normalized to the central value of the self-gravitating
isothermal sphere profile [$\rho_0 = 9 \sigma^2 / (4 \pi G r_{\rm c}^2) = 9.05
\times 10^{-26}$ g cm$^{-3}$]. Inside the core radius, the NFW profile
does not flatten like the BT profile.
In the outer part of the region of interest (above 1 Mpc), agreement
between the two profiles is obtained by increasing the
velocity dispersion and the core radius (or {\it scale radius})
in the NFW profile. Fitting a power law, it can be shown that around
$2.5 \times r_{\rm s} \sim 2$ Mpc
the NFW profile approaches a $r^{-2.4}$ form.
{\bf (Right)} Cumulative fraction of the mass enclosed
within the radius $r / r_{\rm s}$ for a NFW profile and assigned 
concentration of 4, typical for a massive galaxy cluster.
Radii at different overdensities are shown (f.i., the mass
enclosed within $R_{500}$ is about 70\% of $M_{\rm NFW}$ at $R_{200}$).
} \label{fig:nfw} 
\end{figure*}

\begin{figure*}
\vbox{
 \hbox{
 \includegraphics[width=0.45\textwidth]{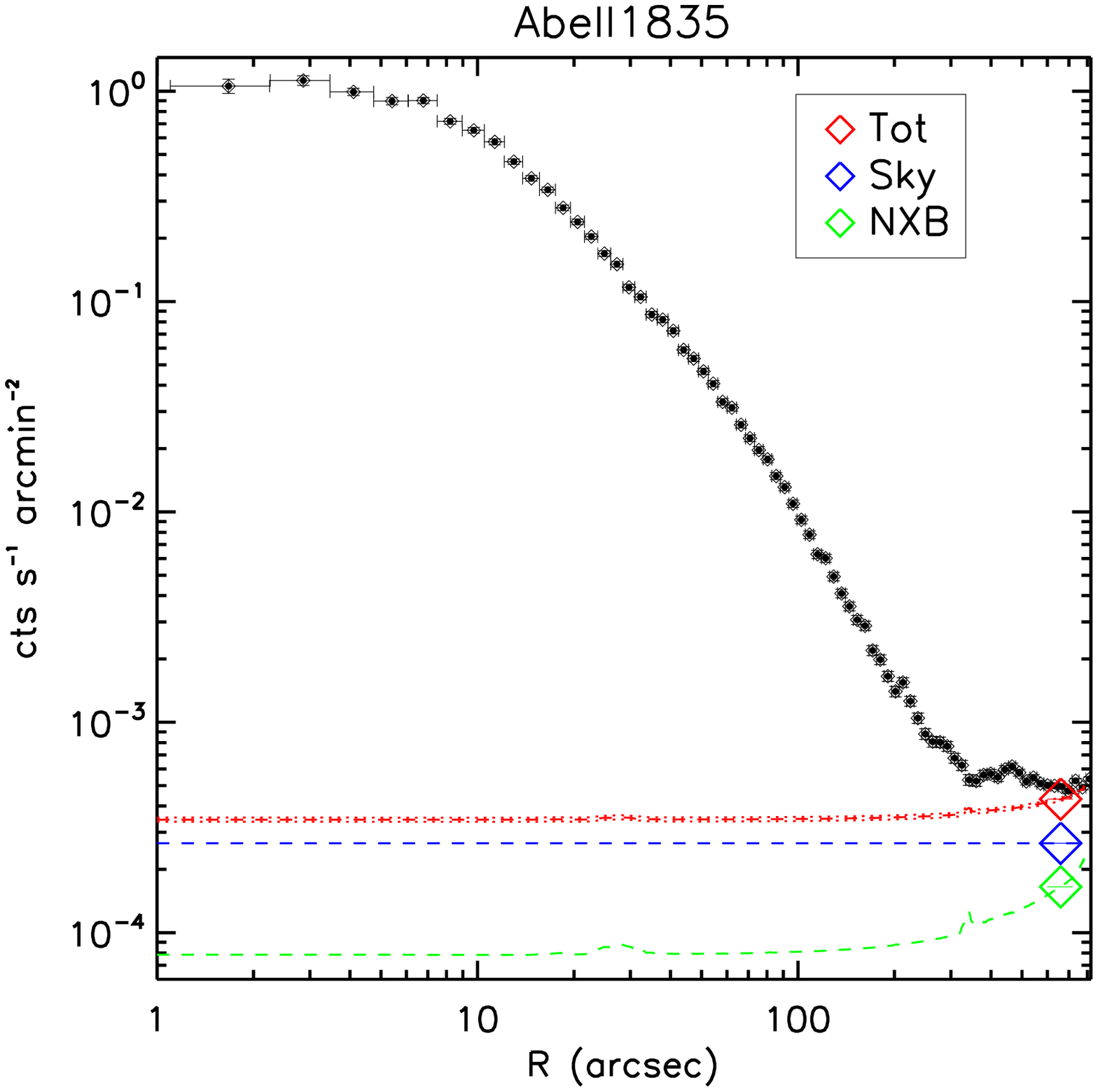}
 \includegraphics[width=0.45\textwidth]{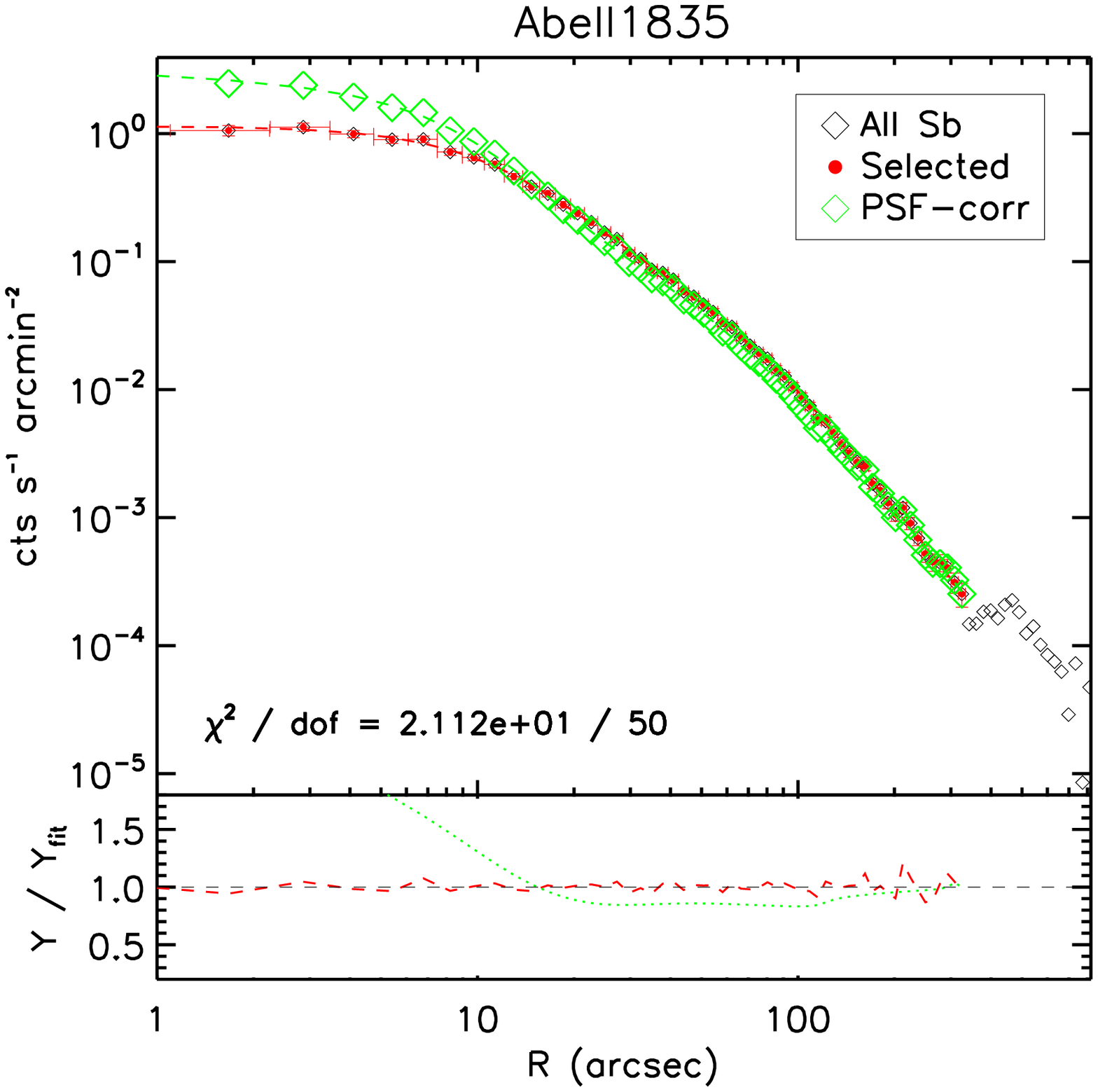}
 } \hbox{
 \includegraphics[width=0.45\textwidth]{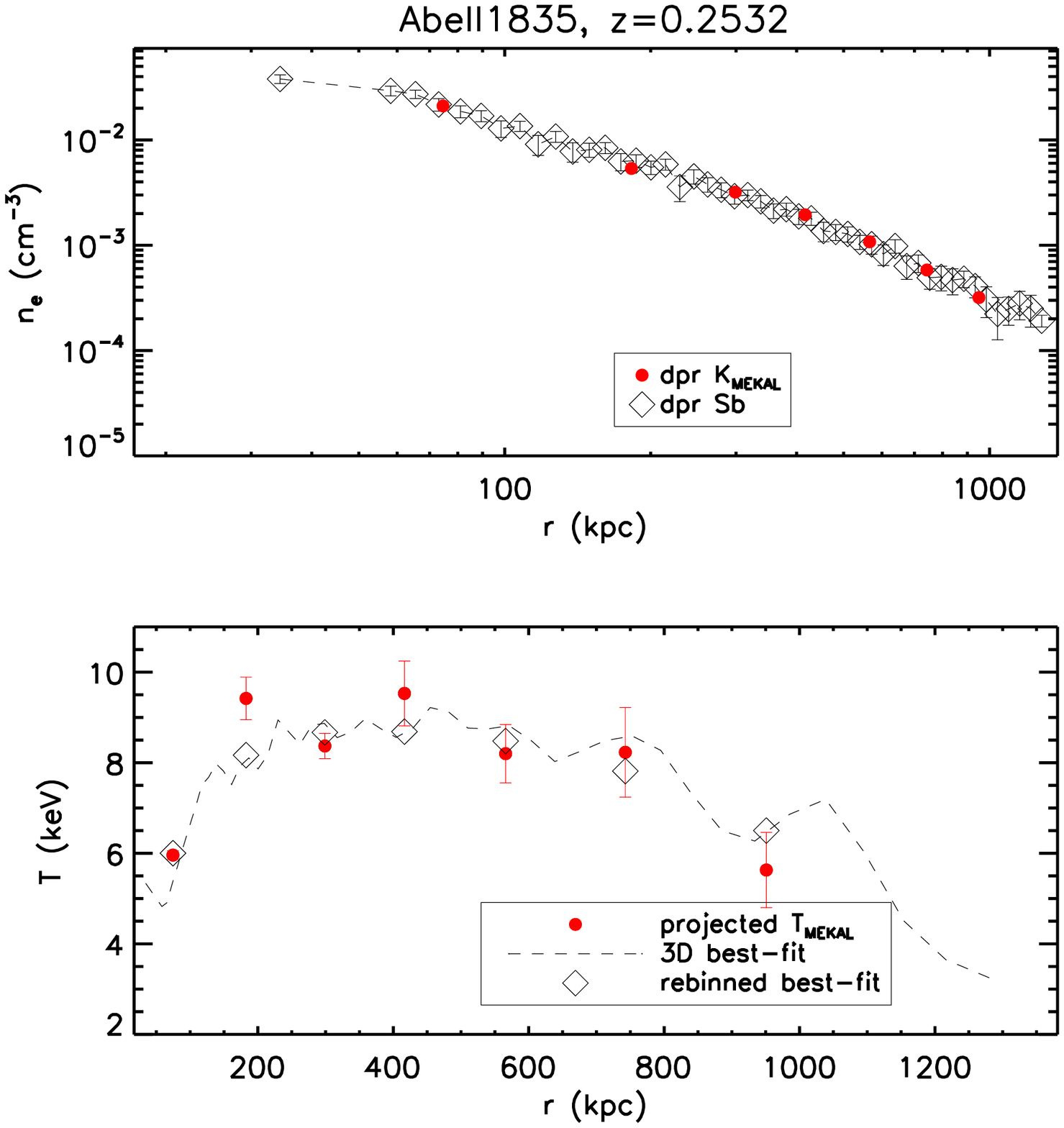}
 \includegraphics[width=0.45\textwidth]{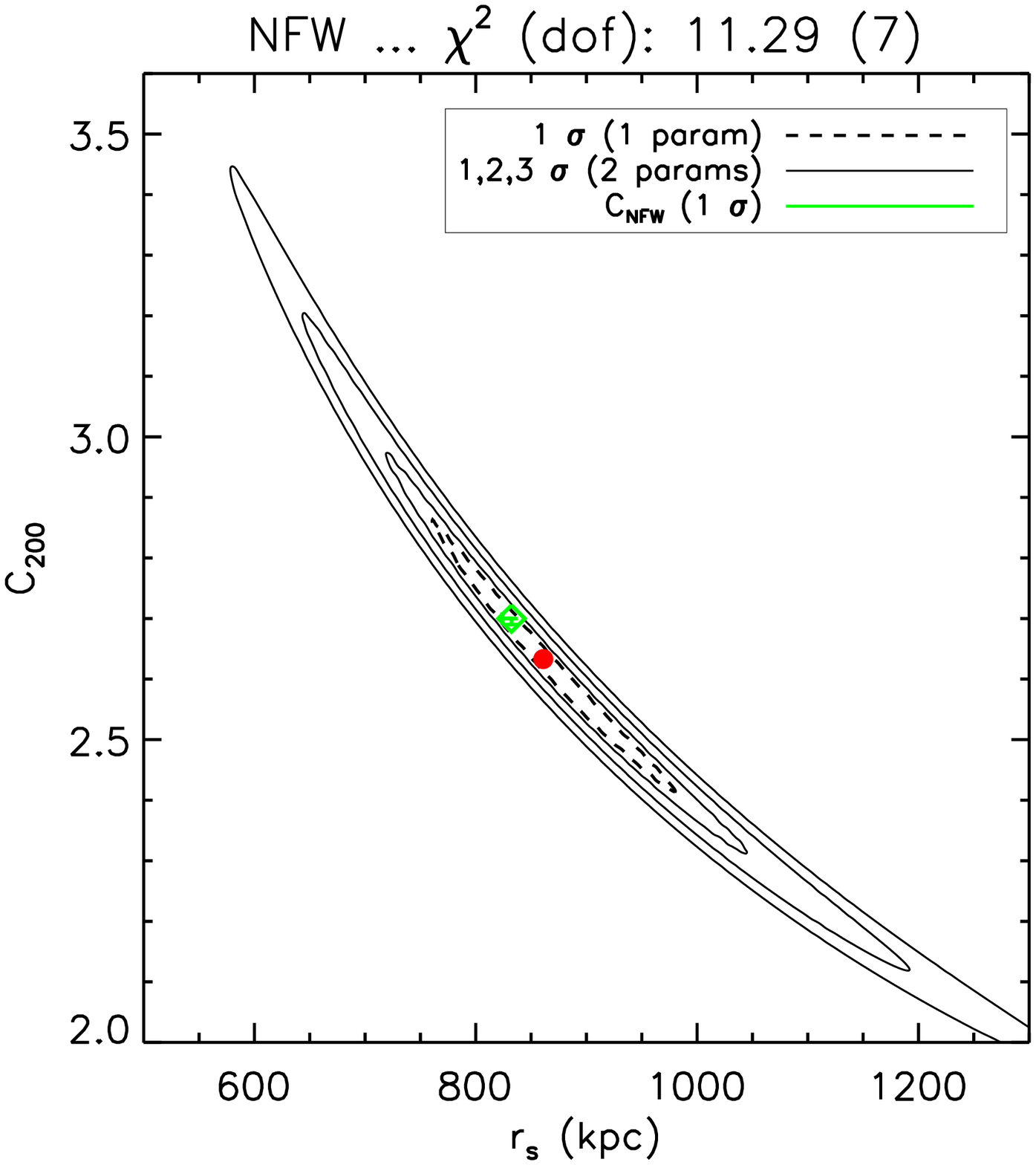}
 }
}
\caption{Procedure to reconstruct the mass profile of Abell1835 (from Ettori et al. 2010).
{\bf (Top, left)}
Surface brightness profile in the $0.7-1.2$ keV band (black filled circles) of Abell1835
compared with the profiles of the background components:
the instrumental component (NXB; green), the photon component (CXB + galactic foregrounds; blue)
and the total background (sky + instrumental; red).
{\bf (Top, right)}
PSF--corrected background--subtracted surface brightness profile.
Abell1835 is one of the objects with the largest smearing effect due to the combination
of the telescope's PSF and the centrally peaked intrinsic profile.
{\bf (Bottom, left)}
Gas density profile as obtained from the deprojection of the
surface brightness profile compared to the one recovered from the deprojection
of the normalizations of the thermal model in the spectral analysis;
observed temperature profile with overplotted the best-fit temperature model.
{\bf (Bottom, right)}
Constraints in the $r_{\rm s}-c$ plane on the best-fit mass model with the prediction
(in green) obtained by imposing the relation $c_{200} = 4.305/(1+z) \times
\left(M_{200}/10^{14} h_{100}^{-1} M_{\odot} \right)^{-0.098}$ from Macci\`o et al. (2008).
} \label{fig:back}
\end{figure*}

When the gas density is assumed to be well described by a $\beta-$model and presents a polytropic dependence
upon the gas temperature, $T_{\rm gas} \propto \rho_{\rm gas}^{\gamma-1}$ with $1 \le \gamma \le 5/3$, 
the total gravitating mass is:
\begin{eqnarray}
M_{\rm tot} (r) 
 & = & \frac{3 \ \beta \gamma \ T_0 \ r_{\rm c}}{G \mu m_{\rm u}}
\frac{x^3}{ (1+x^2)^{B} } \nonumber \\
 & = & 6.728 \times 10^{13} \frac{ \beta \gamma T_0 r_{\rm c}}{\mu} \frac{x^3}{ (1+x^2)^{B} } \ M_{\odot},
\label{app:mtot}
\end{eqnarray}
where the exponent $B = 1.5 \beta (\gamma -1) +1$,
$T_0$ is the central temperature in keV, $r_{\rm c}$ the core radius
in Mpc, $\mu$ is the mean molecular weight in a.m.u. and the numerical values
include the gravitational constant $G$, the mass of the atomic mass unit
$m_{\rm u}$ and all the unit conversions.
This method to determine mass distributions has been applied 
firstly by Bahcall and Sarazin (1977) and Mathews (1978)
and, then, developed extensively by Fabricant et al. (1980, 1984), 
Fabricant and Gorenstein (1983) and extended to the polytropic case by 
Henriksen \& Mushotzky (1986; see also Cowie, Henriksen \& Mushotzky 1987,
Hughes et al. 1988, Ettori 2000).

Nowadays, a more detailed analysis of the gas density and temperature profiles has allowed
to refine the method for the X-ray mass reconstruction.
X-ray surface brightness profiles corrected for vignetting effects  are extracted in the low energy band, e.g. [0.5-2.0]~keV. 
Assuming the cluster spherical symmetry, and chosing the X-ray brightness peak as the cluster centre, profiles are binned from the event files.
The X-ray surface brightness profile are generally PSF-corrected and deprojected into 3D emission measure profiles (e.g. Fabian et al. 1981, Buote 2000, Croston et al. 2006 where a non-parametric regularised method is described; see Fig.~\ref{fig:forw} and \ref{fig:back}). Conversion to a gas density profile is undertaken using the emissivity profile in the energy band under consideration, taking into account the observed temperature profile (note that the emissivity is weakly temperature dependent for $T > 2$~keV). The gas density profile, $n_{\rm gas}$, can thus be estimated from the geometrical deprojection of either the measured X-ray surface brightness or the estimated normalization
of the thermal model fitted in the spectral analysis (see Fig.~\ref{fig:back}).
Temperature profiles are derived from spectra extracted in concentric annuli and modeled with a hot diffuse gas including metals emission lines (e.g. MEKAL or APEC model in XSPEC; Mewe et al. 1985, Arnaud et al. 1985, Smith et al. 2001). 
The measured temperature profiles are then deconvolved from the PSF blurring and either deprojected into 3D physical profiles or modeled with a functional form in 3D projected on the plane of the sky to reproduce the observed quantities.

Three methods, essentially, are adopted to solve the equation of the hydrostatic equilibrium 
under the assumptions of spherical symmetry. We refer to them as (i) observables-modelling $\equiv$ fully parametric $\equiv$ forward process;
(ii) mass-modelling $\equiv$ semi-parametric $\equiv$ backward process; (iii) non-parametric $\equiv$ model-independent approach.
The arguments in favor ({\it pros}) and against ({\it cons}) each method are summarized at the end of the dedicated section (see below).

\begin{enumerate}[(i)]

\item {\bf observables-modelling $\equiv$ fully parametric $\equiv$ forward process}.
Parametric functions are used to model the gas density and temperature radial profiles through
the observed surface brightness and spectral temperature data and, then, propagated through
the HEE to derive the total mass profile
(e.g. Lewis et al. 2003, Buote \& Lewis 2004, Pointecouteau et al. 2005; see e.g. Fig.~\ref{fig:forw}).
The most recent implementation of it (as, e.g., in Vikhlinin et al. 2006;  Nagai et al. 2007
for tests of the technique with high-resolution numerical simulations) 
models the observed X-ray surface brightness profile with a modified and extended version of the $\beta-$model 
and the projected temperature profile with a two-component analytic function, one to reproduce the decline (if any) 
in the cooling region and the other to describe the slowly negative gradient outwards by a broken power-law with a transition region.
Uncertainty intervals for all quantities of interest are generally obtained from the distribution of best-fit parameters produced from thousands of Monte Carlo simulations, in which the full analysis is repeated on simulated data generated by scattering the observed values of the brightness and temperature profiles accordingly to the measured errors.

The mass profile in eq.~\ref{app:mtot} is a particular application of this {\it fully parametric method}
that relies on the $\beta-$model to describe the gas density and on a polytropic relation
between gas density and temperature; a more simplistic assumption would require 
an isothermal gas that could be a convenient approximation for low-quality data, 
as objects at high$-z$, where few hundreds of source counts are available and 
a single temperature measurement is feasible and just a fit with a single $\beta-$model
can be performed on the surface brightness profile.

({\it Pros}) The profiles of the observed quantities are, by definition, smooth and derivable;
({\it cons}) the radial shape is imposed and, often, being many parameters available
(e.g. Vikhlinin et al. 2006: 10 parameters to model $n_{\rm gas}(r)$, 9 for $T_{\rm gas}(r)$), 
strong degeneracy occurs among them, forcing to freeze/constrain some of them to provide
a physical solution on the mass profile;

\item {\bf mass-modelling $\equiv$ semi-parametric $\equiv$ backward process}.
A functional form of the gravitational potential is adopted and used in the HEE 
with, e.g., the geometrical deprojection of the surface brightness profile 
to recover a temperature profile that is fitted to the observed one to minimize a
merit function that depends just on the parameters describing the mass model.
These parameters are stepped over a grid of values and the best-Þt values, and uncertainties, determined via
$\chi^2$ minimization techniques.
For instance, the two parameters $(r_{\rm s}, c_{200})$ describing a NFW profile  
(that is an approximation to the equilibrium configuration produced in simulations 
of collisionless DM particles; see a comparison with the King's approximation to the isothermal sphere in
Fig.~\ref{fig:nfw}) are constrained by minimizing a $\chi^2$ statistic defined as
\begin{equation}
\chi^2_T = \sum_i\frac{ \left(T_{\rm data, i} - T_{\rm model, i} \right)^2}{\epsilon_{T, i}^2}
\end{equation}
where the sum is done over the annuli of the spectral analysis;
$T_{\rm data}$ are the either deprojected or observed temperature
measurements obtained in the spectral analysis; $T_{\rm model}$ are
either the three-dimensional or projected values (following e.g. the recipe
in Mazzotta et al. 2004) of the estimates of $T_{\rm gas}$
recovered from the inversion of the hydrostatic equilibrium equation
(see eq.~\ref{eq:hee}) for a given gas density and total mass profiles;
$\epsilon_T$ is the error on the spectral measurements.
The gas density profile, $n_{\rm gas}$, is estimated from the geometrical deprojection
of either the measured X-ray surface brightness or the estimated normalization
of the thermal model fitted in the spectral analysis (see Fig.~\ref{fig:back}).
This technique has been developed initially as part of the Cambridge
X-ray deprojection code (e.g. White, Jones \& Forman 1997; Schmidt \& Allen 2007), 
then extended to only spectral data in Ettori et al. (2002) and 
is now adopted to recover the mass profiles in recent X-ray studies 
of both observational (e.g. Morandi et al. 2007, Allen et al. 2008, Donnarumma et al. 2009) 
and simulated datasets (e.g. Rasia et al. 2006, Meneghetti et al. 2010)
against which it has been thoroughly tested.
Mahdavi et al. (2007), with the purpose of providing a joint-analysis of cluster observations 
at different wavelengths using X-ray, SZ and lensing data allowing to conduct a full, covariant error analysis 
on all the physical parameters, introduce a method in which
the gravitational potential is composed separately from the gaseous, stellar, and dark matter components
each of which is modeled with a functional form (e.g. a triple $\beta-$model for the gas density and a NFW
for the dark mass). For what concerns the X-ray analysis, an X-ray spectrum is recovered from the 
parametrized total mass and gas density profiles combined through the HEE, projected on the sky, 
distorted with a point-spread-function and convolved with the X-ray instrument response function 
to constrain, finally, the best-fit parameters with a statistical fitting of the observed counts.

({\it Pros}) The gas density is obtained from direct deprojection of the surface brightness
profile and no parameters are needed to model the gas temperature profile; the only parameters
required are the 2--3 requested to constrain the functional form of the mass profile; unlike the {\it forward} one, 
this method does not use parametric fitting functions for the X-ray temperature, gas density or 
surface brightness in measuring the mass that might introduce strong priors 
affecting the interpretation of results and, in particular, leading to possible underestimation of 
the uncertainties (see, e.g. Mantz \& Allen 2011)
({\it cons}) the observed radial profiles of the gas density and temperature are often not smooth enough, 
causing the derivatives  to be problematic;

\item {\bf non-parametric $\equiv$ model-independent approach}.
The HEE is solved by using 
directly the best-fit results on gas temeprature and density of the deprojected spectra 
(as in the pioneering work by Nulsen \& B\"ohringer 1995 on the Virgo cluster;
see also Nulsen et al. 2010; Arabadjis et al. 2002 and 2004; Ettori et al. 2002; Voigt \& Fabian 2006).

({\it Pros}) The only working assumptions are that, in each spherical shell in which the cluster volume is divided, 
the gravitating matter density is constant and the gas is isothermal; there are not 
specific requests on the form of the profiles of either the total mass or the gas density and temperature,
even though, being the gravitating mass free to move between adjacent shells with little impact on the fit,
performance of the method can be improved significantly by requiring the gravitating matter density to be a monotonically 
decreasing function of the radius. This should result in smoother mass profiles and reduces confidence intervals for the mass.
As discussed in Nulsen et al. (2010), a model-dependent prior, such as the assumption that a cluster potential has the 
NFW form, can reduce further the uncertainties on the recovered mass estimates. 
Moreover, considering that the mass can be determined directly in terms of parameters used to fit the X-ray data,
error propagation is more flexible.
({\it Cons}) Due to the lack of assumptions on the radial shape of any component,
results are less tight and stable than the ones obtained from the above-mentioned
methods, and definitely limited only to the regions with high signal-to-noise ratio in the
X-ray counts. For instance, for a sample of 8 local massive galaxy clusters, Nulsen et al. (2010)
find an average range (measured in log space at 90\% confidence level) on the mass estimates of about 2.3, 
that is about 70 per cent higher than the relative errors evaluated with the model-dependent 
technique in Vikhlinin et al. (2006).

\end{enumerate}

In brief, the {\bf fully/semi parametric} processes rely on functional formulae for the radial distribution 
of temperature, surface brightness, or mass that reproduce well the observed and simulated properties
of X-ray luminous galaxy clusters.
The greatest advantage of a parametric treatment is the numerical stability, in particular when 
a derivative must be computed as in the case of the estimate of the gravitating mass profile.
A drawback of these approaches is that (i) it is difficult to quantify the effect of the parameterization 
on the results when strong intrinsic degeneracies among the parameters are present; 
(ii) it is often a cumbersome and computationally-expensive problem
to propagate properly the errors from the observed quantities (like the spectrally-determined values
of temperature and density) to the estimates of the gravitational mass.

In the following sub-sections, we discuss how the estimated mass via the X-ray technique compares to the reconstructed
values obtained through the gravitational lensing (see reviews of  Hoekstra et al., Meneghetti et al., Bartelmann et al. in the present volume) and the application of the virial theorem and of the caustic method on the optically determined distribution of the galaxies in clusters.
In section~\ref{sect:syst}, we revise recent studies on the the systematic uncertainties affecting the X-ray estimates of $M_{\rm tot}$ and $M_{\rm gas}$.

\subsection{Comparison between X-ray and lensing mass estimates}

We will here briefly review some of the most significant recent results on the comparison of galaxy cluster mass measurements obtained with X-ray and lensing analyses.

\begin{figure}
 \includegraphics[width=1.0\textwidth]{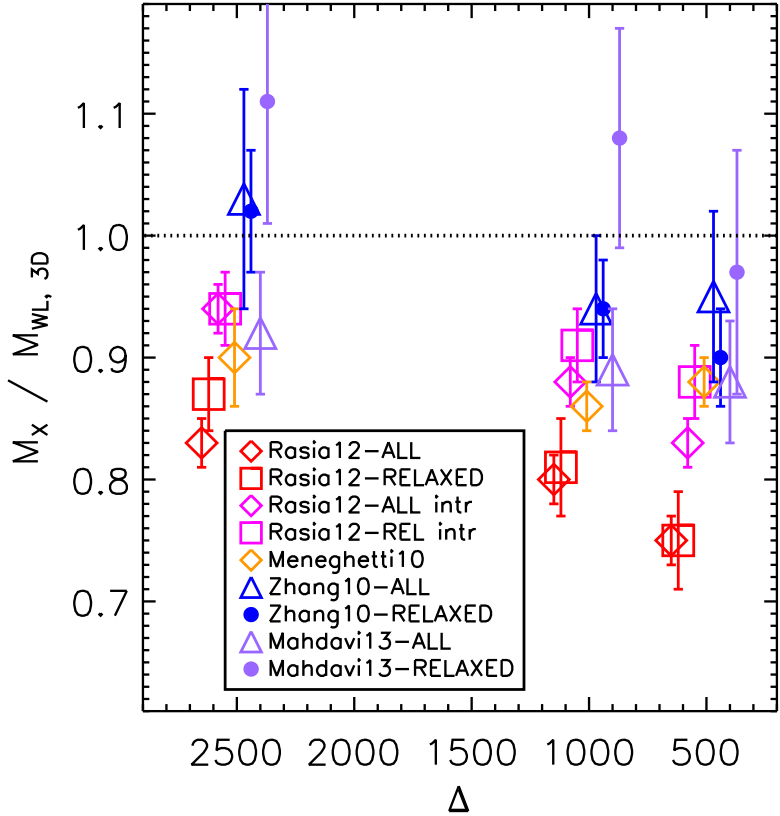}
\caption{Comparison between X-ray and weak lensing mass determinations from recent work based on both observed (Zhang et al. 2010, Mahdavi et al. 2013) and simulated (Meneghetti et al. 2010; Rasia et al. 2012, both as obtained from mock X-ray catalogs and directly from simulations -labelled ``intr'') datasets. This figure is adapted from Table~5 in Rasia et al. (2012). {\it Relaxed} refers to systems with either a not-disturbed X-ray morphological appearance or a relatively low level of the gas entropy in their cores.
} \label{fig:x_lens}
\end{figure}

In the context of the  LoCuSS (Local Cluster Substructure Survey) project, Zhang et al. (2008) have studied the archival \xmm\ exposures of 37 objects in the redshift range 0.14--0.3, 19 of which with weak lensing mass measurements available in the literature. They conclude that, at the same $R_{500}$, weak lensing masses are higher by $9 (\pm 8)$ per cent than X-ray masses.
In a following work, Zhang et al. (2010) compare the hydrostatic masses estimated for 12 galaxy clusters from \xmm\ observations with 
the weak lensing mass estimates obtained from large field of view Suprime-CAM camera $i^{\prime}$- and $V$-band imaging data at the \textit{Subaru} telescope. The background  galaxy catalog was selected by considering only faint galaxies with colors that are redder and bluer (by a minimum color offset) than the red-sequence of cluster galaxies. At  $R_{500}$, they measure $M^{\rm X}/M^{\rm WL}=0.99 \pm 0.07$ for the whole sample, $M^{\rm X}/M^{\rm WL} = 0.91 \pm 0.06$ and $1.06 \pm 0.12$ (errors at 68\% confidence level) for the undisturbed (5 objects) and disturbed (7 objects, including A1914, the most extreme of the disturbed systems) samples, respectively, as defined by analyzing the cluster X-ray morphology.
They claim that the discrepancy between the X-ray hydrostatic and the weak lensing masses for the undisturbed sample can be explained with an  additional non-thermal pressure support, while they state that the (in)consistency between the X-ray  and the weak-lensing masses for the disturbed cluster sample is more difficult to assess, due to large scatter in the mass results for this subsample. However, they note that a competing effect associated with adiabatic  compression and/or shock heating could lead to overestimate X-ray hydrostatic masses for disturbed galaxy clusters. 
In the undisturbed sample, they also detect an improving agreement  between $M^{\rm X}$ and $M^{\rm WL}$ as a function of increasing over-density, $M^{\rm X}/M^{\rm WL}=(0.908 \pm 0.004)+(0.187 \pm 0.010)  \cdot \log_{10} (\Delta/500)$, as expected. 

Following a similar approach, Mahdavi et al. (2008) perform a uniform analysis of \chandra\ X-ray data and CFHT weak lensing data for a sample of 18 galaxy clusters.  The authors find an excellent agreement between lensing and X-ray mass estimates at $R_{2500}$, with the ratio $M_X/M_L$ between X-ray and lensing masses being $1.03 \pm 0.07$. Conversely, they observe a significant decrease in the X-ray to lensing ratio when larger radii are considered ($M_X/M_L = 0.78 \pm 0.09$ at $R_{500}$), that becomes significant at $3\sigma$ when accounting for correlations between the mass estimates. Even correcting for a systematic overestimate of the weak lensing masses due to correlated large scale structures, the trend persists. It appears to be consistent with hydrodynamical simulations of galaxy clusters in which non-thermal pressure support is present, thus causing a systematic underestimate of the X-ray cluster mass not correlated with the presence or absence of a cool core. 

Recently, Mahdavi et al. (2013) have extended the previous work to 50 massive systems, also including \xmm\ exposure combined with the \chandra\ ones. They conclude that hydrostatic masses underestimate weak lensing masses by 10 per cent, on average,  at $R_{500}$ as determined from weak-lensing analysis. However, cool-core clusters, characterized by a more relaxed dynamical state, are consistent with no bias, while non-cool-core clusters have a large and constant bias of about 15--20 per cent between $\Delta=2500$ and $500$. They also find that the bias correlates well with the ellipticity of the Brightest Central Galaxy.

Overall, independent analyses of samples of galaxy clusters show that some tension between X-ray and (mostly weak) lensing mass estimates appears at lower overdensities (see Fig.~\ref{fig:x_lens}) in more disturbed objects, where the disturbance is here mainly characterized through X-ray analyses (like, e.g., evidence of centroid shifts and increasing values of the power-ratios at greater multipoles in the X-ray surface brightness; see, e.g., B\"ohringer et al. 2010, Cassano et al. 2010). On the contrary, when the comparison is performed on a given object with data that overlap in the radial distribution, the differences tend to be within the statistical error, in particular when the object in exam is round in its X-ray appearance and dynamically relaxed or, at least, with no major merging in action (see, e.g., Donnarumma et al. 2011, Umetsu et al. 2012). Project like {\it CLASH} (Postman et al. 2012) will provide a much clear view on any mismatch in the cluster mass distribution as mapped with different probes.

\subsection{Comparison with methods based on the velocity distribution of the galaxies}

Zwicky (1933, 1937) first used the virial theorem applied to the observed distribution of galaxies to estimate the mass of the Coma cluster. With some modifications, notably a correction term for the surface pressure (The \& White 1986), the virial theorem remains in wide use (e.g., Girardi et al. 1998 and references therein). Jeans analysis incorporates the radial dependence of the projected galaxy velocity dispersion (e.g., Carlberg et al. 1997; van der Marel et al. 2000; Biviano \& Girardi 2003 and references therein) and obviates the need for a surface term.
Jeans analysis and the caustic method are closely related. 
Both use the phase-space distribution of galaxies to estimate the cluster mass profile. The primary difference is that the Jeans method assumes that the cluster is in dynamical equilibrium; the caustic method does not. The Jeans method depends on the width of the velocity distribution of cluster members at a given radius, whereas the caustic method calculates the edges of the velocity distribution at a given radius. 
Katgert et al. (2004), for a combination of 59 nearby clusters from the ESO Nearby Abell Cluster Survey, reported a best-fit with a NFW with concentration of 4 ($^{+2.7}_{-1.5}$ at 1 $\sigma$ level), permitting mass models with a core only if the core radius is sufficiently small ($r_{\rho_0/2} \le 0.13 R_{200}$ at the 99\% CL). 
Rines \& Diaferio (2006) confirmed that cluster infall regions are well fitted by NFW and Hernquist profiles and poorly fitted by singular isothermal spheres. This much larger sample enables new comparisons of cluster properties with those in simulations. The shapes (specifically NFW concentrations) of the mass profiles agree well with the predictions of simulations. The mass in the infall region is typically comparable to or larger than that in the virial region. Specifically, the mass inside the turnaround radius is on average $2.19\pm0.18$ times that within the virial radius. This ratio agrees well with recent predictions from simulations of the final masses of dark matter halos (see also Biviano \& Salucci 2006).
Wojtak \& Lokas (2010) analyzed the kinematic data of 41 nearby ($z < 0.1$) relaxed galaxy clusters and found that, if the total mass distribution is approximated by a NFW profile, the concentration is $6.9^{+0.6}_{-0.7}$ at the virial mass of $5 \times 10^{14} M_{\odot}$. They demonstrated that less evolved clusters have shallower mass profiles and their galaxy orbits are more radially biased at the virial sphere. 
Ettori et al. (2002), using X-ray based mass profiles from \sax\ data and optical determinations from Girardi et al. (1998), measured a median deviation of about 1.2 and 0.8 $\sigma$ in mass and velocity dispersion estimates, respectively, in a sample of 16 objects. Moreover, they obtained evidence that larger deviations are present in the subsample of no-cool-core, not-relaxed systems (2.7 and 2.0 $\sigma$ deviation in mass and velocity dispersion, respectively, for NCC objects; 1.0 and 0.7 $\sigma$ for CC clusters), suggesting that mergers affecting them propagate to the optical determinations of the velocity dispersion and to the validity of the hydrostatic assumption made in the process of the estimation of the X-ray mass.

\subsection{Systematic uncertainties on the mass measurements} \label{sect:syst}

All the X-ray mass measurements are affected from systematic uncertainties.
For instance, different X-ray detectors, even on board of the same satellite, provide different estimate of the 
spectral temperature and flux of the same region of the ICM. This is mainly due to the present limits
in cross-calibrating the energy dependence and normalization of the effective area of the X-ray instrument.
Nevalainen et al. (2010) provide the most recent effort to compare the spectral estimates with different instruments 
like \xmm\ EPIC-pn and EPIC-MOS, \chandra\ ACIS-S and ACIS-I, and \sax\ MECS.
Their figure~20 and table~11 summarize the state of the art as of December 2009: with respect to EPIC-pn, 
the temperature estimates in the soft (0.5--2 keV), hard (2--7 keV) and wide (0.5--7 keV) band can be recovered,
on average, with a relative accuracy better than 5\%, as for the fluxes in the soft and hard band; however, a clear tension 
is shown between ACIS and EPIC-pn, with weighted mean of the relative difference of 18 and 14 per cent in the measurements
of the temperature in the soft and wide band, and of 11 per cent in the measurement of the flux in the hard band.
These mismatches have to be considered when mass estimates for the same objects are obtained from different
detectors.
Moreover, if corrections to the hydrostatic equilibrium equation are required for bulk motions of the ICM or non-thermal pressure support, 
the X-ray total mass would be underestimated, whereas if clumpiness is present in the ICM that is assumed to be 
smoothly distributed (e.g. Mathiesen et al. 1999), the gas mass would be overestimated and the total mass underestimated 
(because of the smaller gradient entering the HE).

Galaxy clusters are dynamically young systems. They accrete cosmic  materials through mergers
that can make inaccurate the assumption of the hydrostatic equilibrium between the X-ray emitting plasma
and the underlying gravitational potential. 
Recent cosmological simulations probed that the reconstructed total mass profiles can be biased
low by 10--20 \% when the spherical symmetry and the hydrostatic equilibrium are assumed.
The assumption of spherical symmetric distribution of the ICM has been shown to induce differences 
in the order of few per cent at large distance from the cluster center for compressed/elongated shapes 
of the ICM (e.g. Piffaretti et al. 2003; see more details in Limousin et al. in the present volume). 
Buote \& Humphrey (2012a, b) demonstrate that the mass enclosed within an ellipsoidal gravitational potential 
in hydrostatic equilibrium is defined in the same way as in equation~\ref{eq:mhe} once a corrective factor $\eta = 
\frac{q_b q_c}{3} \left( 1 +1/q_b^2 +1/q_c^2 \right) \approx 1$, where $q_b = b/a$ and $q_c = c/a$ 
are the axis ratios satisfying the relation $0 < q_b \le q_c \le 1$, is considered.

On the other hand, the assumption of the hydrostatic equilibrium itself can be limited from the presence of residual bulk motions in the ICM due to major mergers that can provide a non-thermal contribution to the dynamical stability of even relaxed objects
(see, e.g. Evrard, Metzler, Navarro 1996; Schindler 1996; Bartelmann \& Steinmetz 1996; Balland \& Blanchard 1997;
Kay et al. 2004; Rasia et al. 2006; Hallman et al. 2006; Nagai, Vikhlinin, Kravtsov 2007;
Fang et al. 2009; Lau et al. 2009; Meneghetti et al. 2010; Nelson et al. 2012; Rasia et al. 2012).
As illustrated in equation~\ref{eq:euler}, if the ICM bulk velocity is not null, some extra components, apart from the thermal gas, are expected to support the total gravitational potential. Suto et al. (2013) identify three other effective mass terms to be added to the thermal one if the gas motion is not negligible: a rotational term produced from the centrifugal force, a term related to the streaming speed, and an acceleration term that contributes with a positive/negative addendum on the total mass when the bulk of the gas is decelerating/accelerating.

In recent years, the uncertainties affecting the reconstructed X-ray masses have been addressed
by applying the observational techniques to mock observations. In one of the first attempts of this kind,
Rasia et al. (2006), using some mock \chandra\ observations of hydrodynamical simulations of massive galaxy clusters,  have found that the mass profile obtained via a direct application of the HEE is strongly dependent upon the measured temperature profile, with irregular radial distributions and associated large errors inducing a significant scatter on the reconstructed mass measurements.
The poorness of the $\beta-$model in describing the gas density profile makes the evaluated masses to be underestimated by $\sim$40 per cent with respect to the true mass, both with an isothermal and a polytropic temperature profile, in particular when some extrapolation is required from the narrow field-of-view with respect to the total cluster X-ray emission. In comparison with it, a {\it  mass-modelling} ({\it backward process}; see sect.~\ref{sect:mass}) method has been shown to be more robust in reconstructing the total mass.

\begin{figure}
 \hbox{
 \includegraphics[width=0.5\textwidth]{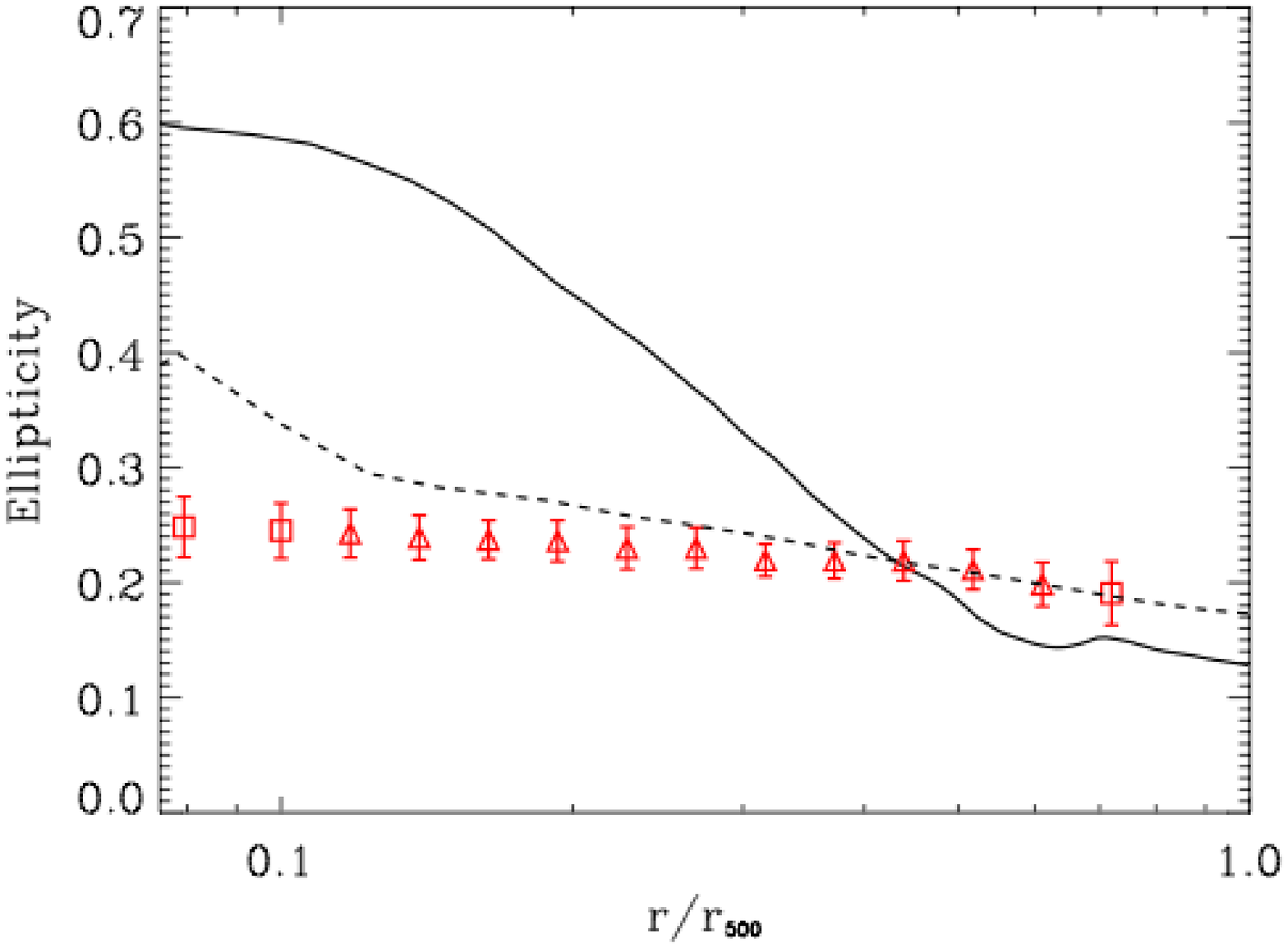}
 \includegraphics[width=0.45\textwidth]{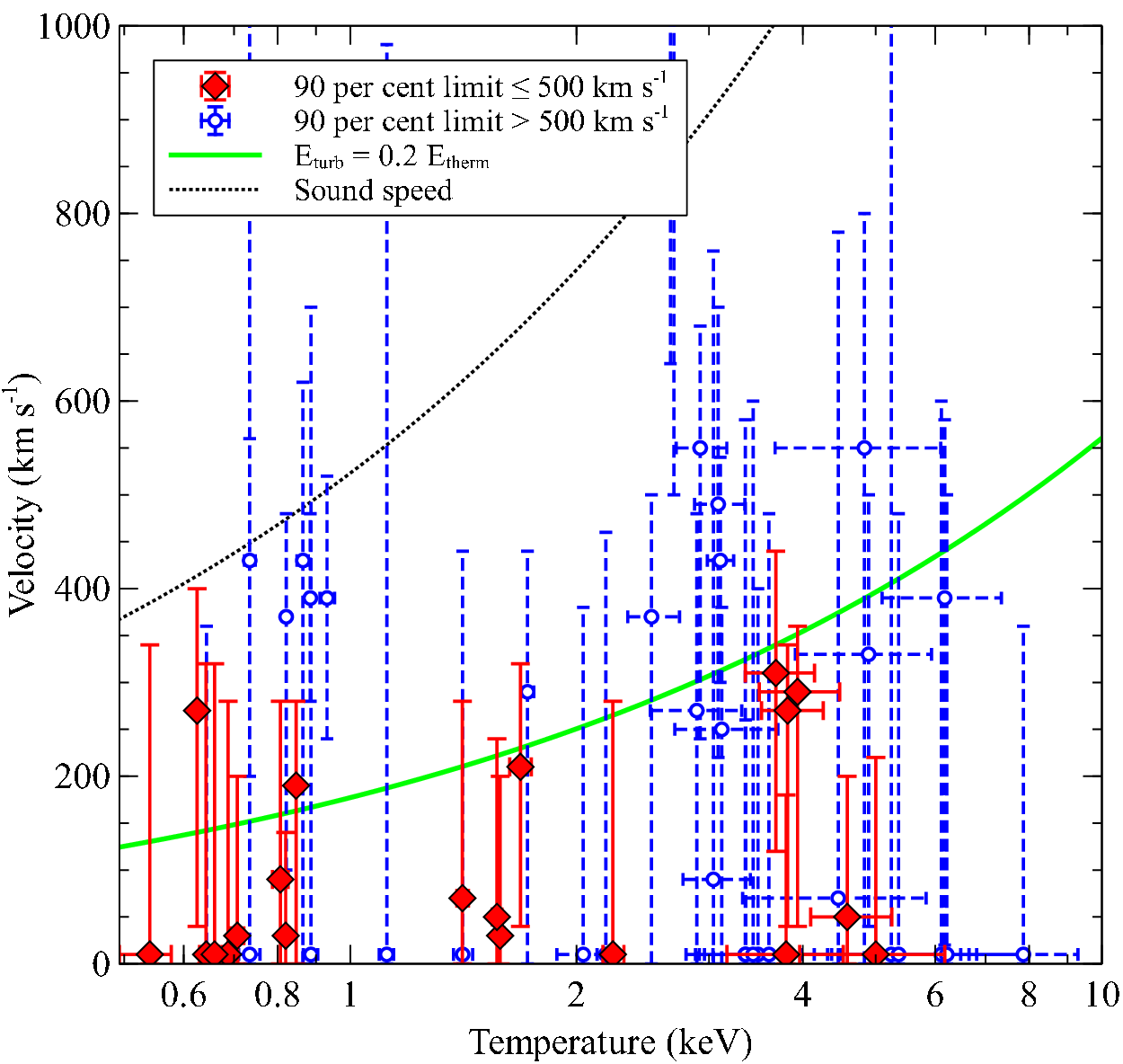}
}
\caption{Observational limits on the ICM turbulence.
{\bf (Left; from Fang et al. 2009)}
Average X-ray ellipticity profiles measured in ÒrelaxedÓ simulated CDM clusters (solid black; dashed line: from an object simulated without radiative cooling or star formation) and in the observed clusters (dashed red; from \chandra\ and \rosat\ PSPC data for the inner and outer parts, respectively). 
{\bf (Right; from Sanders \& Fabian 2013)}
Upper limits on velocity broadening as a function of the ICM temperature measured with \xmm\ RGS. The sound speed as a function of temperature and the energy density in random motions corresponding to 20 per cent of the thermal energy density are also plotted. 
} \label{fig:turbu}
\end{figure}

From mock \chandra\ exposures of high-resolution Eulerian simulations of 16 galaxy clusters, Nagai et al. (2007) 
concluded that  the X-ray hydrostatic estimate of the total  mass is biased low by $\sim$ 5\%--20\% within the virial region, mainly for the  additional non-thermal  pressure support provided by subsonic bulk motions in the ICM. They  observed that this  bias increases toward cluster outskirts and is dependent on the cluster dynamical state, affecting however   also  relaxed clusters.  
The effect of residual subsonic bulk and random gas motions on the X-ray mass and concentration estimates was also evaluated by Lau et al. (2009), using the same set of simulated clusters presented in Nagai et al. (2007). They estimated that the gas motions can provide up to 5\%-15\% of the total pressure support in relaxed clusters, while this relative contribution  increases for unrelaxed systems. They concluded that this effect can lead to underestimate the cluster total mass, and it is more significant at larger radii where the ICM is less relaxed (and thus the contribution of gas motions to the pressure is larger). 
In a following work, Nelson et al. (2012) examined specifically the effect of mergers on the hydrostatic mass estimate 
and claimed it becomes negligible 4 Gyrs after the major event, finding that, at $R_{500}$, the contribution from non-thermal pressure support peaks at about 30\% of the total pressure during the merger and quickly decays to 10--15\% as a cluster relaxes.
Overall, simulators emphasized the role played from random turbulent gas motion to the additional pressure support required in their simulated galaxy clusters, presenting it as the main contributor to the systematic errors budget in X-ray mass estimates. 
Fang et al. (2009), instead, have presented convincing evidence that the support of the ICM in relaxed simulated objects from rotational and streaming motions is comparable to the support from the random turbulent pressure out to $\sim 0.8 R_{500}$. This should translate in large ellipticities of the X-ray isophotes, that are not observed in a sample of 9 clusters observed with \rosat\ PSPC and \chandra\ (see Fig.~\ref{fig:turbu}). They conclude, thus, that observed clusters are, on average, much rounder and have a distinctly different radial variation in ellipticity than the simulated objects from which a systematic bias in the hydrostatic mass has been estimated. 

Another way to detect and measure the ICM rotational motion is through the Doppler shifts and broadening of the emission lines present in X-ray spectra (e.g. Inogamov \& Sunyaev 2003). While the low resolution spectra from CCD on board of \asca\, \chandra\ and \suzaku\ have been used to look for changes in the ICM bulk velocity as a function of position on, e.g., Centaurus cluster (e.g. Dupke \& Bregman 2006, Ota et al. 2007), the only direct limit on the ICM turbulence has been obtained from Sanders \& Fabian (2013) with the \xmm\ Reflection Grating Spectrometer (RGS) spectra of the the X-ray emitting gas from the central regions of about 60 galaxy systems. The authors find an upper limit of 500 km s$^{-1}$ for more than a third of these objects. Massive galaxy clusters like A1835 and MACSJ2229.7--2755 have limits on the velocity width close to 300 km s$^{-1}$. 
Overall, about half of the elliptical galaxies, galaxy groups and clusters studied show limits that are consistent with hosting less than 20 per cent of the thermal energy density in the form of random motions in the central regions mapped through this analysis (see Fig.~\ref{fig:turbu}).

\begin{figure}
\hspace*{-0.8cm}
\vbox{
 \hbox{
 \includegraphics[width=0.55\textwidth]{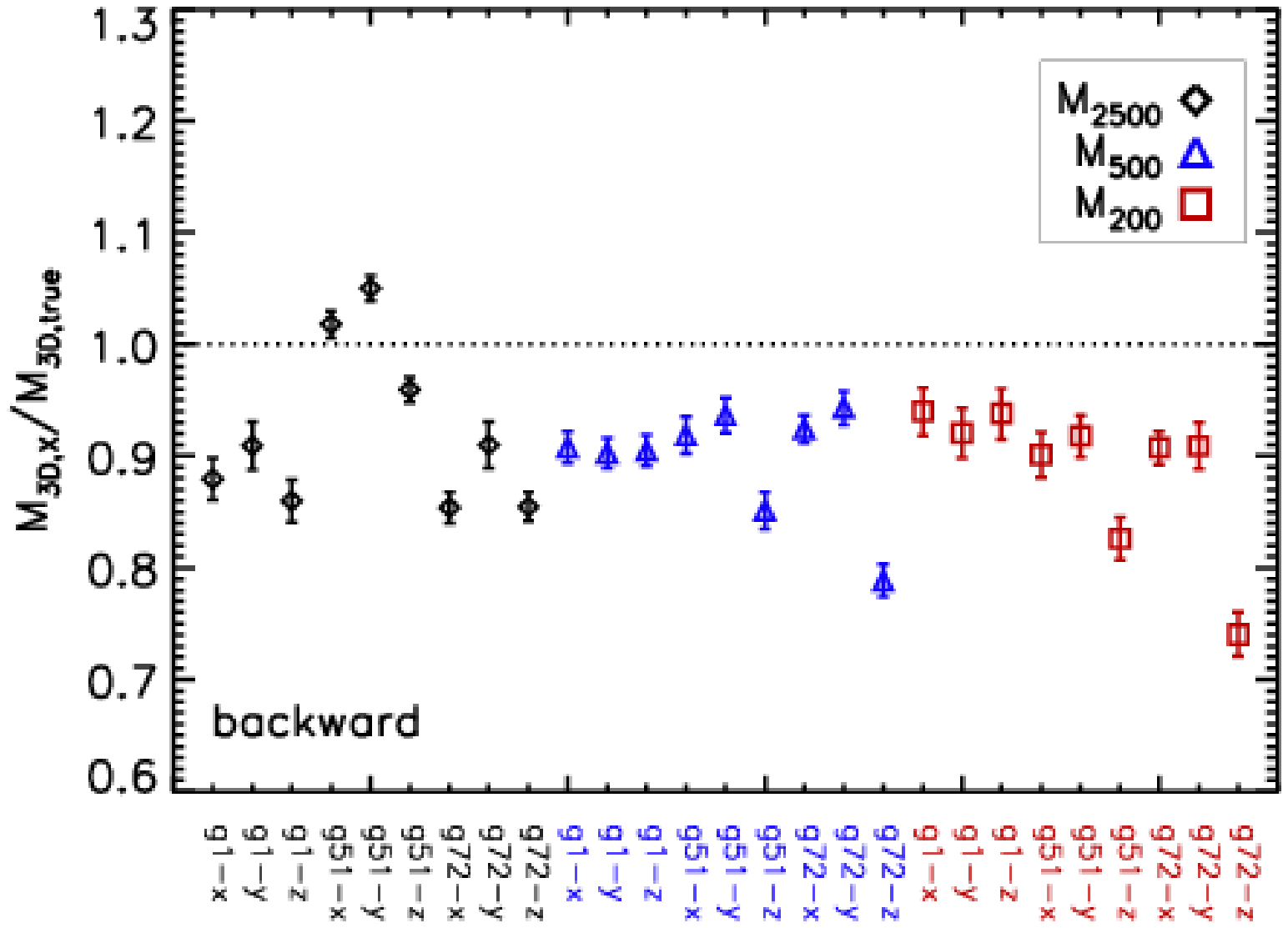}
 \includegraphics[width=0.55\textwidth]{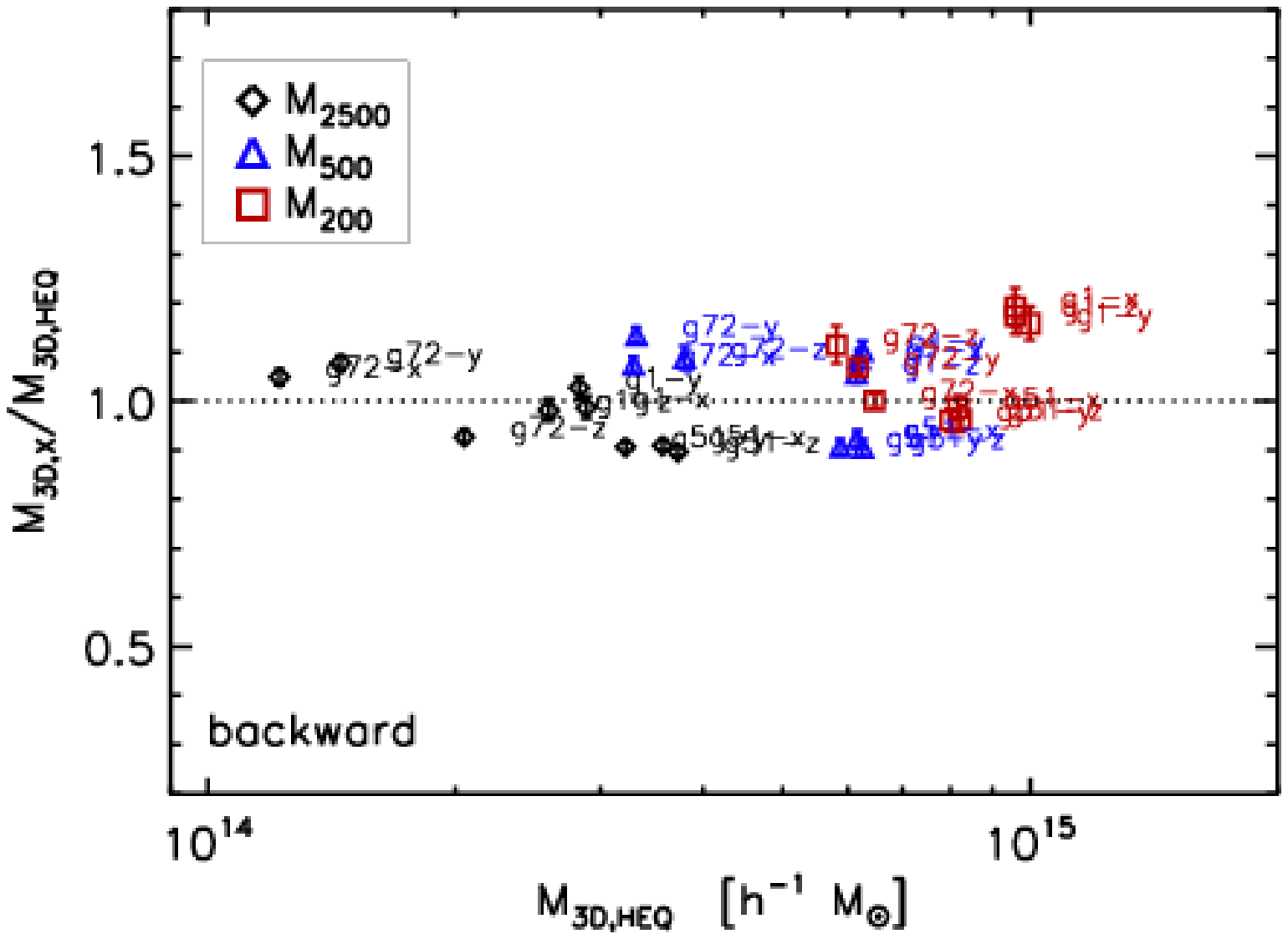}
 } 
\hbox{
 \includegraphics[width=0.55\textwidth]{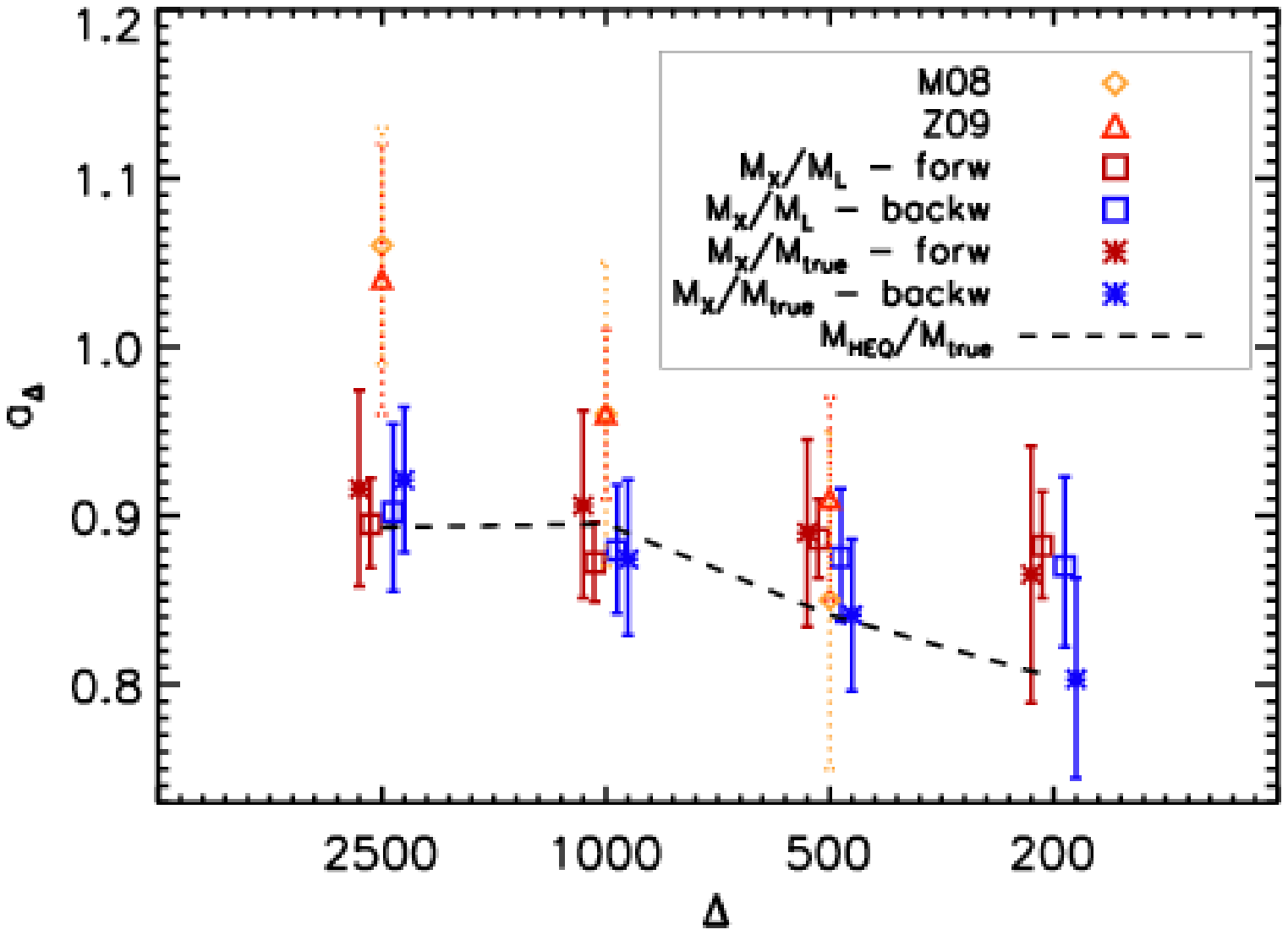}
 \includegraphics[width=0.55\textwidth]{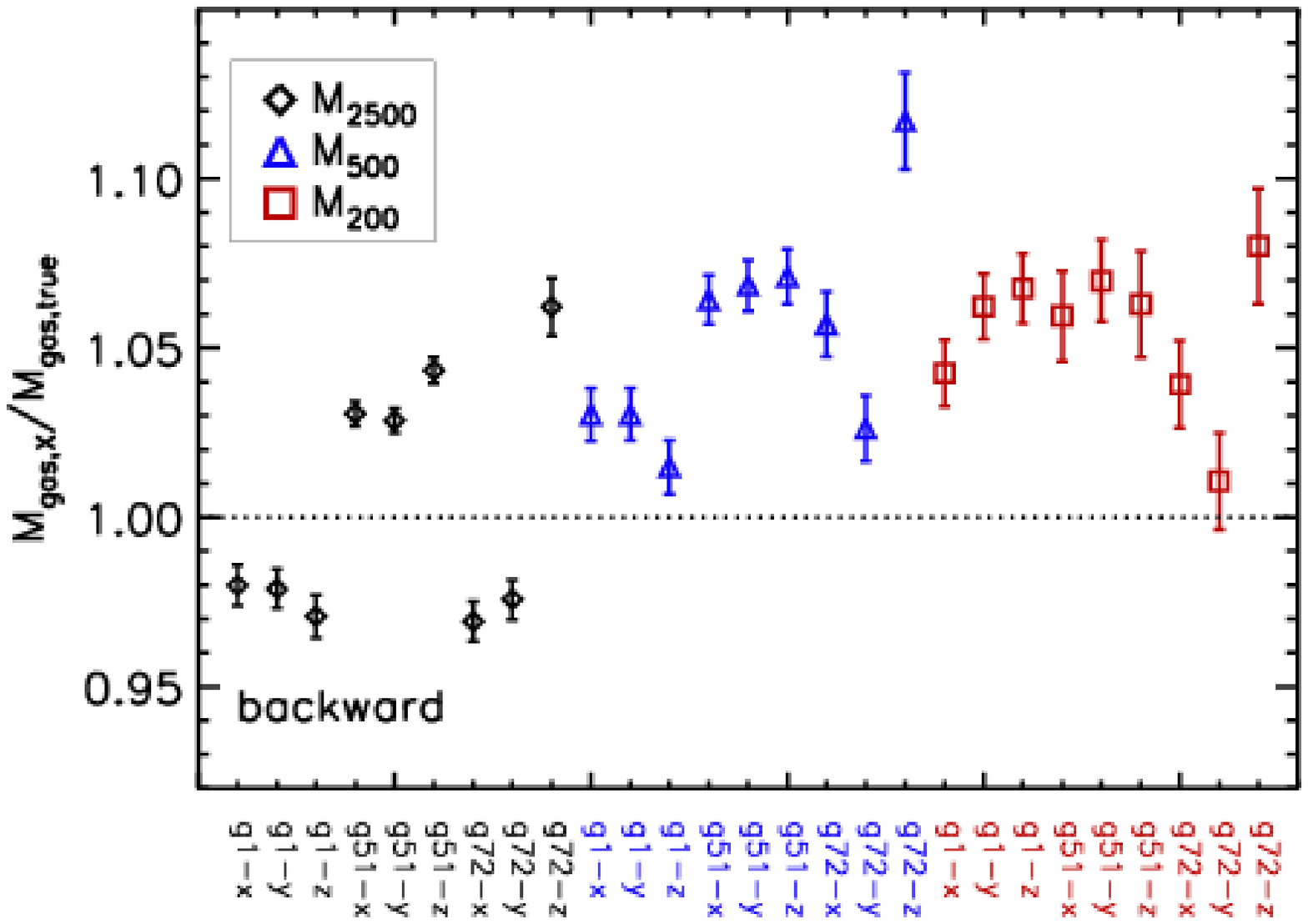}
 }
}
\caption{Reconstructed mass profiles from hydrodynamical simulations (from Meneghetti et al. 2010).
{\bf (Top panels)} Ratios between $M_{\rm tot}$ as obtained from the {\it backward}
method (see Sect.~\ref{sect:mass}) in the X-ray analysis and the true masses of the 3 simulated objects seen
along 3 different line-of-sight. On the right, the comparison with masses estimated 
by applying the HEE is shown.
{\bf (Bottom panel, left)} Ratio between X-ray and lensing masses as a function of the
overdensity $\Delta$ (squares). The results are shown for the X-ray masses
obtained with the {\it forward} (red) and {\it backward} (blue) methods
and for the lensing masses obtained with the SL+WL method. For
comparison, they also show the ratios between the X-ray masses and the
true masses of the clusters (asterisks) and the ratios between the masses
determined via the hydrostatic equilibrium equation using the true gas
density and temperature profiles and the true masses (dashed line). The
diamonds and the triangles show the results published by Mahdavi et al. (2008, M08) 
and Zhang et al. (2010, Z09).
{\bf (Bottom panel, right)} Comparison between estimated and true gas mass.
} \label{fig:m10}
\end{figure}

The work on simulations has been extended recently from X-ray to the strong and weak lensing mass reconstruction
in Meneghetti et al. (2010), where the mass distribution in 3 massive ($M_{200}
\sim 6.8-11.4 \times 10^{14} M_{\odot}$) objects obtained from N-body / hydrodynamical simulations 
has been studied through observational techniques applied
to the realistic mock observations along three different line-of-sight.
They concluded that strong lensing models can be trusted over a limited region around the cluster core. 
Extrapolating the strong lensing mass models to outside the Einstein ring can lead 
to significant biases in the mass estimates. Weak lensing mass measurements can be largely
acted by substructures, depending on the method implemented to convert the shear 
into a mass estimate. Using non-parametric methods which combine weak and strong lensing data, 
the projected masses within $R_{200}$ can be constrained with a precision of about 10\%. 
Deprojection of lensing masses increases the scatter around the true masses by more 
than a factor of two due to cluster triaxiality. X-ray mass measurements present a bias that is 
radial dependent and is entirely ascribable to bulk motions of the simulated gas.
Using the lensing and the X-ray masses as proxies for the true and the hydrostatic 
equilibrium masses of the simulated clusters, by averaging over the cluster sample 
Meneghetti et al. (2010) were able to measure the lack of hydrostatic equilibrium in the systems 
investigated (see Fig.~\ref{fig:m10}), similarly to what has been obtained by comparing observational
constraints on the weak-lensing and X-ray masses 
($M_{500, X} / M_{500, WL} \approx 0.9 \pm 0.1$ in a sample of 19 clusters observed with 
{\it Subaru} and \xmm\ and studied in Zhang et al. 2008; $M_{X} / M_{WL} =
1.03\pm0.07$ and $0.78\pm0.09$ at $R_{2500}$ and $R_{500}$, respectively, 
for the 18 objects analyzed in Mahdavi et al. 2008). 
An extension of this work to 20 simulated objects with an extensive analysis of the dependence of the hydrostatic bias on cluster morphology, environment, temperature inhomogeneity and mass is presented in Rasia et al. (2012).
They conclude that the X-ray mass bias grows from the inner to the outer regions of the clusters and is strongly correlated with temperature inhomogeneities and, more weakly, with some morphological parameters like the the centroid shift and third-order power ratio. The increase of the bias in the cluster outskirts is due to a more dramatic lack of hydrostatic equilibrium and to a flattening of the gas density profile due to gas clumping (e.g. Nagai \& Lau 2011). The amount of the bias estimated by the evaluation of the hydrostatic mass using the mass-weighted temperature and the intrinsic gas density of the simulated objects is of about 15\% and is consistent with the estimates from previous works (e.g. Jeltema et al. 2008, Piffaretti \& Valdarnini 2008, Ameglio et al. 2009, Lau et al. 2009; see Fig.~\ref{fig:x_lens}). A further 10-15\% is induced from temperature inhomogeneities particularly high in this set of simulations where the thermal conduction is set to zero, not allowing to this process, which is very effective in hot systems, to make more homogeneous the ICM thermal structure.

We can parametrize the uncertainties on the measurements of the
total gravitating mass and gas mass through the factors
$B$ and $C$, respectively: 
$M_{\rm tot, obs}  = M_{\rm tot, true} \times B$; 
$M_{\rm gas, obs}  = M_{\rm gas, true} \times C$.
These factors go in the direction to rise the total mass
estimates (i.e. $B<1$) if corrections to the hydrostatic
equilibrium equation are required for bulk motions of the ICM
or non-thermal pressure support, and to lower the true gas mass (i.e. $C>1$)
if clumpiness is present in the ICM that is assumed to be smoothly distributed.

The factor $B$, which parametrizes the uncertainties on $M_{\rm tot}$ is expected
to be between 0.8 and 1 from the cluster mass profiles recovered
from both X-ray and lensing data, and it seems to depend on the overdensity
at which it is measured, decreasing at lower $\Delta$, where the assumption 
of the hydrostatic equilibrium is less tenable (e.g. Mahdavi et al. 2008 and 2013, Zhang et al. 2010; 
see Fig.~\ref{fig:x_lens} and \ref{fig:m10}).

The factor $C$ represents the level of clumpiness that affects the estimate of $M_{\rm gas}$ in X-ray analysis
and that simulations show to be lower than 1.2 (Mathiesen et al. 1999;
see also recent observational constraints in Simionescu et al. 2011, Eckert et al. 2013a, 2013b).
Meneghetti et al. (2010; see also Nagai et al. 2007 and panel at the bottom of Fig.~\ref{fig:m10}) compare the gas mass derived from an X-ray analysis and its true value in numerical simulations. They conclude that different X-ray methods provide results consistent within $2 \sigma$. 
Moreover,  the gas mass is recovered, on average, to better than 1 per cent at $R_{2500}$ and at 7 per cent (with a 3 per cent scatter) at $R_{500}$.
At larger radii, a trend to slightly overestimate $M_{\rm gas}$  is detected as consequence of the inhomogeneities (i.e. clumpiness, but also asymmetry --Vazza et al. 2011, Eckert et al. 2012) present in the gas distribution.

\section{Mass profiles as cosmological proxies}
 
Galaxy clusters are really powerful cosmological probes in constraining 
the geometry and the relative amounts of the matter and energy constituents of the 
Universe, in particular through the normalization, slope and evolution of the mass
function (see, e.g.,  review in Allen, Evrard \& Mantz 2011).
We refer to the review by Giodini et al. in the present volume
to appreciate which X-ray observables, and at which level of accuracy,
can be associated to the integrated values of total and gas mass.

In this section, we present how the distribution of the total and baryonic mass
in galaxy clusters can be used to validate the scenario of structure
formation in a CDM Universe. 
We discuss here the two cases that provide the most stringent limits nowadays
(along with the cluster mass function discussed in e.g. Allen , Evrard \& Mantz 2011) 
(i) the gas mass fraction, and (ii) the concentration-mass relation.
Before that, we describe the cosmological framework in which we operate.

\subsection{The cosmological model}

The Friedmann model is the simplest model of the Universe based on
the cosmological principle that the matter distribution is
{\it isotropic} (i.e. the same in all directions) and
{\it homogeneous} (i.e. independent of location).
From these assumptions, the equations of Einstein can be solved
introducing the time derivatives of the scale factor $a$, the energy
density ($\rho c^2$) and pressure ($P$) of the perfect fluid of which
the energy-momentum tensor is adopted, the curvature parameter $k$
and the cosmological constant $\Lambda$:
\begin{eqnarray}
H^2 = \left( \frac{\dot{a}}{a} \right)^2 = \frac{8 \pi G \rho}{3}
- \frac{k c^2}{a^2} + \frac{\Lambda c^2}{3}, \nonumber \\
\dot{H} + H^2 = \left( \frac{\ddot{a}}{a} \right) =
-\frac{4 \pi G}{3} \left( \rho +\frac{3P}{c^2} \right)
+ \frac{\Lambda c^2}{3}.
\label{eq:fried}
\end{eqnarray}
These relations are
known as {\it Friedmann equations} and are related
one to each other once the adiabatic expansion of the Universe
is taken into account, i.e. $d(\rho c^2 a^3) = -3P a^2 da$.
To solve this system of equations and determine the time evolution
of the cosmic scale factor $a(t)/a(0) = (1+z)^{-1}$,
we need to specify the equation of state $w=P/(\rho c^2)$.
In the present Cold Dark Matter (CDM) model, the universe starts
as dominated from relativistic particles ($w=1/3$)
and is presently filled with cold matter ($w=0$).

Introducing the density parameter $\Omega = \rho / \rho_{\rm c}$,
where $\rho_{\rm c} = 3 H_0^2/ (8 \pi G)$ is the {\it critical density}
at $z=0$, $H_0$ is Hubble's constant and $G$ is the gravitational
constant, equation~\ref{eq:fried} can be rewritten in a more convenient form as
\begin{eqnarray}
 \frac{H^2_z}{H^2_0} & = & E(z)^2 =
\Omega_{\rm m} (1+z)^3 +\Omega_{\rm k} (1+z)^2 +
\Omega_{DE} \lambda(z)    \nonumber \\
\lambda(z) & = & \exp\left(3 \int_0^z \frac{1+w(z)}{1+z} dz\right)\,.
\label{eq:ez}
\end{eqnarray}
Note that the above equations consider the dependence upon the ratio $w$
between the pressure and the energy density in
the equation of state of the dark energy component
(see a review in Peebles \& Ratra 2003).
In particular, the case of a cosmological constant,  $\Omega_{DE} = \Omega_{\Lambda}$, requires $w=-1$.
When a flat ($k=0$) cosmology with matter density $\Omega_{\rm m}$ is considered 
(which is the cosmological model usually adopted for reference),
$E(z) = \left[\Omega_{\rm m} (1+z)^3 + 1 - \Omega_{\rm m} \right]^{1/2}$.

Hereafter, we refer to $\Omega_{\rm b}$ as the {\it baryon matter} density parameter,
to $\Omega_{\rm m}$ as the {\it total matter} density parameter
(i.e. $\Omega_{\rm m} = \Omega_{\rm b} + \Omega_{\rm c}$,
where $\Omega_{\rm c}$ is the {\it cold dark matter} component),
to $\Omega_{DE}$ (or  $\Omega_{\Lambda}$ when $w=-1$) as the {\it dark energy} density parameter.
We neglect (i) the energy associated with
the cosmic radiation, $\Omega_{\rm r} \approx 4.16 \times 10^{-5}
(T_{\rm CMB}/2.726K)^4$, and (ii) any possible contributions from
light neutrinos, $\Omega_{\nu} h_{70}^2 = \sum {\rm m}_{\nu} / 45.5%
{\rm eV}$, that is expected to be non-zero, but less than $0.01$ for
a total mass in neutrinos, $\sum {\rm m}_{\nu}$, lower than $\sim 1$ eV
(note that the neutrino thermal history is very different from that of
the CDM with very distinct signature on cosmic structure;
in particular, cosmology is sensitive to the total energy density in
neutrinos, which for non-relativistic neutrinos is simply proportional
to $\sum {\rm m}_{\nu}$; see review on Hannestad 2010).
The Einstein equation can be written then in the form 
$\Omega_{\rm m} + \Omega_{DE} +\Omega_{\rm k} = 1$,
where $\Omega_{\rm k}$ accounts for the curvature of space.
From, e.g., Carroll, Press \& Turner (1992, cf. eq.~25),
we can then write the angular diameter distance as
\begin{eqnarray}
d_{\rm ang} = & \frac{d_{\rm lum}}{(1+z)^2} =
\frac{c}{H_0 (1+z)} \frac{S(\omega)}{|\Omega_{\rm k}|^{1/2}},   \nonumber  \\
\omega = & |\Omega_{\rm k}|^{1/2} \int^z_0 \frac{d \zeta}{E(\zeta)},
\label{eq:dang}
\end{eqnarray}
where $d_{\rm lum}$ is the luminosity distance,
$S(\omega)$ is sinh$(\omega)$, $\omega$, $\sin(\omega)$ for
$\Omega_{\rm k}$ greater than, equal to and less than 0, respectively.

\subsection{The gas mass fraction as a cosmological probe}

\begin{figure}[ht]
 \vbox{
  \hbox{
  \includegraphics[width=0.5\columnwidth]{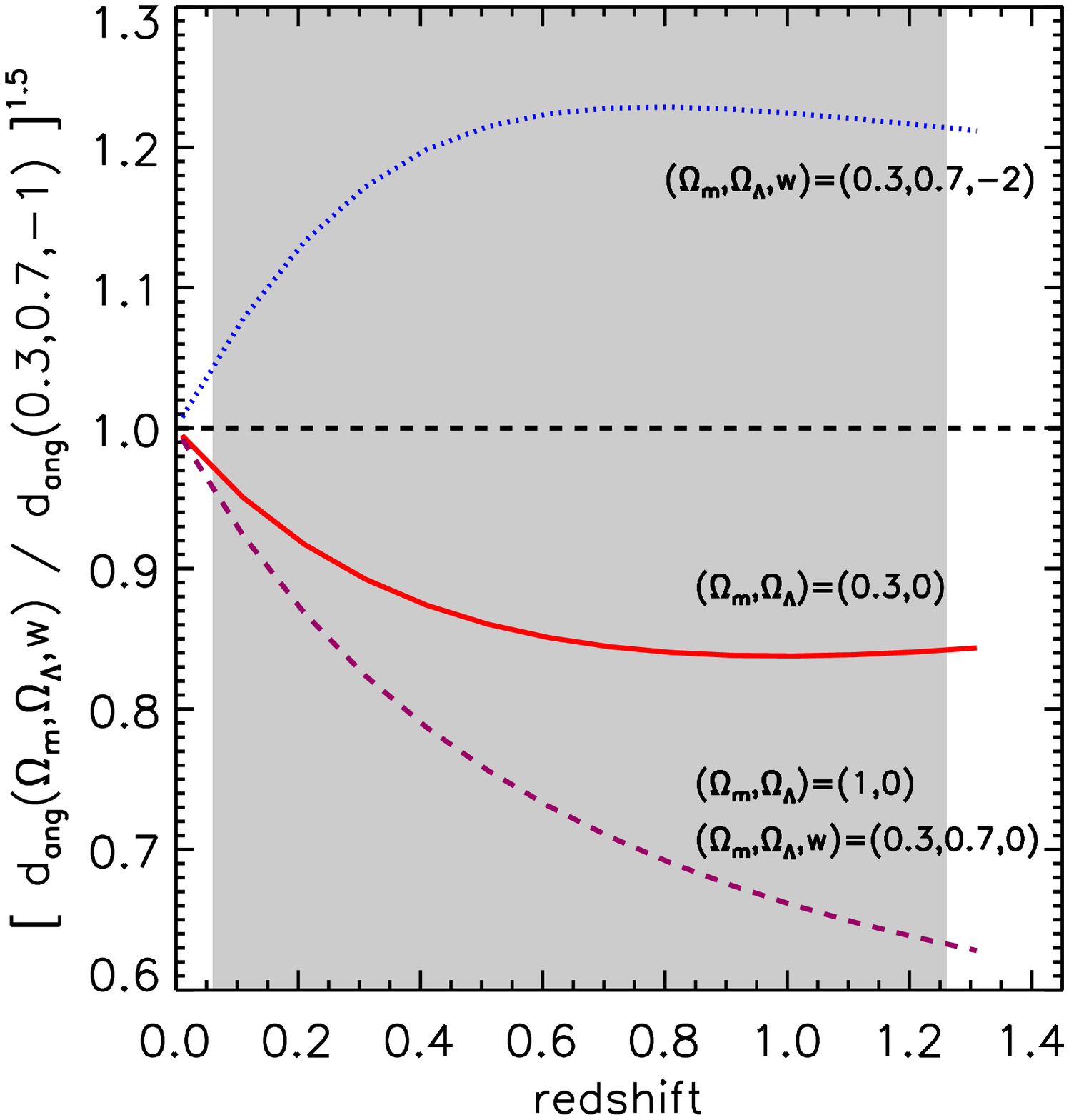}
  \includegraphics[width=0.5\columnwidth]{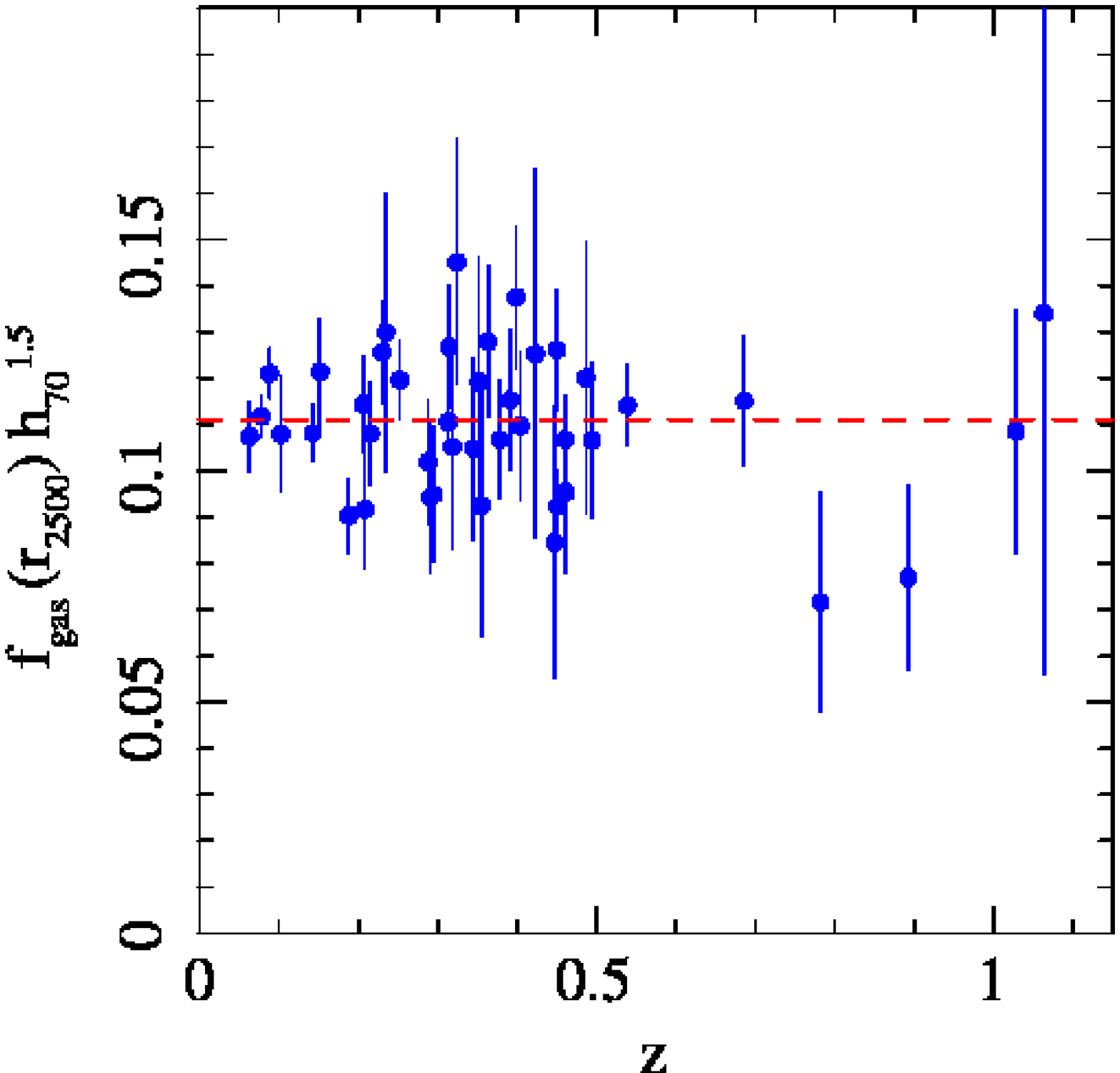}
  }
  \hbox{
  \includegraphics[width=0.5\columnwidth]{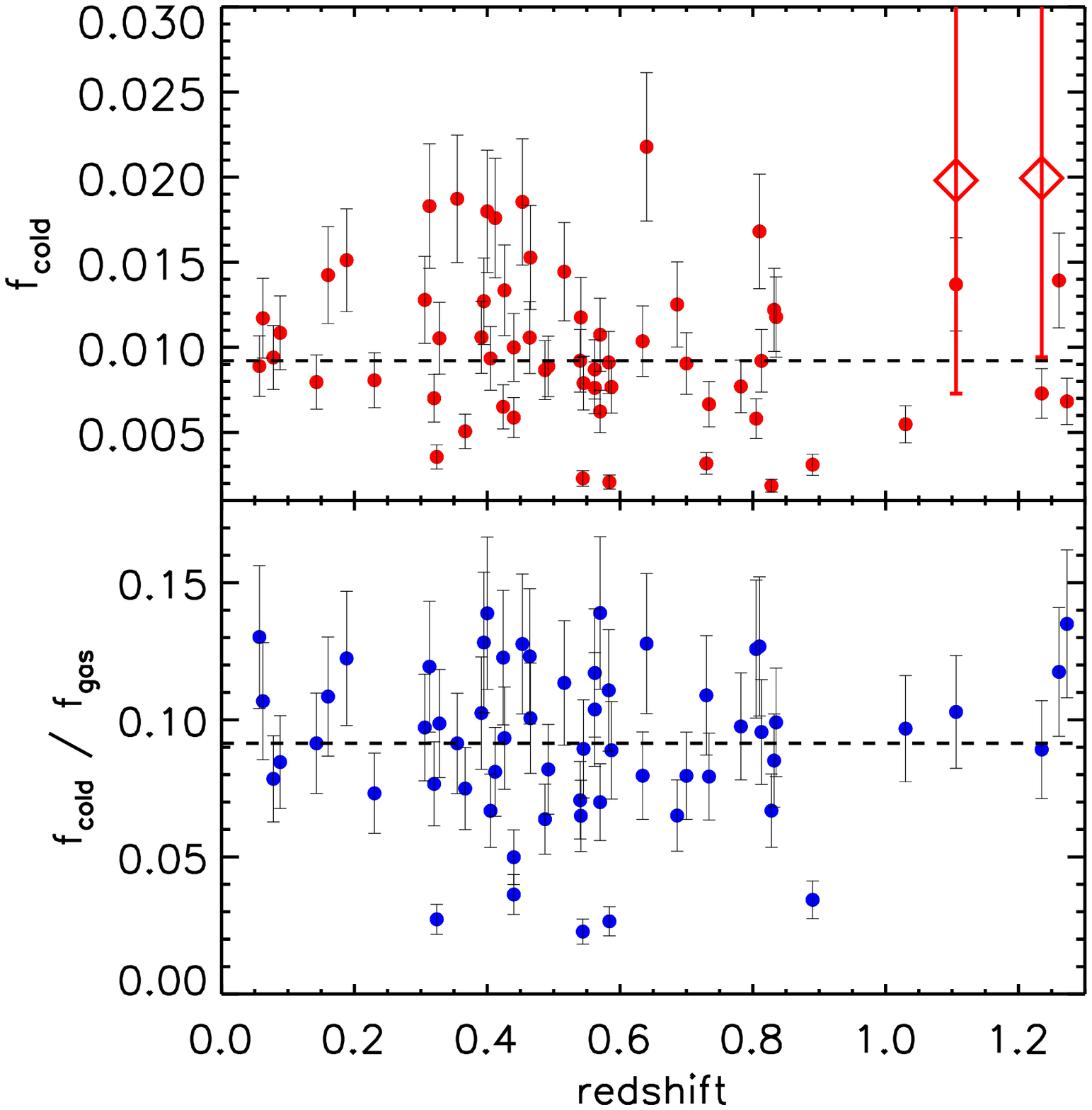}
 \includegraphics[width=0.5\columnwidth]{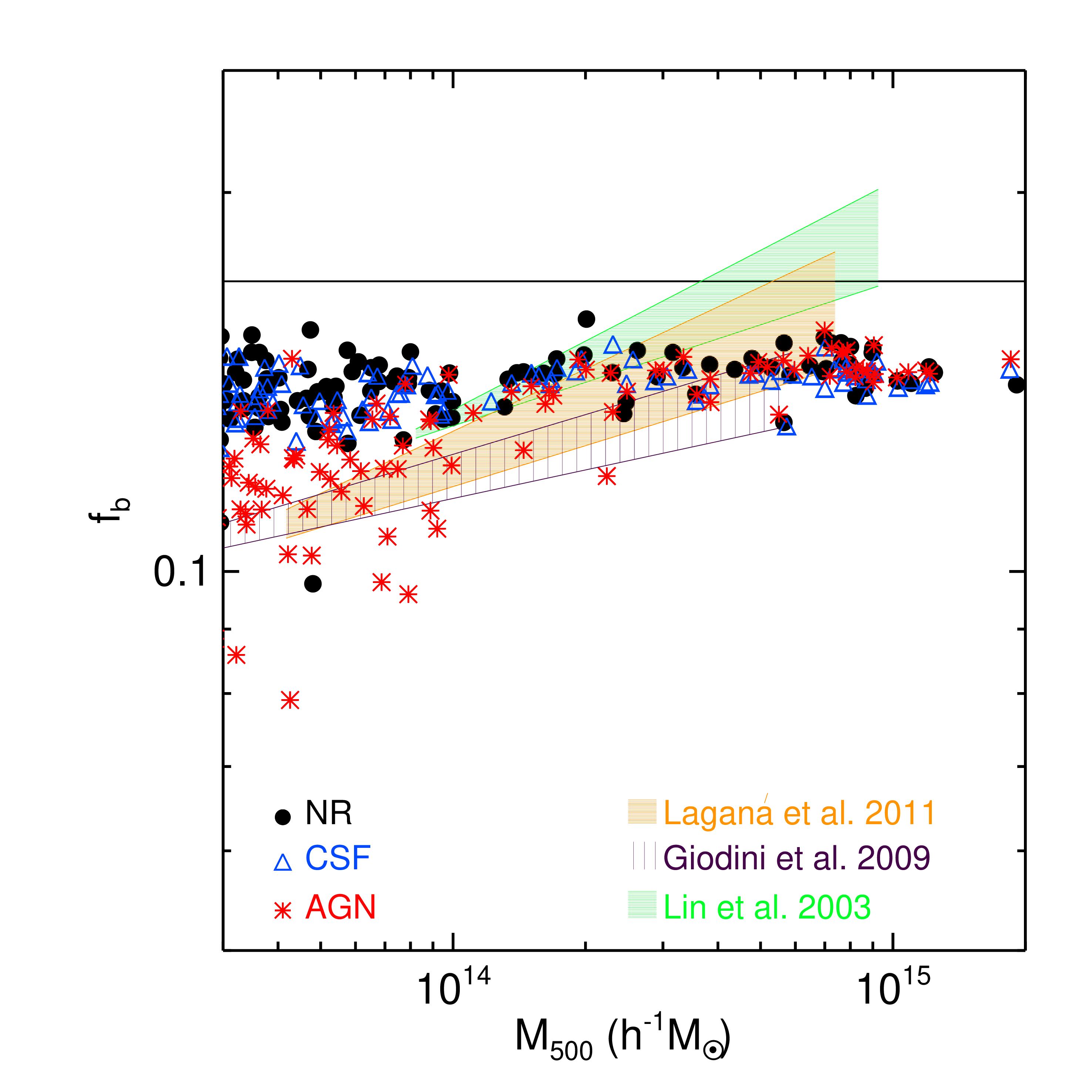}
  }
 }
\caption{{\bf (Top, left)}
Sensitivity of the cluster baryon fraction method to the variation
of the cosmological parameters. The shaded region indicates the redshift
range considered in the present studies.
{\bf (Top, right)} Distribution of the X-ray gas mass fraction
measured within $R_{2500}$ as a function of redshift in a
$\Lambda$CDM model (from Allen et al. 2008).
{\bf (Bottom, left)}
Stellar mass fractions and ratios between their values and the estimated
gas mass fraction at $R_{500}$ as a function of redshift.
Dashed lines indicate the median value:
$f_{\rm cold}=0.009$ and $f_{\rm cold}/f_{\rm gas}=0.091$.
The two diamonds indicate the values for RDCS-J0910 and RDCS-J1252 with
the stellar masses estimated from the near-infrared luminosity function
in Strazzullo et al. (2006).
A cosmology of $(H_0, \Omega_{\rm m}, \Omega_{\Lambda}) = (70$ km s$^{-1}$ Mpc$^{-1},
0.3, 0.7)$ is adopted here.
{\bf (Bottom, right)}
Baryon fraction measured in SPH hydrodynamical simulations (from Planelles et al. 2013)
compared with some observational constraints (Lagan\'a et al. 2011, Giodini et al. 2009, Lin et al. 2003). 
The solid line represents the cosmic baryon fraction assumed in the simulations.
} \label{fig:fgas} \end{figure}

The gas mass fraction, $f_{\rm gas} = M_{\rm gas}/M_{\rm tot}$, as inferred from X-ray observations
of clusters of galaxies uses and combines two independent methods 
to constrain the cosmological parameters:
(i) the relative amount of baryons with respect to the total mass 
observed in galaxy clusters is compared to the cosmic baryon fraction
to provide a direct constraint on $\Omega_{\rm m}$
(this method was originally adopted by White et al. 1993 to show the limitation
of the standard cold dark matter scenario in an Einstein-de Sitter Universe), 
(ii) the parameters that describe the geometry of the
Universe (specifically $\Omega_{\Lambda}$ or $w$) are limited by assuming that the gas fraction is constant 
in time, as first suggested by Sasaki (1996) and Pen (1997).

Starting from these pioneering works, many studies have followed this approach
to constrain the cosmological parameters (see, e.g., 
Allen et al. 2008, Ettori et al. 2009 and references therin).

No selection effect is expected to occur in the application of the
gas mass fraction method once the clusters are selected
to ensure (i) the use of the hydrostatic equilibrium equation
to recover the total mass, and (ii) a negligible contribution from non-gravitational energy
in the region of interest to allow the use of cross--calibration
with numerical simulations, such as the estimate of the depletion parameter (see below).
The selection of X-ray morphologically round, relaxed, hot, massive systems 
dominated energetically by gravitational collapse satisfies both these conditions.

To run this cosmological analysis, one needs the estimates of the gas mass fraction, $f_{\rm gas} = M_{\rm gas}/M_{\rm tot}$, 
provided from equations~\ref{eq:mhe} and \ref{eq:mgas}, and few more ingredients.
In detail, they are the following:

\begin{itemize}
\item $f_{\rm bar} = f_{\rm gas}+f_{\rm cold}$, where the gas mass
  fraction $f_{\rm gas}$ is directly measured from X-ray
  observations and depends upon the cosmological parameters through the
  angular diameter distance, $d_{\rm ang}$, defined in
  equation~\ref{eq:dang}, being $f_{\rm gas} = M_{\rm gas} / M_{\rm
    tot} \propto n_{\rm gas} R^3 / R \propto d_{\rm ang}^{5/2} /
  d_{\rm ang} \propto d_{\rm ang}(\Omega_{\rm m}, \Omega_{\Lambda}, w)^{3/2}$
  (see Fig.~\ref{fig:fgas} for the relative dependence upon different cosmologies), 
while the mass fraction of cold baryons, $f_{\rm cold} = M_{\rm cold} / M_{\rm tot}$, 
is defined as the sum of the stellar component and the intracluster
light (see, e.g. Gonzalez et al. 2007, Lagan\'a et al. 2008) divided by the total mass, 
and is generally estimated through a statistical approach lacking often specific 
information on the single cluster analyzed (Fig.~\ref{fig:fgas}).
The error on $f_{\rm bar}$, $\epsilon_{\rm bar}$, is the sum in quadrature
of the uncertainties on $f_{\rm gas}$, on $f_{\rm star}$ and on the assumed
value of Hubble's constant $H_0$ (see below) propagated through the
following dependence: $f_{\rm gas} \propto H_0^{-1.5}$ and
$f_{\rm cold} \propto H_0^{-1}$.

\item The depletion parameter $b = f_{\rm bar} / (\Omega_{\rm b}/\Omega_{\rm m})$ 
(with error $\epsilon_b$)
represents the fraction of cosmic baryons that fall in the cluster dark matter halo
and is estimated from hydrodynamical simulations (e.g. Kravtsov et al. 2005, 
Ettori et al. 2006, Planelles et al. 2013; see Fig.~\ref{fig:fgas}). 
In massive galaxy clusters at $z=0$, 
$b(<R_{2500})$ is estimated in the range 0.77--0.85 and $b(<R_{500})$ is equal to 
$0.85 \pm 0.05$. The latter value (as estimated from Planelles et al.) is confirmed 
to be nearly independent of the physical processes considered in recent SPH hydrodynamical 
simulations and characterized by a negligible redshift evolution.
It is worth mentioning that, recently, Eckert et al. (2013b), 
by combining \rosat\ PSPC gas density profiles and \planck\ 
pressure profiles for a sample of nearby X-ray luminous galaxy clusters, have reconstructed 
the gas mass fraction distribution out to $R_{200}$, making a direct measure, for the first time, of the 
depletion parameter associated to the gas only. Using as reference for the cosmic baryon budget
the constraints from \wmap7 (Komatsu et al. 2011), they measure $b_{\rm gas} (<R_{500}) =
0.76 \pm 0.02 \times (\Delta/500)^{-0.2} \times (T_{\rm gas}/7 \; {\rm keV})^{0.5}$, which is	
marginally consistent with the present limits from hydrodynamical simulations (see e.g. 
Planelles et al. 2013).

\item The cosmic baryon density $\Omega_{\rm b}$ and the Hubble constant $H_0$ have to be
assumed from independent probes, like e.g. Primordial Nucleosythesis calculations
(see, e.g., Steigman 2006) or analysis of the power spectrum of the temperature anisotropy measured 
in the Comic Microwave Background (e.g. Hinshaw et al. 2013) 
and calibrations of Cepheid distance scale (Riess et al. 2011, Freedman et al. 2012), respectively.
\end{itemize}

These quantities are then combined to evaluate which is the best cosmological set of parameters that reproduces the observed baryonic mass fraction in galaxy clusters. Starting from some reference values of $f_{\rm gas}$ estimated for an assumed cosmological model (e.g. a $\Lambda$CDM) and at given overdensity (e.g. $\Delta=2500$ or $500$), their values $f_{\rm gas}^{\rm cor}$ for the ``correct'' cosmology will be the ones that satisfy at the same time the following equations (e.g. Allen et al. 2008) 
\begin{eqnarray}
f_{\rm gas}^{\rm cor}(z) & = & f_{\rm gas}^{\Lambda CDM}(z) \big( \frac{d_{\rm ang}^{cor}}{d_{\rm ang}^{\Lambda CDM}} \big)^{3/2} \nonumber \\
& = & k \frac{b(z)}{1 + s(z)} \frac{\Omega_{\rm b}}{\Omega_{\rm m}}, 
\end{eqnarray}
where $k$ is a factor that includes a bias from any non-thermal pressure support and any residual uncertainty in the accuracy of
the instrument calibration and X-ray modeling and $s(z)$ is the observed ratio of the mass in stars (both in galaxies and intracluster light) to the X-ray emitting gas mass. Similarly,  we can write these conditions in a single merit function that has to be minimized (e.g. Ettori et al. 2009)
\begin{eqnarray}
\chi_f^2 & = & \sum_{i=1}^{N_{\rm dat}}
\frac{(f_{{\rm bar}, i} / b_i - \Omega_{\rm b}/\Omega_{\rm m})^2}
{\epsilon_{{\rm bar}, i}^2/b_i^2 +(f_{{\rm bar}, i} \epsilon_{b_i} / b_i^2)^2 +
\epsilon_{\Omega_{\rm b}}^2/\Omega_{\rm m}^2}.
\label{eq:chi2f}
\end{eqnarray}
The distribution of the values of this $\chi^2$ provides the constraints
on the set of cosmological parameters
more sensitive to the gas fraction distribution, namely
$\Omega_{\rm m}$ and $\Omega_{\Lambda}$ (or $w$).

\begin{figure}
\begin{center}
  \hbox{
  \includegraphics[height=.25\textheight]{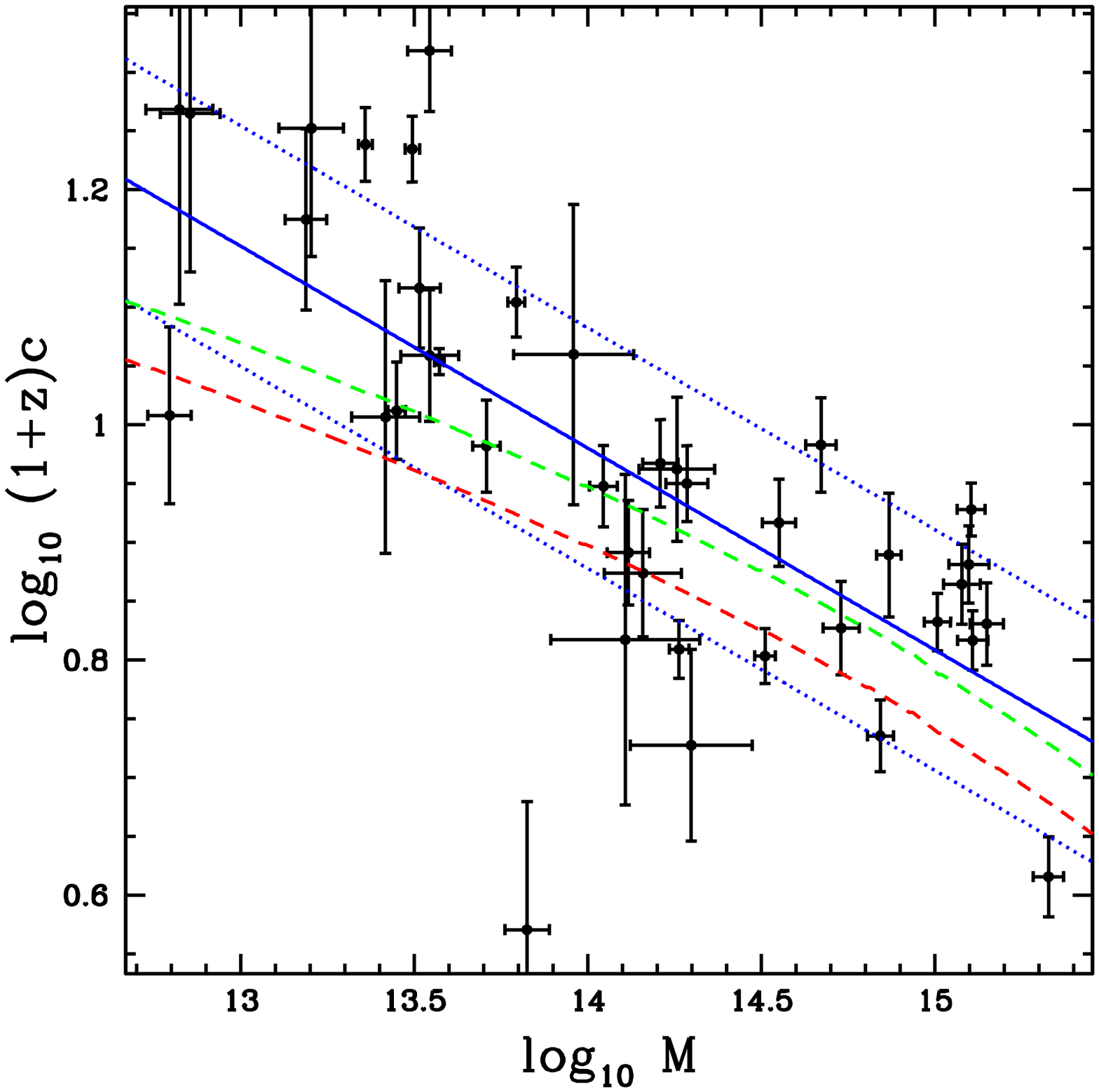}
  \includegraphics[height=.25\textheight]{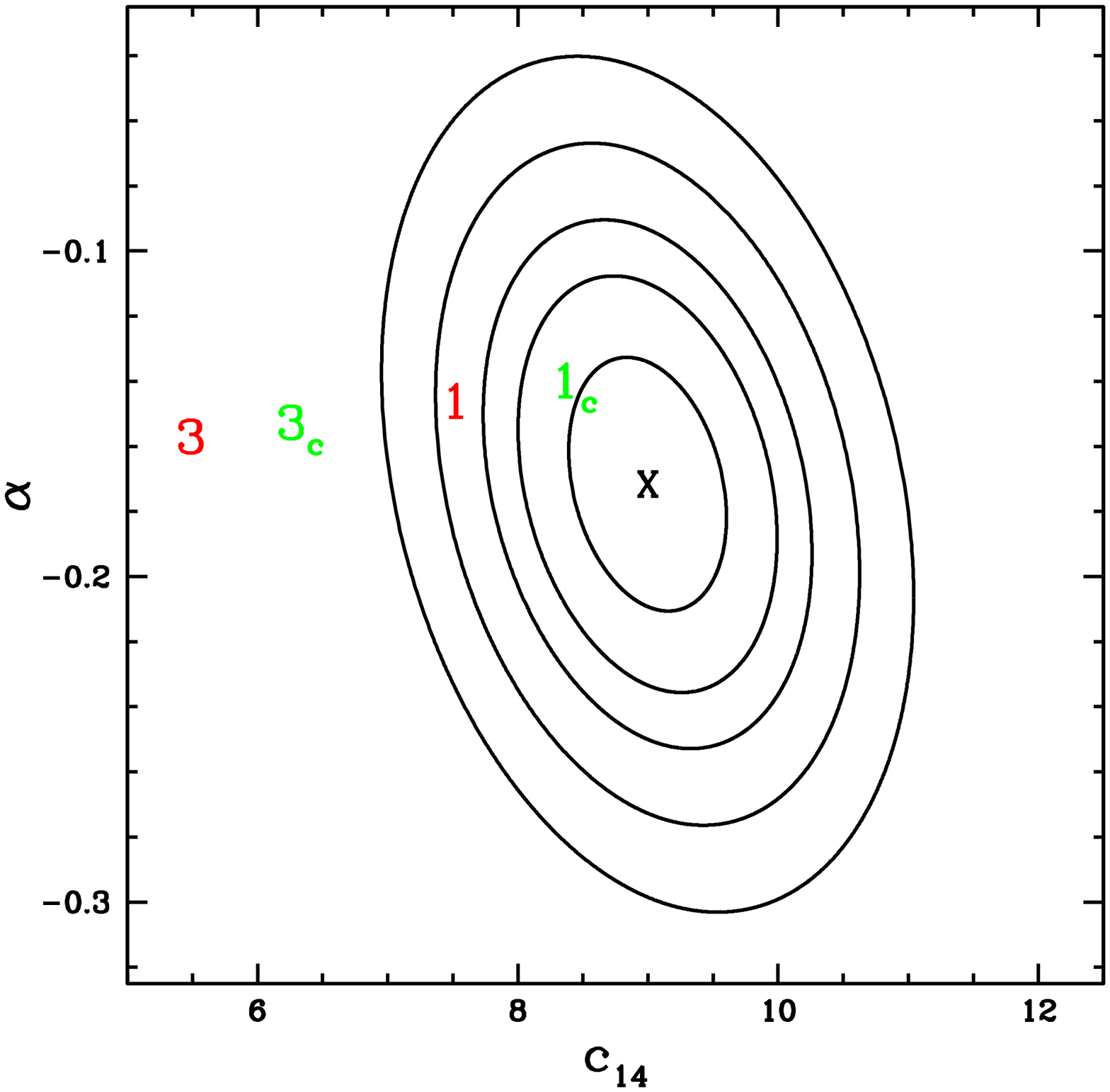}
  }
 \end{center}
 \vspace*{-.7cm}
 \begin{center}
  \hbox{
  \includegraphics[height=.3\textheight]{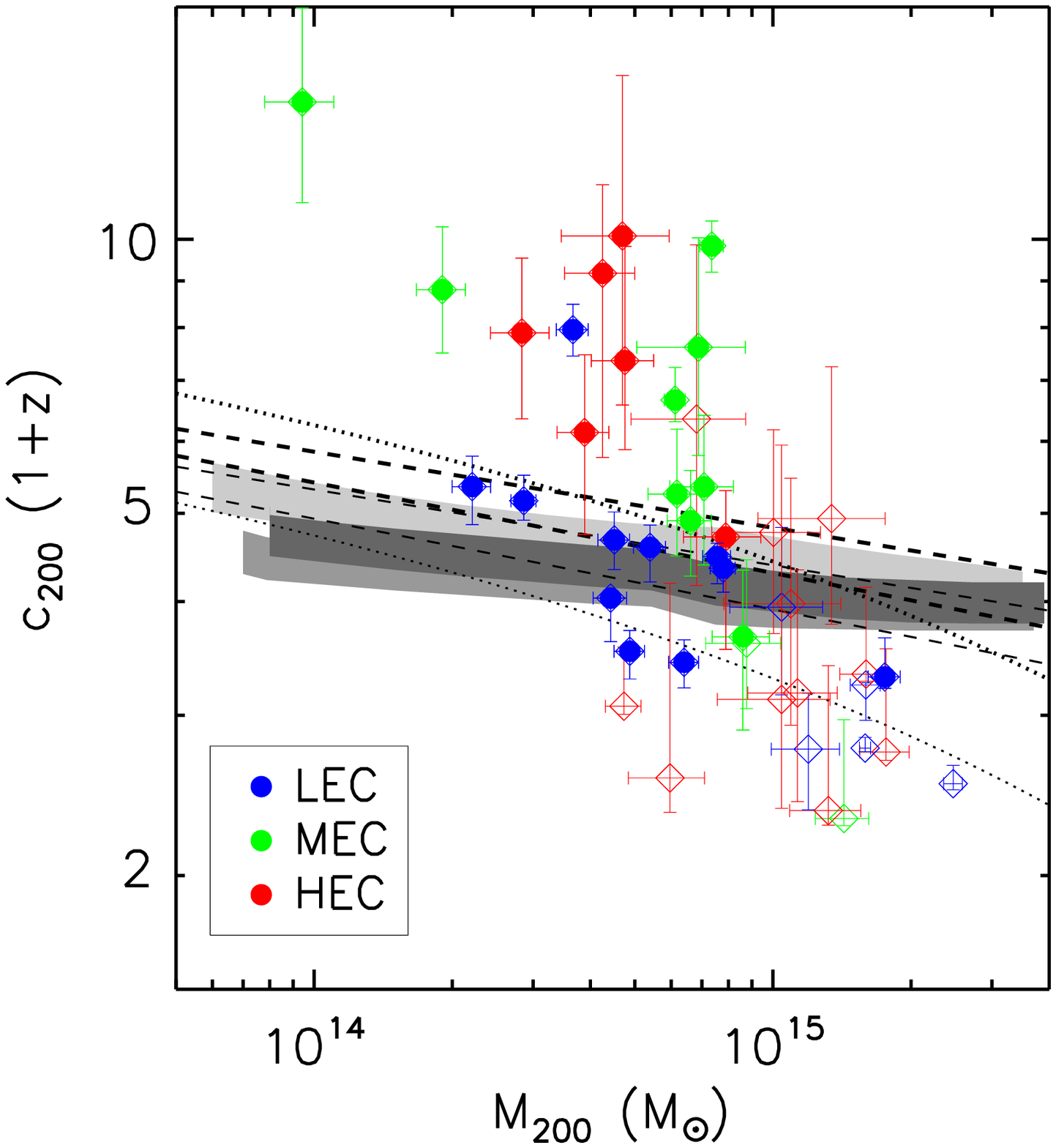}
  \includegraphics[height=.3\textheight]{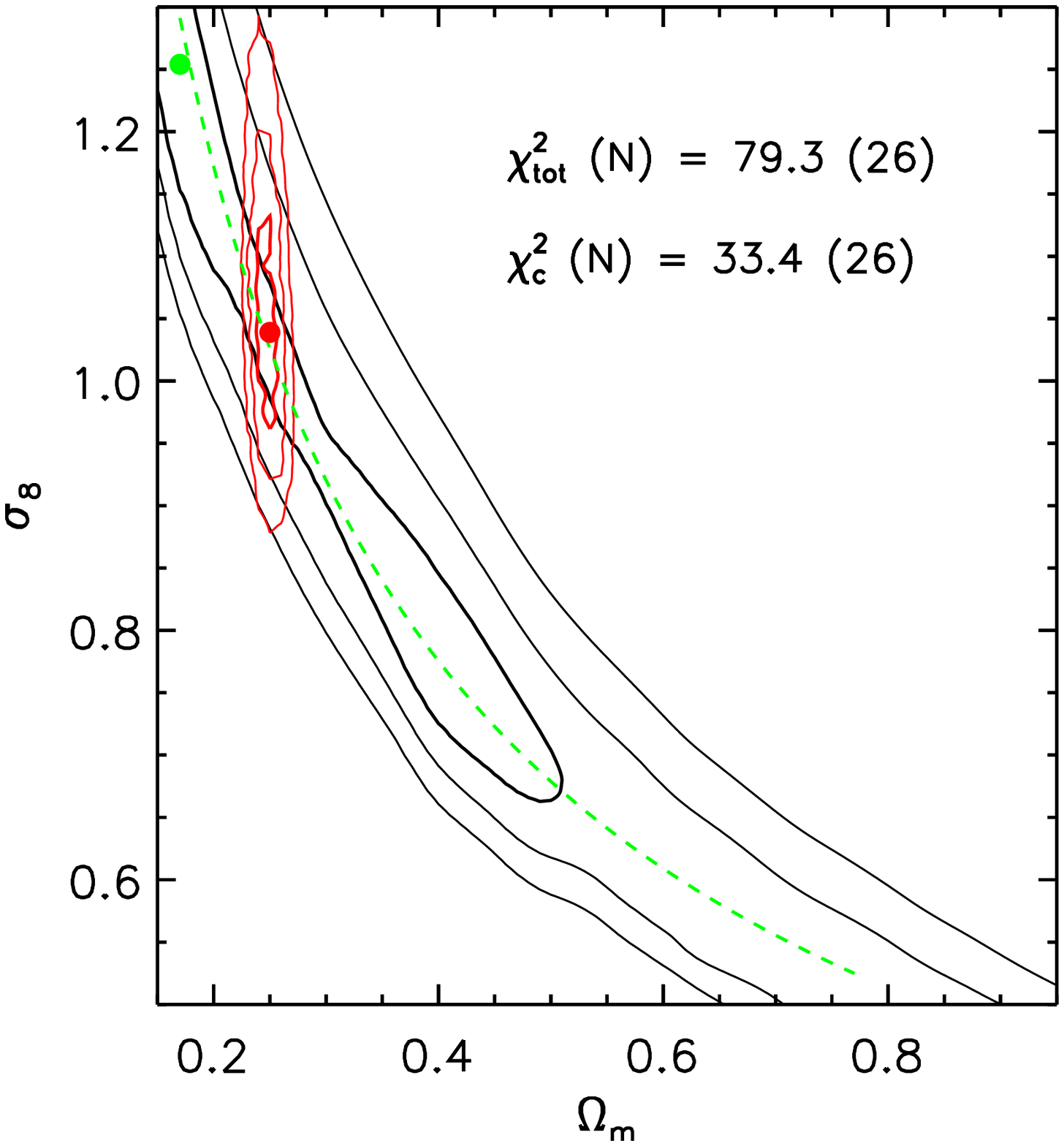}
  }
 \end{center}
 \vspace*{-0.7cm}
\caption{({\bf Top}; from Buote et al. 2007)
{\it (Left)} $c-M$ relation for the sample of 39 galaxy systems analyzed in Buote et al. (2007).
The lines indicate the best-fit result (solid), the intrinsic scatter (dotted) and the predictions from 
different  $\Lambda$CDM models.
{\it (Right)} Best-fitting estimates (black X) and confidence contours (68\%, 95\%, 99\%, 99.9\%, and 99.99\%) 
for the best-fit parameters of the relation $c = c_{14} / (1+z) \times (M/M_{14})^{\alpha}$ with, overplotted, 
the expected values from different cosmological models.
({\bf Bottom}; from Ettori et al. 2010) 
{\it (Left)} Data in the plane $(c_{200}, M_{200})$ used to constrain the cosmological
parameters $(\Omega_{\rm m}, \sigma_8)$. The colour-code refers
to the dynamical state of the clusters accordingly to the level of entropy measured in their cores
(from LEC = {\it low entropy core}, more relaxed systems to HEC = {\it high entropy core}, more disturbed, objects,
with MEC = {\it medium entropy core} clusters lying between the two).
The dotted lines show the predicted relations from Eke et al. (2001)
for a given $\Lambda$CDM cosmological model at $z=0$ (from top to bottom:
$\sigma_8=0.9$ and $\sigma_8=0.7$).
The shaded regions show the predictions in the redshift range $0.1-0.3$
for an assumed cosmological model in agreement with \wmap\ 1,
5 and 3 year (from the top to the bottom, respectively; 
see footnote~4 for the cosmological values associated to these datasets) from
Bullock et al. (2001; after Macci\`o et al. 2008).
The dashed lines indicate the best-fit range at $1\sigma$ obtained for
relaxed halos in a \wmap\ 5-year cosmology from Duffy et al.
(2008; thin lines: $z=0.1$, thick lines: $z=0.3$).
{\it (Right)}
Cosmological constraints in the $(\Omega_{\rm m}, \sigma_8)$ plane
obtained from equations~\ref{eq:chi2c} and \ref{eq:chi2f}
by using predictions from the model by Eke et al. (2001).
The confidence contours at $1, 2, 3 \sigma$ on 2 parameters
(solid contours) are displayed.
The combined likelihood with the probability distribution provided from the
cluster gas mass fraction method is shown in red.
The dashed green line indicates the power-law fit
$\sigma_8 \, \Omega_{\rm m}^{0.6} = 0.45$.
} \label{fig:cm} \end{figure}

\subsection{The concentration-mass relation}

Within a CDM model of the Universe, the $N-$body simulations of structure formation 
indicate that dark matter halos aggregate with a typical mass density profile
characterized by only 2 parameters, the concentration $c$ and 
the scale radius $r_{\rm s}$ (e.g. Navarro, Frenk \& White 1997).
The product of these two quantities fixes the radius within which the mean cluster
density is 200 times the critical value at the cluster's redshift
[i.e. $R_{200} = c_{200} \times r_{\rm s}$ and the cluster's volume
$V = 4/3 \pi R_{200}^3$ is equal to $M_{200} / (200 \rho_{c,z})$,
where $M_{200}$ is the cluster gravitating mass within $R_{200}$].
With this prescription, the structural properties of DM halos from galaxies to
galaxy clusters are dependent on the halo mass, with systems at higher masses
less concentrated. Moreover, the concentration depends upon the properties
of the cosmological background at the assembly redshift 
(e.g. Bullock et al. 2001, Neto et al. 2007 and reference therein),
which happens to be later in cosmologies with lower matter density,
$\Omega_{\rm m}$, and lower normalization of the linear power spectrum
on scale of $8 h^{-1}$ Mpc, $\sigma_8$. Under these conditions of formation, 
less concentrated DM halos at given mass are expected.
The concentration -- mass relation, and its evolution in redshift, is therefore
a strong prediction obtained from CDM simulations of structure formation
and is quite sensitive to the assumed cosmological parameters
(NFW; Bullock et al. 2001; Eke, Navarro \& Steinmetz 2001;
Dolag et al. 2004; Neto et al. 2007; Macci\`o et al. 2008).
In this context, NFW, Bullock et al. 2001 (with revision after
Macci\`o et al. 2008) and Eke et al. 2001 have provided simple
and powerful models that match the predictions from numerical
simulations and allow comparison with the observational measurements.

Recent X-ray studies (Pointecouteau et al. 2005; Vikhlinin et al. 2006;
Voigt \& Fabian 2006; Zhang et al. 2006)
have shown good agreement between observational constraints at low
redshift and theoretical expectations.
By fitting 39 systems in the mass range between early-type galaxies up to
massive galaxy clusters, Buote et al. (2007; see fig.~\ref{fig:cm}) 
confirm with high significance that the concentration decreases with increasing mass, as predicted
from CDM models, and require a $\sigma_8$, the dispersion of the mass
fluctuation within spheres of comoving radius of 8 $h^{-1}$ Mpc,
in the range $0.76-1.07$ (99\% confidence), definitely in tension with
the lower constraints obtained, for instance, at that time from the analysis
of the \wmap\ data (3-year results in Spergel et al. 2007; as a further example
of the complementarity of the cosmological constraints via independent 
methods see also the effect of the WMAP 3-year results 
on the galaxy cluster X-ray luminosity-gravitational mass relation
in Reiprich 2006)\footnote{We quote here the best-fit values for $(\Omega_m h^2, \sigma_8)$,
with relative uncertainty at 68\% confidence levels,
for the power-law flat $\Lambda$CDM model using the different releases of WMAP data only:
1-year (Spergel et al. 2003): $(0.14\pm0.02, 0.9 \pm 0.1)$; 
3-year (Spergel et al. 2007): $(0.128 \pm 0.008, 0.76 \pm 0.05)$; 
5-year (Komatsu et al. 2009): $(0.1326 \pm 0.0063, 0.796\pm0.036)$; 
7-year (Komatsu et al. 2011): $(0.1345 \pm 0.0056, 0.811\pm0.030)$; 
9-year (Hinshaw et al. 2013): $(0.1364 \pm 0.0045, 0.821 \pm 0.023)$.
}. 
Since they are based upon a selection of the most relaxed systems, 
these results assumed a 10\% upward early formation bias in the 
concentration parameter for relaxed halos.
Using a sample of 34 massive, dynamically relaxed galaxy clusters
resolved with \chandra\ in the redshift range $0.06-0.7$,
Schmidt \& Allen (2007) highlight a possible tension between the observational
constraints and the numerical predictions, in the sense that either 
the relation is steeper than previously expected or some redshift evolution 
has to be considered.
Comerford \& Natarajan (2007) compiled a large dataset of observed
cluster concentration and masses, finding a normalization
higher by at least 20 per cent than the results from simulations.
In the sample, they use also strong lensing measurements of the concentration
concluding that these are systematically larger than the ones
estimated in the X-ray band, and 55 per cent higher, on average, than
the rest of the cluster population.
Recently, Wojtak \& {\L}okas (2010) analyze kinematic data of 41 nearby ($z<0.1$)
relaxed objects and find a normalization of the concentration -- mass relation
fully consistent with the amplitude of the power spectrum $\sigma_8$
estimated from \wmap\ 1-year data and within $1 \sigma$ from 
the constraint obtained from \wmap\ 5-year.
Ettori et al. (2010) recover the total and gas mass profiles for a sample of 
44 X-ray luminous galaxy clusters located in the redshift range $0.1 - 0.3$, 
to constrain the cosmological parameters
$\sigma_8$ and $\Omega_{\rm m}$ through the analysis of the measured distribution
of $c_{200}$, $M_{200}$ and baryonic mass fraction in the mass
range above $10^{14} M_{\odot}$.
This dataset allows to resolve the temperature profiles up to about $0.6-0.8 R_{500}$
and the gas density profile, obtained from the geometrical deprojection
of the PSF--deconvolved surface brightness, up to a median radius of $0.9 R_{500}$.
Beyond this radial end, the estimates are the results of an
extrapolation obtained by imposing a NFW profile for the total mass
and different functional forms for $M_{\rm gas}$.
They estimate a dark ($M_{\rm tot} - M_{\rm gas}$) mass within $R_{200}$
in the range (1st and 3rd quartile) $4-10 \times 10^{14} M_{\odot}$,
with a concentration $c_{200}$ between 2.7 and 5.3, and a gas mass fraction
within $R_{500}$ between 0.11 and 0.16.

The $c_{200}-M_{200}$ relation is constrained to have a normalization
$c_{15} = c_{200} \times (1+z) \times \left( M_{200}/10^{15} M_{\odot} \right)^{-B}$
of about $2.9-4.2$ and a slope $B$ between $-0.3$ and $-0.7$ (depending on the
methods used to recover the cluster parameters and to fit the linear
correlation in the logarithmic space),
with a relative error of about 5\% and 15\%, respectively.
Once the slope is fixed to the expected value of $B=-0.1$, the normalization,
with estimates of $c_{15}$ in the range $3.8-4.6$, agrees with
results of previous observations and simulations for calculations
done assuming a low density Universe.
A total scatter in the logarithmic space of about 0.15 is measured
at fixed mass. This value decreases to 0.08 when the subsample
of clusters more dynamically relaxed and hosting a cooling core 
is considered. For this sample, a slightly lower normalization and
flatter distribution are measured.
This is consistent in a scenario where disturbed systems have an
estimated concentration through the hydrostatic equilibrium equation
that is biased higher (and with larger scatter)
than in relaxed objects up to a factor of 2 due to the action of the ICM motions
(see e.g. Lau et al. 2009).

To constrain the cosmological parameters of interest, $\sigma_8$ and $\Omega_{\rm m}$,
Ettori et al. (2010) calculate first the concentration $c_{200, ijk} =
c_{200}(M_i, \Omega_{\rm m, j}, \sigma_{8, k})$
predicted from the model investigated at each cluster redshift
for a given grid of values in mass, $M_i$,
cosmic density parameter, $\Omega_{\rm m, j}$,
and power spectrum normalization, $\sigma_{8, k}$.
Then, (i) a new mass $M_{200, j}$ and concentration $c_{200, j}$ 
are estimated for a given $\Omega_{\rm m, j}$; 
(ii) a linear interpolation on the theoretical prediction of $c_{200, ijk}$
is performed to associate a concentration $\hat{c}_{200, jk}$ to the new mass $M_{200, j}$
for given $\Omega_{\rm m, j}$ and $\sigma_{8, k}$;
(iii) the merit function $\chi_c^2$ is evaluated
\begin{equation}
\chi_c^2 = \chi_c^2(\Omega_{\rm m, j}, \sigma_{8, k}) =
\sum_{{\rm data}, i} \frac{\left(c_{200, i} - \hat{c}_{200, jk}\right)^2}
{\epsilon_{200, i}^2 +\sigma_c^2},
\label{eq:chi2c}
\end{equation} 
where $\epsilon_{200, i}$ is the $1 \sigma$ uncertainty related to the measured $c_{200, i}$
and $\sigma_c$ is the scatter intrinsic to the mean predicted value $\hat{c}_{200, jk}$
as evaluated in Neto et al. (2007).

To represent the observed degeneracy in the ($\sigma_8, \Omega_{\rm m}$) plane,
a power-law fit $\sigma_8 \, \Omega_{\rm m}^{\gamma} = \Gamma$ can be adopted
to obtain $\gamma = 0.60 \pm 0.03$ and $\Gamma=0.45\pm0.02$ (at $2\sigma$ level).
By using the gas mass fraction method described in the previous section, this degeneracy can be 
further broken to measure $\sigma_8 = 1.0 \pm 0.2$ and $\Omega_{\rm m}=0.26 \pm 0.01$
(at $2\sigma$ level; statistical only and for relaxed objects; see Fig.~\ref{fig:cm}).

However, because a calibration is needed in mapping the observed distribution of the 
concentrations with the expected one, for a given mass, as a function of
$\sigma_8$, $\Omega_{\rm m}$ and redshift, 
they have also noted how the cosmological constraints depend upon the models adopted 
to relate the properties of a DM halo to the background cosmology.
In particular, to make this technique more reliable and robust,
$N-$body simulations produced with different input cosmological parameters over cosmological volumes 
large enough to sample massive ($>10^{14} M_{\odot}$) DM halos are required.

\begin{figure*}
\begin{center}
\hbox{
 \includegraphics[height=.3\textheight]{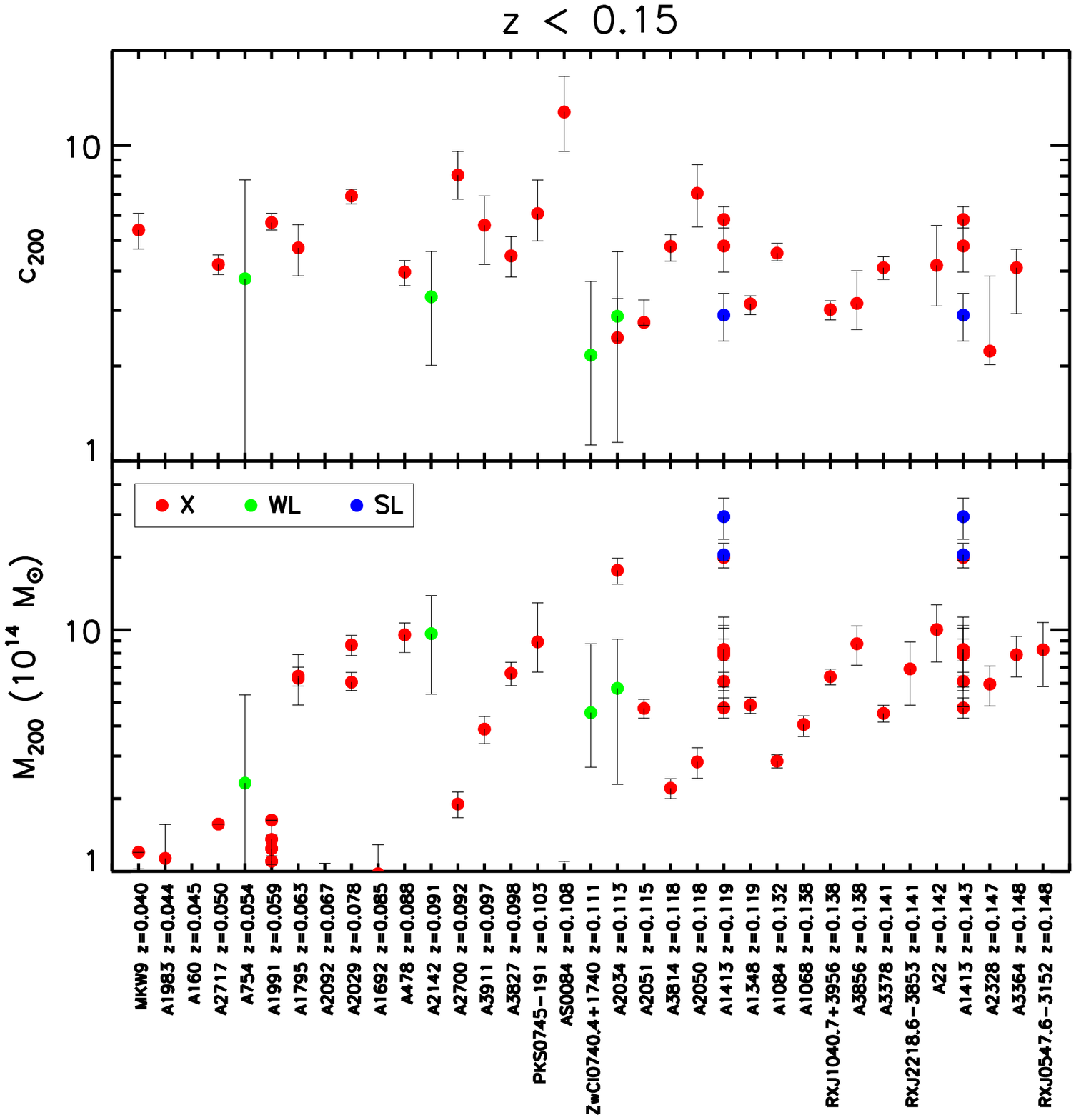}
 \includegraphics[height=.3\textheight]{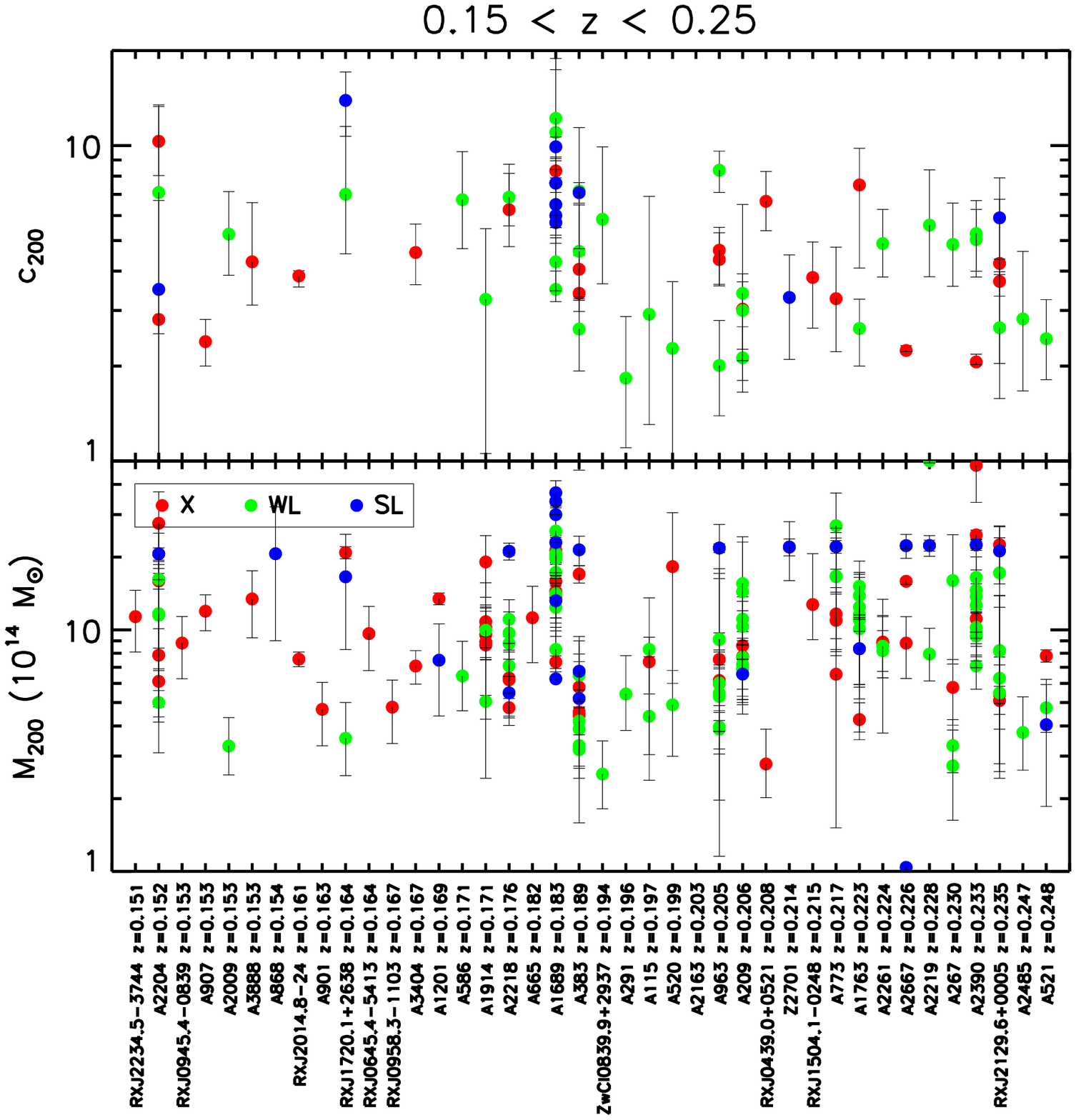}
  }
\hbox{
 \includegraphics[height=.3\textheight]{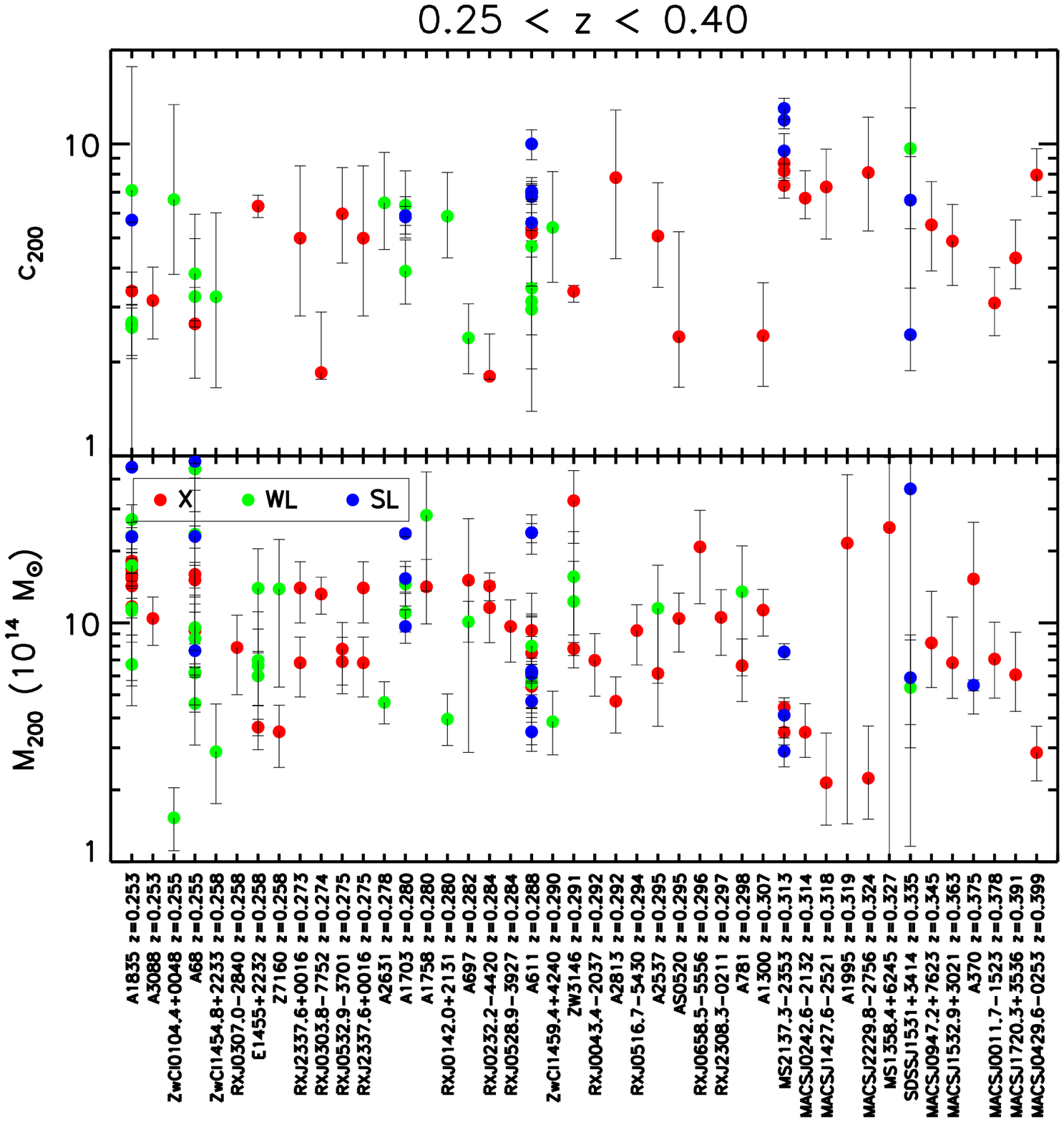}
  \includegraphics[height=.3\textheight]{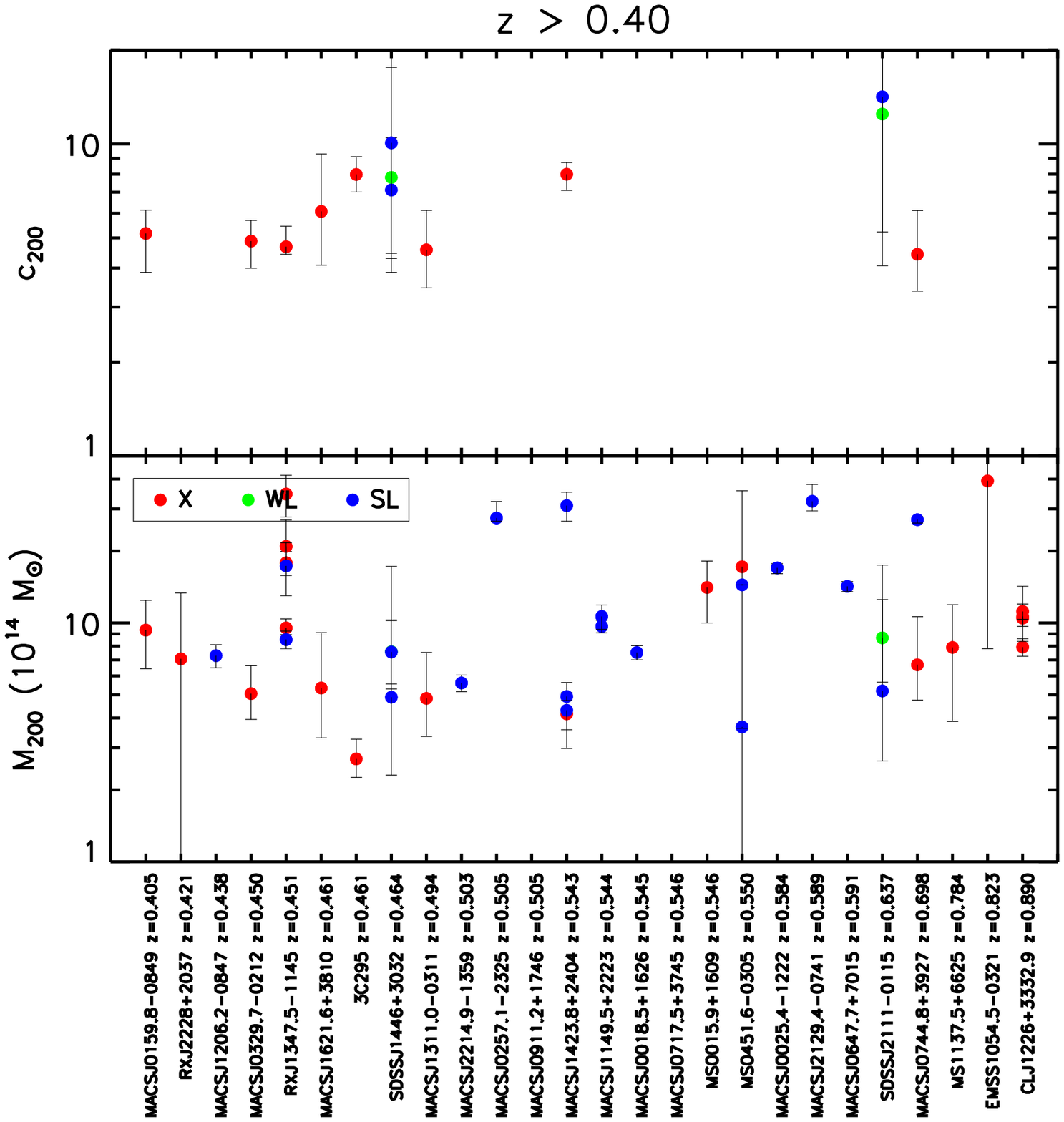}
  } 
\includegraphics[height=.4\textheight]{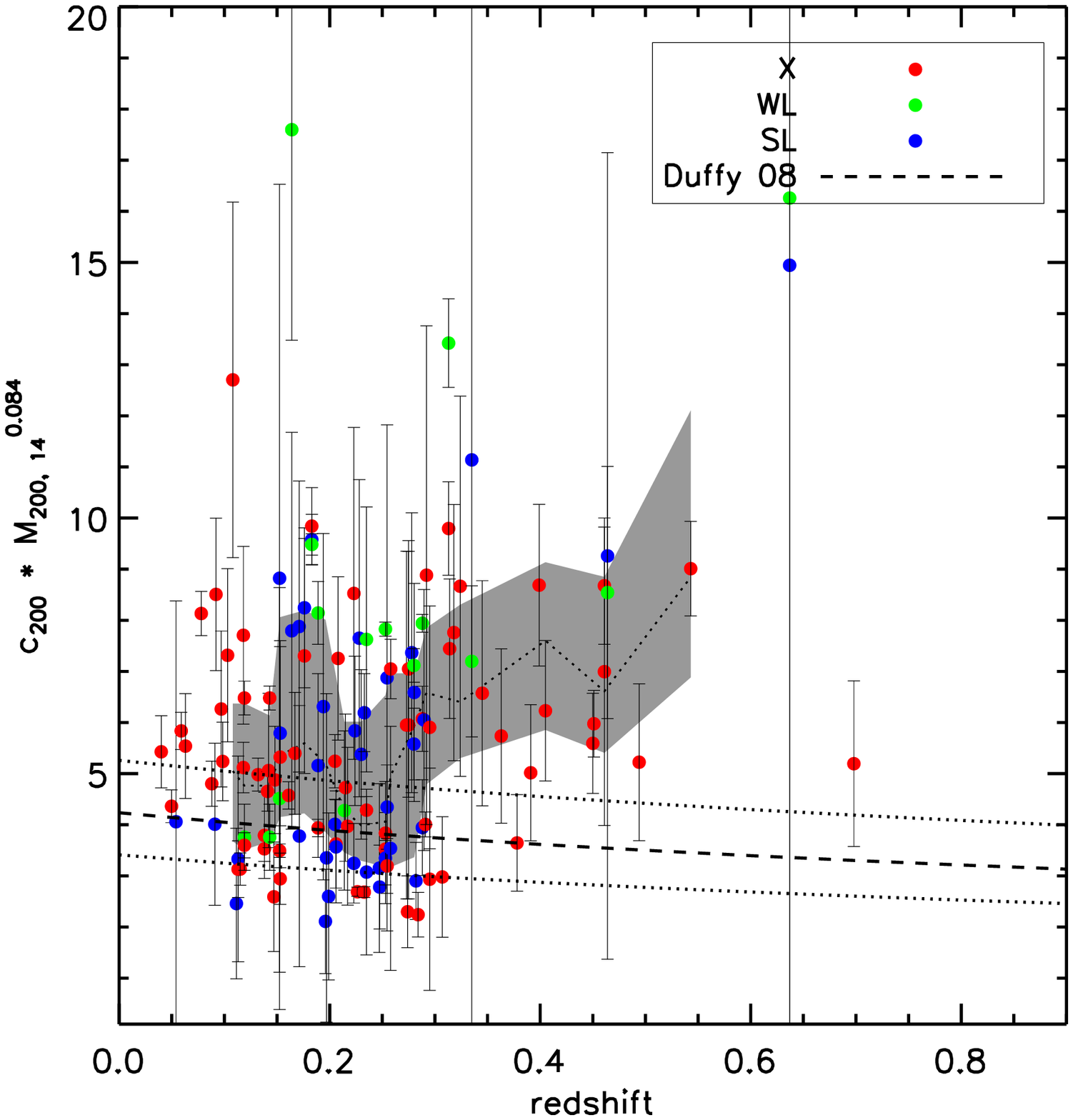}
\end{center}
\caption{Estimates of concentration and mass obtained through X-ray, strong and weak lensing analysis and available in literature since 2005.
The top 4 panels collect the measurements in 4 redshift bins ($z<0.15$, $0.15 < z < 0.25$, $0.25< z < 0.4$, $z>0.4$) and contain 35, 40, 44 and 26 clusters,
respectively.
The panel at the bottom shows the distribution of the published values of $c_{200}$ and $M_{200}$ as a function of the clusters' redshift.
These observed values are compared with the expected distribution from numerical simulations in Duffy et al. (2008) and relative scatter ({\it dashed} and {\it dotted} lines). The shaded region indicates the lower/upper quartiles computed in redshift's bins with 24 measurements and after excluding recursively the 8 at at the lowest redshifts. } 
\label{fig:cmz}
\end{figure*}

\subsubsection{A collection of estimates of concentration and mass through X-ray and lensing techniques}

We present in figure~\ref{fig:cmz} a collection of measurements of the values of concentration and mass 
as estimated through X-ray and lensing techniques since the year 2005.
Descriptions of the mass reconstruction methods through gravitational lensing distortion of the optical light from background galaxies around a galaxy cluster are presented in Hoekstra et al., Meneghetti et al., Bartelmann et al. in the present volume.

\begin{figure}
\includegraphics[width=0.5\columnwidth]{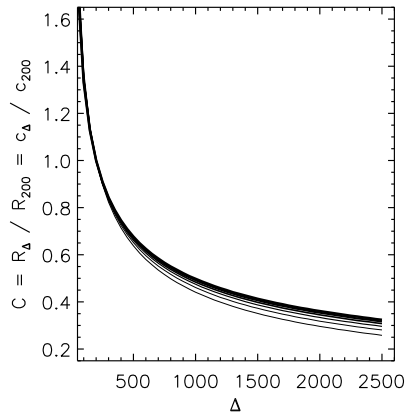}
\caption{
Fraction of $R_{200}$ mapped within a given overdensity $\Delta$
for an assumed NFW profile with $c_{200}$ in the range 3--8 (from the thinnest to the thickest line)
} \label{fig:cnfw} 
\end{figure}

To convert these estimates to a common value of the overdensity, we proceed in the following way.
The total mass within a given overdensity $\Delta$ with respect to the critical density of the Universe 
is described by eq.~\ref{eq:mdelta} and implies that $M_{\Delta} / (R_{\Delta}^3 \Delta)$ is constant. 
By definition of the NFW mass profile, the radius $R_{\Delta}$ of the spherical region that encloses $M_{\Delta}$, 
the concentration $c_{\Delta}$ and the scale radius $r_{\rm s}$ 
of the mass distribution are related by the equation $R_{\Delta} = c_{\Delta} r_{\rm s}$.
Hereafter, we assume as overdensity of reference the value $\Delta=200$.
Then, we can write
\begin{equation}
\frac{M_{\Delta}}{M_{200}} =  \frac{c_{\Delta}^3}{c_{200}^3} \frac{\Delta}{200} = C^3  \frac{\Delta}{200} , 
\label{eq:mdelta_c}
\end{equation}
where $C = c_{\Delta} / c_{200}$ and the values of concentrations are related through the 
NFW mass density profile
\begin{equation}
C^3 \frac{\Delta}{200} = \frac{\ln (1+c_{200} C) - c_{200} C /(1+c_{200} C) }
{\ln (1+c_{200})-c_{200}/(1+c_{200})}.
\label{eq:rdelta}
\end{equation}
This function is monotonic and can be easily solved numerically to estimate
$C$, that is a quantity that depends mostly on $\Delta$ and only marginally on the guessed $c_{200}$ (see fig.~\ref{fig:cnfw}).
For instance, for $\Delta = 178 \Omega_{\rm m, z}^{0.45}$, which
indicates the virial overdensity predicted from the spherical collapse model
in a flat Universe with a contribution from dark energy (Eke et al. 2001),
$C = 1.34$ and $1.11$ at $z=0$ and $z=1$, respectively, for $\Omega_{\rm m}=0.3$ and $c_{200}=4$,
with deviations within 2\% in the range of $c_{200} = 3-8$.
In case the overdensity is referred to the background density of the
Universe, $\rho_b = \Omega_{\rm m, z} \, \rho_{c, z}$, it is straightforward
to correct $\Delta$ by $\Omega_{\rm m, z} = \Omega_{\rm m} (1+z)^3 / H_z^2$
to recover the definition in equation~\ref{eq:mdelta}.

Thus, equation~\ref{eq:mdelta} and the value of $C$ as obtained from equation~\ref{eq:rdelta} are used
to convert from $\left\{ c_{\Delta}, M_{\Delta} \right\}$ to $\left\{ c_{200}, M_{200} \right\}$.

In Fig.~\ref{fig:cmz}, we present the collection of about 400 measurements obtained for
145 single objects in the redshift range 0.04--0.89. Eighty per cent of these values 
lie at intermediate redshift (0.1--0.5). About 47 per cent originates from X-ray analysis, 
21 and 32 per cent from strong and weak lensing technique, respectively.
By using a Bayesian linear regression method (coded in the {\it LINMIX\_ ERR} IDL routine; Kelly 2007) to fit the distribution plotted in Fig.~\ref{fig:cmz}, in which the predicted mass dependence of the $c-M$ relation from N-body simulations is assumed, we measure $c_{200} \times M_{200}^{0.084} = 4.58 (\pm 0.44) \times (1+z)^{0.78 \pm 0.44}$.
However, given the heterogeneous origin of these quantities, we do not speculate further on, e.g., some tension between 
their distribution in the concentration-mass-redshift plane and the one predicted from numerical simulations at $z>0.3$ or 
on the comparison between mass and concentration as estimated for the same object with different methods.

We postpone a more detailed analysis of the best constraints on the distribution of the values in the concentration-mass-redshift plane 
in a dedicated forthcoming study.

\section{Conclusions}

Galaxy clusters form through the hierarchical accretion of cosmic matter.
The end products of this process are virialized structures that
feature, in the X-ray band, similar radial profiles of the surface brightness
$S_{\rm b}$ (e.g. Croston et al. 2008, Eckert et al. 2012)
and of the plasma temperature $T_{\rm gas}$
(e.g. Allen et al. 2001, Vikhlinin et al. 2005, Leccardi \& Molendi 2008a).

In recent years, measurements of the spatially-resolved X-ray properties
of galaxy clusters have definitely improved thanks to the arcsec resolution and 
large collecting area of the present X-ray satellites, like \chandra\ and \xmm.
For instance, central luminous regions have shown 
{\it fronts}, i.e. sharp contact discontinuities between 
regions of gas with different densities.
The classic {\it bow shocks} are driven by infalling subclusters.
The {\it cold fronts} are found in mergers as well as around 
the central density peaks in relaxed clusters and
are caused by motion of cool, dense gas clouds in the ambient 
higher-entropy gas. These clouds are either remnants of the infalling 
subclusters, or the displaced gas from the cluster's own cool cores
(see, e.g., review in Markevitch \& Vikhlinin 2007).
More difficult is to characterize properly the cluster outskirts where 
the X-ray surface brightness is comparable to the level of the fore/background.
Noticeable progress has occurred recently thanks to \suzaku\ exposures, 
also in combination with revised analyses of \rosat\ PSPC observations and
data of the Sunyaev-Zeldovich signal provided from \planck\
(see Reiprich et al. in the present volume).

To facilitate comparisons on the properties of the mass determinations,
we suggest the adoption of a {\it checklist} to be followed during the analysis and
production of new estimates based on X-ray observations:
\begin{itemize}
\item quote the adopted background cosmology
\item use the latest calibration files for the data reduction and specify them in the text
\item mask point-sources in the soft X-ray image both for spectral and spatial analysis. This is particularly relevant at high redshift, where the combined effects of the dimming of the X-ray surface brightness and the observed increases in the population of X-ray bright AGNs in clusters' regions (e.g. Martini et al. 2009) could biases significantly the measured quantities. For example, Branchesi et al. (2007) quantify the contamination on the gas temperature and luminosity estimates in an increase by about 13 and 17 per cent, that becomes of 24-22 per cent respectively at $z>0.7$.
\item mask clumps and mergers localized in the soft X-ray image both for spectral and spatial analysis;
remind that hydrostatic equilibrium holds locally, i.e. look for relaxed regions also in merging systems
\item define and quote the X-ray center (for instance, the peak or the centroid
in the X-ray image) used for both spatial and spectral analysis
\item define the binning adopted to construct the spectra and the surface brightness profiles
\item describe in a reproducible way the strategy adopted for the background correction (e.g. subtraction of a background determined locally --i.e. in the same field of the target--; subtraction of a background estimated from other deep blank fields; modelling of the spectral emission of the background)
\item prefer a C statistic in the spectral fit, in particular when the counts statistic is low,
 and quote the goodness of the spectral and spatial fits
\item apply both a PSF correction -if required- and a deprojection (or projection of your model) of the temperature profile
\item several methods are available to solve the HEE (see Section~\ref{sect:mass}):  they have different pros and cons, but they do not appear affected from systematic differences (see e.g. Meneghetti et al. 2010) and, more importantly, they are not missing any contribution from the thermalized ICM
\item quote the values of the measured mass with the relative statistical
 and systematic errors and the physical radius at which these values are estimated
\item try to avoid any strong extrapolation and, in any case, state clearly a measurement done on actually measured profiles and, if the estimates are extrapolated, where is the observational limit.
\end{itemize}


\begin{acknowledgements}
Part of this work was done during visits to the International Space
Science Institute (ISSI) in Bern and we acknowledge ISSI's hospitality.
We thank Monique Arnaud, Silvano Molendi and Gabriel Pratt for their comments.
SE and AD acknowledge the financial contribution from contracts ASI-INAF
I/023/05/0 and I/088/06/0.
THR acknowledges support by the German Research Association
(DFG) through Heisenberg grant RE 1462/5 and grant RE 1462/6.
\end{acknowledgements}

%
%
%

\newcommand{\noopsort}[1]{}

\end{document}